\documentclass[a4paper,11pt]{article}
\usepackage[a4paper,top=3.5cm,bottom=3cm,left=3cm,right=3cm]{geometry}

\usepackage{natbib}
\usepackage{setspace}

\usepackage[utf8]{inputenc} % allow utf-8 input
\usepackage[T1]{fontenc}    % use 8-bit T1 fonts
\usepackage{hyperref}       % hyperlinks
\usepackage{url}            % simple URL typesetting
\usepackage{booktabs}       % professional-quality tables
\usepackage{amsfonts,amsmath,amssymb}       % blackboard math symbols
\usepackage{amsthm}
\usepackage{nicefrac}       % compact symbols for 1/2, etc.
\usepackage{microtype}      % microtypography
\usepackage{algorithm,algorithmic}
\usepackage{graphicx}
\usepackage{pdflscape}
\usepackage{longtable}

\usepackage{latexsym}

\usepackage{color}

\newcommand{\bs}{\boldsymbol}
\newcommand{\indep}{\perp \!\!\! \perp}

\newtheorem{definition}{Definition}
\newtheorem{proposition}{Proposition}

\newtheorem{lemma}{Lemma}

\usepackage{lineno,hyperref}
\newcommand*\patchAmsMathEnvironmentForLineno[1]{
  \expandafter\let\csname old#1\expandafter\endcsname\csname #1\endcsname
  \expandafter\let\csname oldend#1\expandafter\endcsname\csname end#1\endcsname
  \renewenvironment{#1}
     {\linenomath\csname old#1\endcsname}
     {\csname oldend#1\endcsname\endlinenomath}}
\newcommand*\patchBothAmsMathEnvironmentsForLineno[1]{
  \patchAmsMathEnvironmentForLineno{#1}
  \patchAmsMathEnvironmentForLineno{#1*}}
\AtBeginDocument{
\patchBothAmsMathEnvironmentsForLineno{equation}
\patchBothAmsMathEnvironmentsForLineno{align}
\patchBothAmsMathEnvironmentsForLineno{flalign}
\patchBothAmsMathEnvironmentsForLineno{alignat}
\patchBothAmsMathEnvironmentsForLineno{gather}
\patchBothAmsMathEnvironmentsForLineno{multline}
}
\modulolinenumbers[5]

\title{False discovery rate control for confounder selection using mirror statistics}
\author{
    Kazuharu Harada \thanks{Tokyo Medical University, Japan. e-mail: \texttt{haradak (at) tokyo-med.ac.jp}}
        \and
    Masataka Taguri \thanks{Tokyo Medical University, Japan. e-mail: \texttt{taguri (at) tokyo-med.ac.jp}}
    }
\date{\today}

\begin{document}
\maketitle

\begin{abstract}
While data-driven confounder selection requires careful consideration, it is frequently employed in observational studies. Widely recognized criteria for confounder selection include the minimal-set approach, which involves selecting variables relevant to both treatment and outcome, and the union-set approach, which involves selecting variables associated with either treatment or outcome. These approaches are often implemented using heuristics and off-the-shelf statistical methods, where the degree of uncertainty may not be clear. In this paper, we focus on the false discovery rate (FDR) to measure uncertainty in confounder selection. We define the FDR specific to confounder selection and propose methods based on the mirror statistic, a recently developed approach for FDR control that does not rely on p-values. The proposed methods are p-value-free and require only the assumption of some symmetry in the distribution of the mirror statistic. It can be combined with sparse estimation and other methods that involve difficulties in deriving p-values. The properties of the proposed methods are investigated through exhaustive numerical experiments. Particularly in high-dimensional data scenarios, the proposed methods effectively control FDR and perform better than the p-value-based methods.
\end{abstract}

% keywords can be removed
% \keywords{FDR \and Causal Inference \and Confounder Selection}

% \linenumbers
\section{Introduction}
In observational studies, it is necessary to adjust for confounding to estimate a causal effect of treatment without bias. Confounder selection is a procedure to choose variables that can cause confounding bias among many collected variables. In principle, it is recommended that confounder selection is based on subject-matter knowledge before data analysis. If we knew the true causal structure of all relevant variables, including unobserved ones, the theory of directed acyclic graphs (DAG) would inform us whether the causal effect can be identified and which variables should be adjusted for identification. However, as one can easily imagine, it is often difficult to know the complete causal structure of relevant variables. \citet{VanderWeele2019-qw} have reviewed criteria for confounder selection under incomplete knowledge of the causal structure. 
Typical approaches are based on two criteria: the common cause criterion, which selects direct causes of both outcome and treatment, and the disjunctive cause criterion \citep{VanderWeele2011-wo}, which selects direct causes of either outcome or treatment. While these criteria are not perfect in that they sometimes make incorrect recommendations for confounder selection depending on the true causal structure, they are still useful as they often provide a reasonable set of variables based on available knowledge.

Recently, however, even these compromised criteria are not always applicable, especially as large datasets, such as those from registries and omics studies, are increasingly in demand. High-dimensional potential confounders make it difficult for an analyst to have (partial) knowledge of the true causal structure.
Regarding efficiency in estimation, it is undesirable to use all available variables without examining the necessity for adjustment \citep{De_Luna2011-dx}. In practice, data-driven confounder selection is often adopted. Although data-driven confounder selection needs a somewhat strong assumption that the screened variable set does not contain variables that can introduce bias in the estimate, it is widely discussed and applied due to practical demands \citep{Brookhart2006-zk, Schisterman2009-cw, Patrick2011-ao, Witte2019-hr, VanderWeele2019-qw, Staerk2024-co}.

\citet{Witte2019-hr} reviewed various approaches currently used for data-driven confounder selection, discussing situations in which they are effective or fail. Table \ref{tab:approaches} briefly summarizes the popular approaches, which generally mimic the confounder selection criteria under partial knowledge. Various off-the-shelf statistical methods, including regression, tests, and sparse estimation, are used to measure statistical association. As will be discussed in detail in Section 4, many heuristics are used, but many of them lack guarantees regarding uncertainty in confounder selection, which motivates us to develop a method with theoretical preference. Data-driven confounder selection is also discussed in numerous articles, such as \citep{Robins1986-hb, Greenland2008-qw, Greenland2015-vu, Vansteelandt2012-pi}.
\begin{table}[ht]
    \centering
    \caption{Major approaches for statistical confounder selection. The names of the approaches follow \citet{Witte2019-hr}. Other approaches, the causal search and the change-in-estimate are omitted because they are less relevant to this paper.}
    \begin{tabular}{cc}
        Approaches & Included variables  \\ \hline
        Pre-treatment & All variables obtained temporally before the treatment. \\
        Outcome & Strong predictors of the outcome. \\
        Treatment & Strong predictors of the treatment. \\
        Minimal set & Strong predictors of both the outcome and the treatment. \\
        Union set & Strong predictors of either the outcome or the treatment.
    \end{tabular}
    \label{tab:approaches}
\end{table}

In this paper, we suggest the false discovery rate (FDR) control in confounder selection, especially in low- and high-dimensional settings. In the context of ordinary variable selection, the FDR is defined as the expected proportion of variables falsely identified to be relevant to the outcome among all selected variables. Controlling the FDR below the pre-specified level is crucial for reducing discoveries of non-reproducible statistical associations and is widely applied in exploratory studies \citep{Benjamini2010-wu}. 

For confounder selection, we first extend the FDR to the union-set and the minimal-set approaches. Specifically, the FDR for the union-set approach can be defined as the expected proportion of variables that are associated with neither the outcome nor the treatment among the selected variables. We define the FDR for confounder selection and introduce novel methods to ensure FDR control for the union-set and the minimal-set approaches. We apply {\it the mirror statistic}, which has been developed as a variable selection method with FDR control, to confounder selection. To the best of our knowledge, this is the first attempt to control FDR in confounder selection.

This paper is structured as follows: Section 2 outlines the problem of causal effect estimation and introduces the mirror statistic, which is a key concept for FDR control in this paper. In Section 3, we propose novel methods to control FDR for the union-set and the minimal-set approaches. Since there are two ways of construction, we present them separately. Section 4 introduces various methods of confounder selection associated with our proposed methods and compares them. Section 5 details exhaustive numerical experiments. In Section 6, we report on the application of our proposed method to a real-world dataset. Finally, Section 7 concludes the paper, summarizing our contributions and discussing the remaining issues.

% De_Luna2011-dx: 因果グラフの考察に基づき，最小の調整変数集合を同定するアルゴリズムを提案．実行には条件付独立性の検定を用いる（Marginal co-ordinate test, Dennis_Cook2004-we）．
% Patrick2011-ao: データ駆動型アプローチの有効性を示す．曝露だけに関連する変数の選択を避けるように明示．
% Brookhart2006-zk: PSモデルに含めるべき変数を指南．特に，Yに関連する変数を含め，Aにだけ関連する変数は外すよう指摘．

% Wang2012-jl: Bayesian Adjusting Congounder (BAC) を提案，ベイズモデル平均化（BMA）を拡張し，アウトカムモデルと結果モデルの共通性を事前オッズに組み込んだ.
% Wilson2014-nq: ベイズ回帰と信用区間を用いた変数選択？？
% Talbot2015-eq: Bayesian approach, BECC

% Robins1986-hb: 「古典」として目を通しておく
% Judkins2007-qc: 関係はしてるが見なくてよさそう
% Schneeweiss2009-jq: HDPS
% Vansteelandt2012-pi: 総論．ざっと読んでおきたい．
% Rolling2013-xg: 治療効果推定を目的とした情報量規準．純粋に読みたい．
% Schisterman2009-cw: overadjustmentの害について
% Greenland2008-qw: 重要っぽい
% Ertefaie2018-sa: GLiDeRの類っぽく見えるが？

% ##############################################################################################
% ##############################################################################################
% ##############################################################################################

% #################################################################################
\section{Preliminaries}
\subsection{Causal Effect Estimation}
Firstly, we introduce the causal effect and its estimation in observational studies along with Rubin's framework \citep{Rubin1974-zp, Holland1986-ut, Rubin2005-ly}, and state the problem in this paper. Suppose an independent individual $i$ receives the treatment $A_i$, and let $Y_i^{a}$ be the potential outcome under $A_i=a$. The observed outcome $Y_i$ is equal to $Y_i^{a}$ when $i$ actually receives $A_i=a$. For simplicity, we only consider a binary treatment, say $A_i\in\{0,1\}$. The average treatment effect (ATE) is defined as $E[Y_i^{1} - Y_i^{0}]$, and it is interpreted as the average difference in outcomes between the scenario where the entire population receives treatment and the scenario where it does not. In randomized studies, the assumption of unconfoundedness, $(Y_i^{1},Y_i^{0})\indep A_i$ for all $i$, is valid, and ATE is estimated by a simple difference of group average. In observational studies, in contrast, since the treatment is not assigned randomly, the conditional unconfoundedness, $(Y_i^{1},Y_i^{0})\indep A_i\mid X_i$ for all $i$ is often assumed. Then, ATE is identified using some auxiliary functions, such as the conditional expectation of the outcome or the propensity score, the conditional probability of treatment \citep{Rosenbaum1983-ny}. Popular estimators include the regression standardization
\begin{gather*}
    \frac{1}{n} \sum_{i=1}^n E[Y_i\mid A_i=1, X_i] - E[Y_i\mid A_i=0, X_i], 
\end{gather*}
the inverse probability weighting (IPW) estimator
\begin{gather*}
    \frac{1}{n} \sum_{i=1}^n \frac{A_iY_i}{P(A_i=1\mid X_i)} - \frac{(1-A_i)Y_i}{P(A_i=0\mid X_i)},
\end{gather*}
and the locally efficient augmented IPW estimator (only the estimator for the treatment group is shown) \citep{Bang2005-sc}
\begin{gather*}
    \frac{1}{n} \sum_{i=1}^n \frac{A_iY_i}{P(A_i=1\mid X_i)} - \frac{A_i - P(A_i=0\mid X_i)}{P(A_i=1\mid X_i)}E[Y_i\mid A_i=1, X_i]
\end{gather*}
The conditional expectation and the propensity scores are usually substituted by consistent estimators. 
There are enormous estimation methods such as matching and stratification, but we only focus on the above estimators for space limitation.

We assume generalized linear models without interactions to deliver the essence of the proposed methods:
\begin{gather}
    g_Y(E[Y \mid A, X]) = \beta_0 + \tau A + X^T{\bs\beta}, \label{eq:glm_Y} \\
    g_A(P(A=1 \mid X)) = \alpha_0 + X^T{\bs\alpha},  \label{eq:glm_A}
\end{gather}
where $g_Y$ and $g_A$ are link functions and $\tau, \beta_0, \bs\beta, \alpha_0, \bs\alpha$ are regression parameters. Assume that the potential confounders are properly screened based on the domain knowledge and that the entire $X$ is sufficient for the assumption of conditional unconfoundedness. Note that the parameter $\tau$ equals the ATE when $g_Y$ is the identity link, but not necessarily otherwise. Some elements of the true $\bs\beta$ and $\bs\alpha$ are zero, which means some elements of $X$ are not necessary to be adjusted. Our interest is to get a smaller subvector of $X$ to improve efficiency. In addition, the proposed methods aim to provide a reproducible set of confounders with controlled FDR.

\subsection{FDR Control via Mirror Statistics}
False discovery control is one of the popular policies in variable selection mainly for predictive purposes. 
The q-value based on \citet{Benjamini1995-jh} ({\it BHq}) is a method to control FDR by correcting p-values and is widely applied in exploratory studies in many fields like bioinformatics. Since BHq assumes that the p-values are independent of each other, the theoretical guarantee of FDR control is absent for the regression problems, in which estimated coefficients are not usually independent. The extended method proposed by \citet{Benjamini2001-zs} ({\it BYq}) relaxes the independence requirement. BHq and BYq depend on p-values, so it is necessary to approximate the distribution of the estimated regression coefficients or some relevant statistics corresponding to each predictor.

Recently, a novel class of FDR control called the mirror statistic has been developed. Before reviewing the individual methods, we introduce the fundamental idea of FDR control via the mirror statistic for a regression problem. Consider the problem of selecting variables that are truly relevant to the target variable out of the $p$-dimensional predictors. First, we define the mirror statistic.
\begin{definition}[Mirror statistic]\label{def:mirror}
    We call $M_j$ a mirror statistic corresponding to the $j$-th predictor when $M_j$ satisfies the following two properties:
    \begin{enumerate}
        \item $M_j$ is symmetrically distributed around 0 when the $j$-th predictor is not relevant to the target variable.
        \item Large positive value of $M_j$ reflects a strong association between the $j$-th predictor and the target variable.
    \end{enumerate}
\end{definition}
Let $S\subseteq [p]$ be the index set of the truly relevant variables, and let $\hat{S} \subseteq [p]$ be the selected set. Once given the set of mirror statistics, we append the $j$-th variable to $\hat{S}$ if $M_j > t$ for some positive constant $t$. By virtue of the properties of the mirror statistics, the false discovery proportion (FDP) is bounded as
\begin{gather*}
    \mathrm{FDP}(t) := \frac{\#\{j: M_j > t \wedge j\notin S\}}{\#\{j: M_j > t\}\vee 1} \approx
        \frac{\#\{j: M_j < -t \wedge j\notin S\}}{\#\{j: M_j > t\}\vee 1} \le
        \frac{\#\{j: M_j < -t\}}{\#\{j: M_j > t\}\vee 1},
\end{gather*}
where the first approximation is from the property 1 in Definition \ref{def:mirror}. Then, with the designated FDR level $q>0$, $t$ is selected so that
\begin{gather*}
    \frac{\#\{j: M_j < -t\}}{\#\{j: M_j > t\}\vee 1} < q.
\end{gather*}
For the estimated set $\hat{S}$ based on the mirror statistic, the FDR, the expectation of FDP, is controlled below $q$ \citep{Barber2015-lg,Dai2022-jk,Xing2023-mp,Dai2023-ju}.

% Figure \ref{fig:mirror1dim} illustrates the intuition that the mirror statistic can control FDR.

% \begin{figure}
%     \centering
%     \includegraphics[width=0.99\textwidth]{fig/illustration_mirror_1dim.png}
%     \caption{How the mirror statistic controls FDR. We select the variables with $M_j > t$. While the blue circles are correctly selected variables, the orange diamonds with $M_j > t$ are falsely discovered. Since $M_j$s are symmetrically distributed about 0 if $j\notin S$, the number of the orange diamonds with $M_j > t$ approximately equals the number of the orange diamonds with $M_j < -t$. Therefore, we can estimate FDR by $\#\{j: M_j < -t\} / \#\{j: M_j > t\}$.}
%     \label{fig:mirror1dim}
% \end{figure}

To construct a statistic satisfying Definition \ref{def:mirror}, several approaches have been proposed. The knockoff filtering \citep{Barber2015-lg} first introduced the fundamental idea of the mirror statistic. The knockoff filtering focuses on non-asymptotic control of FDR in linear regression and is not strictly the same as the above framework. The knockoff filtering has been expanded to other models including GLM \citep{Candes2016-mi, Huang2020-qg, Poignard2022-cs, Li2023-kv}; however, the extensions of the knockoff filtering often rely on the extremely strong assumptions to ensure FDR control with a finite sample, such as complete knowledge of the joint distribution of the predictors. On the other hand, the Gaussian mirror \citep{Xing2023-mp, Dai2023-ju} and the data-splitting \citep{Dai2022-jk, Dai2023-ju} directly leverage the concept of the mirror statistic. These works allow $M_j$ to satisfy the properties in Definition \ref{def:mirror} asymptotically, and of course, these methods only guarantee the FDR control under the asymptotic situations. Instead of giving up the non-asymptotic guarantee for FDR control, these methods can be applied under moderate assumptions to various problems, including GLM and modeling with high-dimensional data.

The Gaussian mirror (GM) and the data-splitting (DS) share the basic idea of constructing the mirror statistic. For the variables $j \in [p]$, they obtain two independent statistics, $T_j^{(1)}$ and $T_j^{(2)}$, which are typically defined as the scaled estimators of the regression coefficients $\beta_j$. Given the two independent statistics corresponding to the $j$th variable, the mirror statistic is defined as $M_j = \mathrm{sign}\left(T_j^{(1)}T_j^{(2)}\right)f\left(|T_j^{(1)}|,|T_j^{(2)}|\right)$, where $f(u,v)$ is a known function that is non-negative, exchangeable, and monotonically increasing. According to Proposition 2.1 in \citet{Dai2023-ju}, the form $f(u,v) = u + v$ is the most powerful under the hypothesis testing framework, but there is no significant difference in their experiments among other possible forms, such as $2\min(u,v)$ and $uv$. Hereafter, we use $f(u,v) = u + v$ in this paper when using the original mirror statistic.
The GM and the DS procedures produce, in different manners, the independent statistics $T_j^{(1)}$ and $T_j^{(2)}$, which are expected to be symmetrically distributed about zero for $j\notin S$. Briefly speaking, GM obtains $T_j^{(1)}$ and $T_j^{(2)}$ by adding independent noises to $X_j$, and estimates $\beta_j$ twice. Setting the noise level appropriately, the obtained statistics are independent and symmetric when $j\notin S$. The DS procedure, which is much simpler, obtains independent $T_j^{(1)}$ and $T_j^{(2)}$ by splitting the dataset randomly into two parts and estimating $\beta_j$ based on them. For low-dimensional data, the parameters are estimated using the maximum likelihood estimation/-tor (MLE), for example. For high-dimensional data, two methods are discussed. The first one is the debiased lasso \citep[e.g.][]{Van_de_Geer2014-ve}, which corrects the lasso \citep{Tibshirani1996-pc} estimates so that the corrected estimates have consistency and asymptotic normality. The second one is the data-splitting and cross-fitting procedure, which selects essential predictors using the first half of the data via the lasso, and estimates the regression coefficients using another half of the data via the MLE. 

Multiple data splitting (MDS) is also proposed to average the artificial randomness introduced by random splitting. Repeating the DS procedure for $M$ times, then we obtain the inclusion rates $\hat{I}_j = M^{-1}\sum_{m=1}^M \mathbb{I}\{j\in \hat{S}^{(m)}\}/|\hat{S}^{(m)}|\vee 1$ for each predictor. Given the set of inclusion rates and the designated FDR level $q$, we select predictors via the three steps:
\begin{enumerate}
    \item Sort the predictors so that they are in ascending order of their inclusion rates. Let $\hat{I}_{(j)}$ be the inclusion rate of the $j$th sorted predictor.
    \item Find the largest $j'$ that satisfies $\sum_{j\le j'}\hat{I}_{(j)} \le q$.
    \item Select variables as $\hat{S} = \{j: \hat{I}_{(j)} > \hat{I}_{(j')}\}$.
\end{enumerate}
For more details on the mirror statistic, please see the original articles \citep{Dai2022-jk, Xing2023-mp, Dai2023-ju}. 

% #################################################################################
\section{FDR Control for Confounder Selection via Mirror Statistic}
To implement the union-set approach and the minimal-set approach, it is necessary to estimate the sets $S_{\mathrm{OR}}:= S_Y \cup S_A$ and $S_{\mathrm{AND}}:= S_Y \cap S_A$, where $S_Y = \{j:\beta_j \ne 0\}$ in the model \eqref{eq:glm_Y} and $S_A = \{j:\alpha_j \ne 0\}$ in the model \eqref{eq:glm_A}. In this section, we introduce novel methods to control FDR for $S_{\mathrm{OR}}$ and $S_{\mathrm{AND}}$ using the mirror statistic. First, we consider an approach based on two mirror statistics $M^Y_j$ and $M^A_j$, which we call {\it Paired Mirrors}. Next, we construct novel mirror statistics specific to $S_{\mathrm{OR}}$ and $S_{\mathrm{AND}}$, which we call {\it Unified Mirrors}. All proofs of the propositions below are presented in Appendix A.

\subsection{Paired Mirrors}
\subsubsection{Union-set Approach}
Let the mirror statistic $M_j^Y$ for the model \eqref{eq:glm_Y} and $M_j^A$ for the model \eqref{eq:glm_A}, and we select the varaibles with $M^{Y}_j > t$ or $M^{A}_j > t$ to estimate $S_{\mathrm{OR}}$. Then, FDP for the union-set approach is defined as
\begin{align}
    \mathrm{FDP^{OR}_{Paired}}(t) &:= \frac{\#\{j: (M^{Y}_j > t \vee M^{A}_j > t) \wedge j\notin S_{OR}\}}{\#\{j: M^{Y}_j > t \vee M^{A}_j > t\}\vee 1},
\end{align}
and $\mathrm{FDR^{OR}_{Paired}} := \mathbb{E}\left[\mathrm{FDP^{OR}_{Paired}}\right]$. 
In addition to the conditions in Definition \ref{def:mirror}, we assume that $M^{Y}_j$ and $M^{A}_j$ are independent; for example, this condition asymptotically holds for MLE of $\bs\alpha$ and $\bs\beta$, where the $(j,j+p)$ element of the Fisher information matrix is asymptotically zero. Given that $M_j^Y$ and $M_j^A$ are appropriate mirror statistics, the following equation for the union set $S_{OR}$ holds:
\begin{proposition}\label{prp:approx_OR}
    When the mirror statistics $M^{Y}_j$ and $M^{A}_j$ are independent, the following equation holds:
    \begin{gather}
        \#\{j: (M^{Y}_j > t \vee M^{A}_j > t) \wedge j\notin S_{\mathrm{OR}}\} \stackrel{d}{=} \#\{j: (M^{Y}_j < -t \vee M^{A}_j < -t) \wedge j\notin S_{\mathrm{OR}}\}.
    \end{gather}
\end{proposition}
Note that the independence of $M^{Y}_j$ and $M^{A}_j$ for $j\notin S_{OR}$ is somewhat stronger than necessary since we can show Proposition \ref{prp:approx_OR}  assuming that their joint distribution is symmetric about the origin; however, we assume the independence for simplicity of proof and consistency with the proof of the propositions in the minimal-set approach discussed later in Section \ref{sec:unif_min}.

Then, $\mathrm{FDP^{OR}_{Paired}}$ is bounded as
\begin{gather*}
    \mathrm{FDP^{OR}_{Paired}}(t) \stackrel{d}{=} \frac{\#\{j: (M^{Y}_j <-t \vee M^{A}_j < -t) \wedge j\notin S_{OR}\}}{\#\{j: M^{Y}_j > t \vee M^{A}_j > t\}\vee 1} \le  \frac{\#\{j: M^{Y}_j <-t \vee M^{A}_j < -t\}}{\#\{j: M^{Y}_j > t \vee M^{A}_j > t\}\vee 1}.
\end{gather*}
Choose $t$ so that the bound is smaller than the designated FDR $q$, and we obtain the varaible set $\hat{S}_\mathrm{OR} = \{j: M^{Y}_j > t \vee M^{A}_j > t\}$. 
The left plot in Figure \ref{fig:pmirrors} is a visual illustration of the paired mirrors for $S_{\mathrm{OR}}$. Note that the set $\{j: M^{Y}_j <-t \vee M^{A}_j < -t\}$, which estimates the number of false discoveries, can contain relevant variables (see the blue-circle points in the gree area). This means that $\#\{j: M^{Y}_j <-t \vee M^{A}_j < -t\}$ can overestimate $\#\{j: (M^{Y}_j > t \vee M^{A}_j > t) \wedge j\notin S_{OR}\}$ and may decrease statistical power for $S_{\mathrm{OR}}$, but the inequality of FDP still holds, and the most important variables for adjustment, say $S_{AND}$, are unlikely to fall into the green region because they are related to both $Y$ and $A$.

\subsubsection{Minimal-set Approach} \label{sec:unif_min}
In the minimal-set approach, we select the variables with $M^{Y}_j > t \wedge M^{A}_j > t$. Then, FDP for the minimal-set approach is defined as
\begin{align}
    \mathrm{FDP^{AND}_{Paired}}(t) &:= \frac{\#\{j: (M^{Y}_j > t \wedge M^{A}_j > t) \wedge j\notin S_{\mathrm{AND}}\}}{\#\{j: M^{Y}_j > t \wedge M^{A}_j > t\}\vee 1} 
    = \frac{\left|\hat{S}_{\mathrm{AND}}\cap \overline{S_{\mathrm{AND}}}\right|}{\left|\hat{S}_{\mathrm{AND}}\right|\vee 1},
\end{align}
and $\mathrm{FDR^{AND}_{Paired}} = \mathbb{E}\left[\mathrm{FDP^{AND}_{Paired}}\right]$.
In order to bound $\mathrm{FDP^{AND}_{Paired}}$, we first decompose the set $\hat{S}_{\mathrm{AND}}\cap \overline{S_{\mathrm{AND}}}$ and approximate them.

\begin{proposition}\label{prp:approx_AND}
    When the mirror statistics $M^{Y}_j$ and $M^{A}_j$ are independent, the following equation holds:
    \begin{gather}\label{eq:approx_AND}
        \left|\hat{S}_{\mathrm{AND}}\cap \overline{S_{\mathrm{AND}}}\right|
        \stackrel{d}{=}
        \left|\hat{S}_1\cap\overline{S_{\mathrm{AND}}}\right|
        +\left|\hat{S}_2\cap\overline{S_{\mathrm{AND}}}\right|
        -\left|\hat{S}_3\cap\overline{S_{\mathrm{AND}}}\right|,
    \end{gather}
    where
    \begin{align}
        \hat{S}_1 =&~ \{j: M^{Y}_j > t \wedge M^{A}_j < -t\}, \nonumber\\
        \hat{S}_2 =&~ \{j: M^{Y}_j < -t \wedge M^{A}_j > t\}, \nonumber\\
        \hat{S}_3 =&~ \{j: M^{Y}_j < -t \wedge M^{A}_j < -t\}. \nonumber
    \end{align}
\end{proposition}

\begin{figure}[ht]
    \centering
    \includegraphics[width=0.95\textwidth]{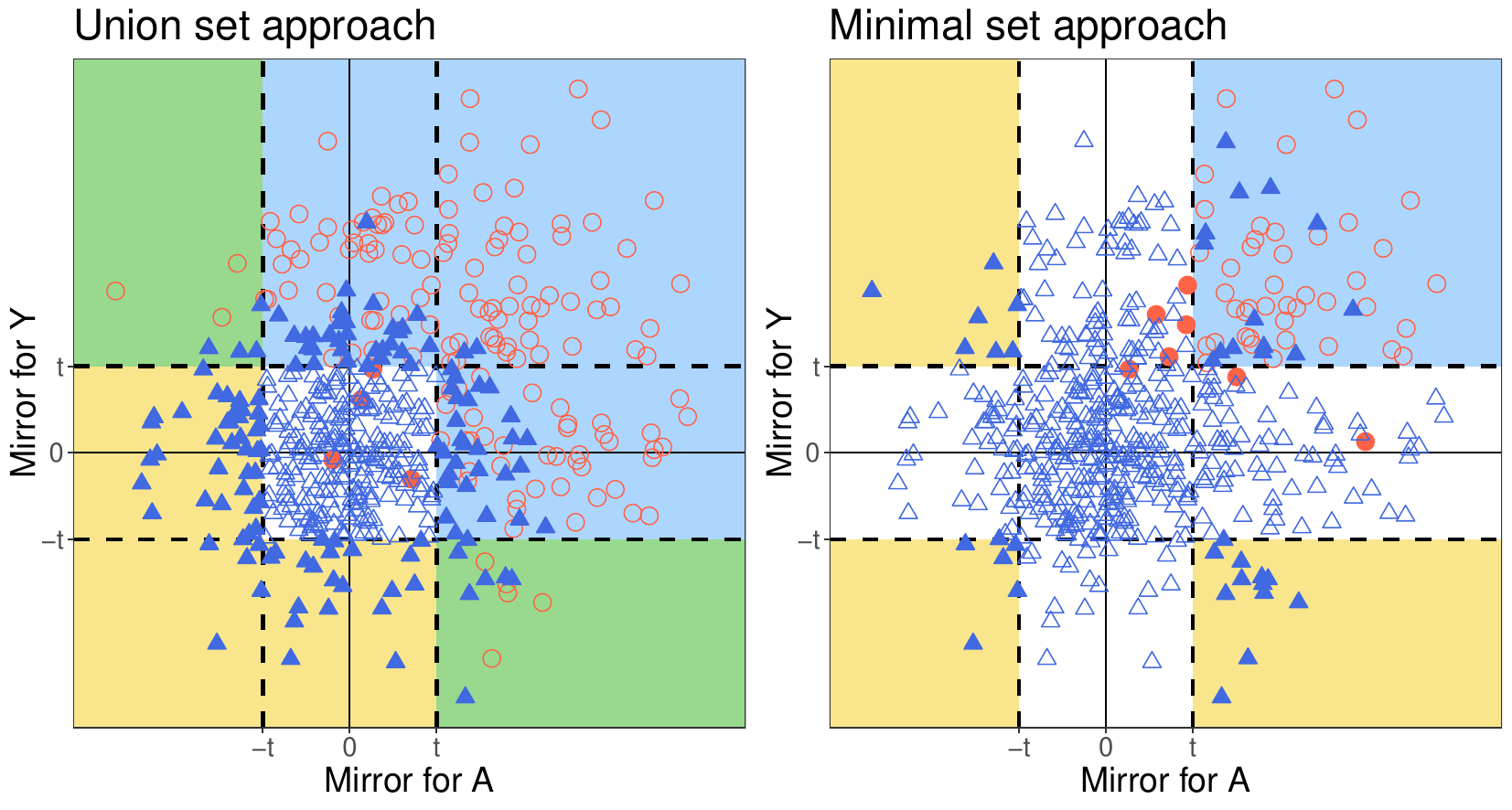}
    \caption{An illustration of the paired mirrors. (Left: the union-set approach) We select the variables with $M^Y_j > t$ or $M^A_j > t$. While the red-hollow circles are correctly selected variables, the blue-filled triangles in the selected area are falsely discovered. The number of the blue-filled triangles in the selected area is estimated by counting the blue-filled triangles in the area of $M^Y_j < -t$ or $M^A_j < -t$. (Right: the minimal-set approach) We select the variables with $M^Y_j > t$ and $M^A_j > t$. While the red-hollow circles are correctly selected variables, the blue-filled triangles in the selected area are falsely discovered. The number of the blue-filled triangles in the selected area is estimated by summing up the blue-filled triangles in the area of $M^Y_j > t \wedge M^A_j < -t$ and $M^Y_j < -t \wedge M^A_j > t$ and subtracting the count in $M^Y_j < -t \wedge M^A_j < -t$.}
    \label{fig:pmirrors}
\end{figure}

Then, similarly to the union-set approach, we want to show the boundedness
\begin{gather}\label{eq:bound_AND}
    \mathrm{FDP^{AND}_{Paired}} \stackrel{d}{\approx}
    \frac{\left|\hat{S}_1\cap\overline{S_{\mathrm{AND}}}\right| + \left|\hat{S}_2\cap\overline{S_{\mathrm{AND}}}\right| - \left|\hat{S}_3\cap\overline{S_{\mathrm{AND}}}\right|}{|\hat{S}_{\mathrm{AND}}|\vee 1} \le  \frac{|\hat{S}_1|+|\hat{S}_2|-|\hat{S}_3|}{|\hat{S}_{\mathrm{AND}}|\vee 1},
\end{gather}
however, this inequality does not hold in general. 
The bound \eqref{eq:bound_AND} holds when the inequality
\begin{align*}
&~ |\hat{S}_1|+|\hat{S}_2|-|\hat{S}_3| -\left(\left|\hat{S}_1\cap\overline{S_{\mathrm{AND}}}\right| + \left|\hat{S}_2\cap\overline{S_{\mathrm{AND}}}\right| - \left|\hat{S}_3\cap\overline{S_{\mathrm{AND}}}\right|\right) \\
=&~ |\hat{S}_1\cap S_{\mathrm{AND}}|+|\hat{S}_2\cap S_{\mathrm{AND}}|-|\hat{S}_3\cap S_{\mathrm{AND}}| \ge 0
\end{align*}
holds.
Then, we prepare the next lemma.
\begin{lemma} \label{lnm:approx_AND2}
    Let $(W,Z)$ be independent random variables, and assume that at least one of the following inequalities hold for all non-negative constant $t$:
    \begin{align*}
        \mathbb{P}(W > t) \ge \mathbb{P}(W < -t),~~~ \mathbb{P}(Z > t) &\ge \mathbb{P}(Z < -t).
    \end{align*}
    Then, we have
    \begin{gather*}
        \mathbb{P}(W > t, Z < -t) + \mathbb{P}(W < -t, Z > t) - \mathbb{P}(W < -t, Z > t) \ge 0.
    \end{gather*}
\end{lemma}
The proof is based on a simple algebra. 
Lenma\ref{lnm:approx_AND2} tells us if either $\mathbb{P}(M_j^Y > t) - \mathbb{P}(M_j^Y < -t) \ge 0$ or $\mathbb{P}(M_j^A > t) - \mathbb{P}(M_j^A < -t) \ge 0$ holds for $j\in S_{\mathrm{AND}}$, the inequality \eqref{eq:bound_AND} is implied with high probability. Since $M_j^Y$ and $M_j^A$ are supposed to take large positive values for $j\in S_{\rm AND}$, the bound for $\mathrm{FDP^{AND}_{Paired}}$ \eqref{eq:bound_AND} is valid with high probability.
Then, by choosing $t$ so that the bound is smaller than the designated FDR $q$, we obtain the varaible set $\hat{S}_\mathrm{AND} = \{j: M^{Y}_j > t \wedge M^{A}_j > t\}$. The right plot in Figure \ref{fig:pmirrors} is a visual illustration of the paired mirror for $S_{\mathrm{AND}}$.

\subsection{Unified Mirrors}
Next, we construct a single mirror statistic using two regression coefficients $\hat\alpha_j$ and $\hat\beta_j$ for the $j$-th variable. We call them the unified mirror.

\subsubsection{Union-set Approach}
Let $\mathbf{T}^{(1)}_j$ and $\mathbf{T}^{(2)}_j$ be two-dimensional independent statistics corresponding to $X_j$ (typically, $\mathbf{T}^{(m)}_j := (\hat{\alpha}^{(m)}_j/SE(\hat{\alpha}^{(m)}_j), \hat{\beta}^{(m)}_j/SE(\hat{\beta}^{(m)}_j))$, $m=1,2$), where $SE$ is the standard error. For $j\notin S_{\rm OR}$, we assume that the components of $\mathbf{T}^{(m)}_j$ are independent and symmetrically distributed about 0. Let $(r_j^{(m)}, \theta_j^{(m)})$ be the polar coordinate representation for $\mathbf{T}^{(m)}_j$. Following the idea of the original mirror statistic, a mirror statistic for the union-set approach is constructed as the product of two terms: concordance of $\mathbf{T}^{(1)}_j$ and $\mathbf{T}^{(2)}_j$ and signal intensity.
Let the mirror statistic for the union-set approach as 
\begin{gather*}
    M_j^{\mathrm{OR}} := f_1\left(\theta^{(1)}_j, \theta^{(2)}_j\right)\cdot f_2\left(\mathbf{T}^{(1)}_j, \mathbf{T}^{(2)}_j\right),
\end{gather*}
where $f_1$ is a concordance function, and $f_2$ is a intensity function. Although we can consider a bunch of forms for $f_1$ to make $M_j^{\mathrm{OR}}$ a mirror statistic, we consider only two forms:
\begin{gather*}
    \cos\left(\theta^{(1)}_j - \theta^{(2)}_j\right),~
    \mathrm{sign}\left(\cos(\theta^{(1)}_j - \theta^{(2)}_j)\right).
\end{gather*}
For $j\in S_{\mathrm{OR}}$, $\mathbf{T}^{(1)}_j$ and $\mathbf{T}^{(2)}_j$ fall far from the origin and are close to each other; the sign of $f_1$ is expected to be positive. On the contrary, the sign of $f_1$ is expected to take $+1$ and $-1$ with equal probability for $j\notin S_{\mathrm{OR}}$. Therefore, $f_1$ represents the concordance of $\mathbf{T}^{(1)}_j$ and $\mathbf{T}^{(2)}_j$. Next, we consider four forms for $f_2$:
\begin{gather*}
    r_j^{(1)} + r_j^{(2)},~~ \max\left(|T_{1j}^{(1)}|,|T_{2j}^{(1)}|\right) + \max\left(|T_{1j}^{(2)}|, |T_{2j}^{(2)}|\right),\\
    r_j^{(1)}r_j^{(2)},~~ \max\left(|T_{1j}^{(1)}|, |T_{2j}^{(1)}|\right) \cdot \max\left(|T_{1j}^{(2)}|, |T_{2j}^{(2)}|\right).
\end{gather*}
These functions take large positive values when the absolute values of $\hat\alpha_j$ or $\hat\beta_j$ are large, which reflects the strength of the association between $X_j$ and $Y$ or $A$. By designing the statistic $M_j^{\mathrm{OR}}$ as described above, $M_j^{\mathrm{OR}}$ is expected to have the two properties of the mirror statistics in Definition \ref{def:mirror}. If we choose $r_j^{(1)}r_j^{(2)}$ for $f_1$ and $\cos\left(\theta^{(1)}_j - \theta^{(2)}_j\right)$ for $f_2$, $M_j^{\mathrm{OR}}$ is nothing but the inner product of $\mathbf{T}^{(1)}_j$ and $\mathbf{T}^{(2)}_j$. In Section 5, we compare the functional forms in the performance of FDP control and power to detect relevant variables. 

Once the mirror statistic is constructed in this way, all that remains is to determine the rejection limit to satisfy the designated FDR level $q$ and to select variables with ${M^{\rm OR}_j > t}$, just as in the same way to the existing works using the mirror statistic.

\subsubsection{Minimal-set Approach}
We need to make two more devices to construct the unified mirror statistic for the minimal-set approach. 
The first one is the form of $f_2$. We use $\min$ instead of $\max$ because both $|\alpha_j|$ and $|\beta_j|$ are large for $j\in S_{\mathrm{AND}}$.
The second one is an additional multiplicative term to obtain sign symmetry. The null set $\overline{S_{\rm AND}}$ is partitioned into three subsets such that the variables they contain have different distributions:
\begin{gather*}
    \overline{S_{\mathrm{AND}}} 
    = (S_{Y}\cap \overline{S_{A}})
    \cup (\overline{S_{Y}}\cap S_{A})
    \cup (\overline{S_{Y}}\cap \overline{S_{A}}).
\end{gather*}
We need to make two more devices to construct the unified mirror statistic for the minimal-set approach. 
The first one is the form of $f_2$. We use $\min$ instead of $\max$ because both $|\alpha_j|$ and $|\beta_j|$ are large for $j\in S_{\mathrm{AND}}$.
The second one is an additional multiplicative term to obtain sign symmetry. The null set $\overline{S_{\rm AND}}$ is partitioned into three subsets such that the variables they contain have different distributions as $\overline{S_{\mathrm{AND}}}  = (S_{Y}\cap \overline{S_{A}})
    \cup (\overline{S_{Y}}\cap S_{A})
    \cup (\overline{S_{Y}}\cap \overline{S_{A}})$.
For $j \in S_{Y}\cap \overline{S_{A}}$, $\mathbf{T}^{(m)}_j$ is distributed around the $Y$-axis but far from the origin; for $j \in \overline{S_{Y}} \cap S_{A}$, $\mathbf{T}^{(m)}_j$ is distributed around the $A$-axis but far from the origin; for $j \in \overline{S_{Y}} \cap \overline{S_{A}}$, $\mathbf{T}^{(m)}_j$ is distributed around the origin. Here, we aim to make the mirror statistic distribute symmetrically about zero for $j\in \overline{S_{\rm AND}}$, we define the unified mirror statistic for the minimal-set approach as $ M_j^{\mathrm{AND}} := f_1\cdot f_2 \cdot f_3\left(\theta^{(1)}_j, \theta^{(2)}_j\right)$, 
where $f_3$ is an additional term. We suggest two forms for $f_3$:
\begin{gather*}
    \sin2\theta^{(1)}_j\cdot\sin2\theta^{(2)}_j, ~~ 
    \mathrm{sign}\left\{
        \sin2\theta^{(1)}_j\cdot\sin2\theta^{(2)}_j
    \right\}.
\end{gather*}
Sign of $\sin{2\theta}$ is positive in the first and third quadrants and negative in the others. For $j \in S_{\mathrm{AND}}$, $\mathbf{T}^{(1)}_j$ and $\mathbf{T}^{(2)}_j$ are expected to fall in the same quadrants, so $f_3$ is expected to be positive; for $j \notin S_{\mathrm{AND}}$, $\mathbf{T}^{(1)}_j$ and $\mathbf{T}^{(2)}_j$ are symmetrically distributed across the quadrants, so the sign of $f_3$ is expected to be $+1$ or $-1$ with equal probability. 

\begin{proposition}\label{prp:sym_unif_AND}
    For $j \notin S_{\mathrm{AND}}$, the sign of $f_3$ takes $+1$ or $-1$ with equal probability.
\end{proposition}

\begin{figure}[ht]
    \centering
    \includegraphics[width=0.7\textwidth]{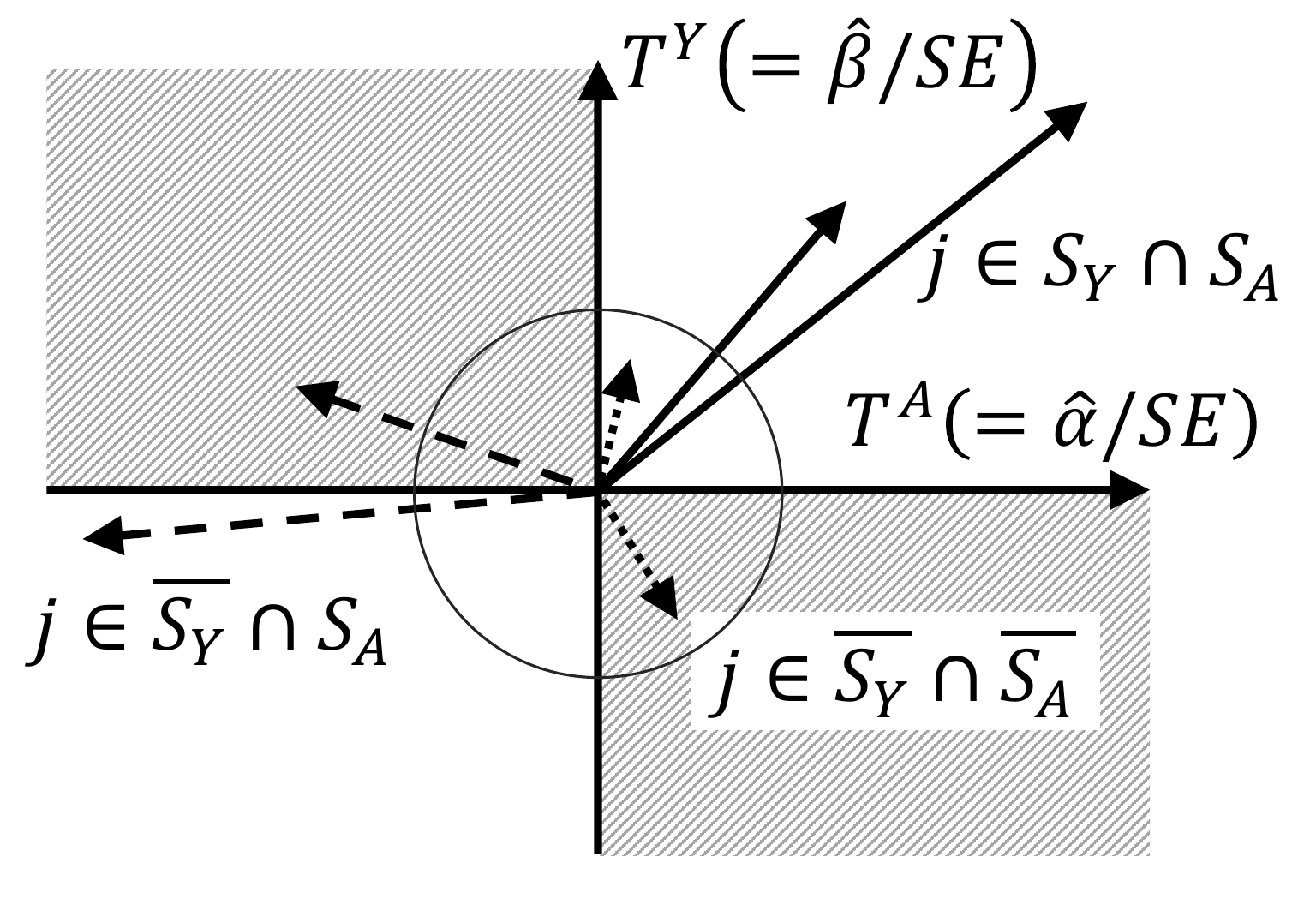}
    \caption{An illustration for the unified mirror statistics. The pair of solid arrows represents $\mathbf{T}^{(1)}_j$ and $\mathbf{T}^{(2)}_j$ for $j\in S_{\rm AND}$. The pairs of dashed and dotted arrows are those for $j\in S_Y \cap \overline{S_A}$ and $j\in \overline{S_Y} \cap \overline{S_A}$, respectively. The angle between the two vectors is small for $j \in S_{\rm OR}$, while not necessarily small for $j\in \overline{S_Y} \cap \overline{S_A}$. The shaded area is the quadrants where the $\sin 2\theta$ is negative and is positive in the other areas. For $j\in S_{\rm AND}$, two vectors tend to fall in the same quadrant.}
    \label{fig:umirrors}
\end{figure}

Figure \ref{fig:umirrors} is a visual illustration of the unified mirrors.
Note that, if we choose the former option for $f_3$, $\sin 2\theta^{(1)}_j\cdot\sin{2\theta^{(2)}_j}$, it is expected that the signal intensity is weakened since $\sin{2\theta}$ takes small absolute values near the axes. 

% #################################################################################
\section{Other Confounder Selection Methods}\label{sec:other_methods}
Although it has never been proposed to the best of our knowledge, a straightforward approach to control FDR in confounder selection is correcting p-values obtained using the union-intersection test (UIT) for the union-set approach and the intersection-union test (IUT) for the minimal-set approach. Suppose $H_{0,1},\ldots, H_{0,K}$ are the null hypotheses to be tested and $H_{1,1},\ldots, H_{1,K}$ are the corresponding alternatives. The union-intersection test (UIT) is a testing procedure of the joint null hypotheses $H_0: \bigcap_{k}H_{0,k}$. Inversely, the intersection-union test (IUT) is a test regarding the joint null of $ H_0: \bigcup_{k}H_{0,k}$. UIT and IUT are mainly discussed in clinical trials and multiple testing. 
In this study, we implement UIT for the union-set approach by testing the hypothesis $\alpha_j = \beta_j = 0$ based on the minimum p-value of the estimated coefficients. For the minimal-set approach, we implement IUT by testing the hypothesis $\alpha_j = 0 \vee \beta_j = 0$ based on the maximum p-value of the estimated coefficients.
Once obtained p-values of the UIT and the IUT for all variables, they are corrected using FDR-control procedures \citep{Benjamini1995-jh, Benjamini2001-zs}. In this way, we can obtain FDR-controlled variable sets. For more details on UIT and IUT, please refer to \cite{Bretz2016-zi}, for example. In comparison with the proposed methods, the UIT/IUT approach requires valid p-values for testing of each hypothesis.

In practice, heuristic approaches are often employed. For example, first obtain the outcome-relevant and treatment-relevant sets using model-wise testing, and take the union or intersection of them. Even if the model-wise testing is done in an FDR-controlling method, the post-hoc set manipulation does not guarantee the entire FDR control of the minimal and union-set approach. 
On the other hand, the proposed methods are unique in that they can control FDR concerning the data-driven confounder selection criteria in epidemiology without p-values. Thanks to the p-value-free construction, they can be easily applied, at least formally, to methods such as the lasso, where the (asymptotic) distribution of the regression coefficients is difficult to derive.

Moreover, a variety of approaches are available, each motivated by different philosophies.
The change-in-estimate (CIE; e.g., \citeauthor{Greenland2016-la}, \citeyear{Greenland2016-la}) selects variables with a large effect on the estimate of the causal effect. In other words, CIE is an approach based on the intuition that variables with a large influence on the estimate of causal effect should be adjusted. The method so-called high-dimensional propensity score \citep{Schneeweiss2009-jq} is an example of CIE, which was suggested in pharmacoepidemiology; this method estimates the marginal bias caused by not adjusting each variable and selects the variables with a large bias to be adjusted. 
\citet{Wang2012-jl}, \citet{Wilson2014-nq}, and \citet{Talbot2015-eq} have employed a Bayesian approach to select confounders, which allows for the estimation of variable selection uncertainty based on posterior probabilities. \citet{De_Luna2011-dx} have discussed the identification of the minimal adjustment set based on DAG and have proposed a confounder selection algorithm by evaluating conditional independence. \citet{Rolling2013-xg} have proposed the treatment effect cross-validation aiming to minimize the bias in treatment effect estimation.
For high-dimensional data, sparse modeling is also employed. Not only the lasso regression with cross-validation but problem-specific methods have also been suggested. The outcome-adaptive lasso \citep{Shortreed2017-rp} incorporates the regression coefficients of the outcome model as weights for the penalty term of the treatment model, selecting confounders without the variables associated only with the treatment. \citet{Koch2018-at} have proposed a simultaneous estimation of the outcome model and the treatment model using a group-lasso penalty, which enables us to select confounders associated with either the outcome or treatment. \citet{Ertefaie2018-sa} also discussed the penalization approach for confounder selection.

Although these approaches are all interesting, the FDR control is not considered. Since this paper is not intended to be an exhaustive comparison of confounder selection methods, they are not involved in our experiments. However, it is still an important issue to know when these methods are valid in confounder selection and when they fail.

% #################################################################################
\section{Simulations}
In this section, we report the results of numerical experiments. 
First, we compare many options for the function form of the unified mirrors, and we see the large-sample behavior of the proposed methods. Next, the proposed methods are compared to the q-value-based FDR control methods in terms of FDR, power, and the bias induced into the ATE estimates. Details of experiments are presented in each subsection and Section \ref{app:settings} in Appendix B. 

\subsection{Choice of Functional Forms and Large-sample Behavior}\label{sec:chouceFF}
In this section, we demonstrate that increasing the sample size improves the performance of the proposed methods and conduct an exhaustive comparison of various options for the functional forms of the unified mirrors. We simulate 100 dimensions of potential confounders, drawn from a multivariate normal distribution with a Toeplitz covariance structure in every 10 dimensions. Among the 100 potential confounders, 15 variables are exclusively related to the outcome, 15 to the treatment only, and 15 to both the outcome and the treatment. The treatment variable is binary, and the outcome variable is set to be continuous or binary. We estimate the FDR and power for the union-set and the minimal-set approaches across 100 simulations, spanning sample sizes of 500, 1000, 2000, 4000, and 8000.

Firstly, we compared the various options for the functional forms of the unified mirrors  (please see the spreadsheet listed in Appendix D). Overall, the FDP control tended to fail at $N=500$ when using MLE for parameter estimation. This is probably because the sample size of 500 for 100-dimensional covariates did not necessarily constitute a low-dimensional setting, and the asymptotic approximation may not have worked well. 

In the union-set approach, the FDR was effectively controlled in sample sizes exceeding 1000. When using the product of radii for $f_1$ and $\cos ({\theta_j^{(1)} - \theta_j^{(2)}})$ for $f_2$, the statistical power was higher compared to other options. No significant difference was observed whether we used the sum or product of radii for $f_1$. The differences among the choices for $f_1$ and $f_2$ almost vanished when the sample size was sufficiently large. Consequently, we chose the product of the radii for $f_1$ and $\cos (\theta_j^{(1)} - \theta_j^{(2)})$ for $f_2$ for subsequent analyses.

In the minimal-set approach, the results varied significantly depending on the functional forms. Notably, FDR was not controlled with $N=4000$ when $\cos(\theta_j^{(1)} - \theta_j^{(2)})$ was used as $f_2$. Among the options with moderate FDR violations, some that used the sign of $\cos(\theta_j^{(1)} - \theta_j^{(2)})$ for $f_2$ showed better statistical power. In these options, the pairs of $(f_1, f_3)$ were either $\min(|T_j^{(1)}|,|T_j^{(2)}|)$ and $\sin{2\theta_j^{(1)}}\sin{2\theta_j^{(2)}}$, or the radii and the sign of $\sin{2\theta_j^{(1)}}\sin{2\theta_j^{(2)}}$. The difference between using the sum or product for $f_1$ was negligible. Consequently, we chose $f_1$ to be the product of the minimum absolute values, $f_2$ to be the sign of $\cos(\theta_j^{(1)} - \theta_j^{(2)})$, and $f_3$ to be the sign of $\sin{2\theta_j^{(m)}}$.

\begin{figure}[ht]
    \centering
    \includegraphics[width=\textwidth]{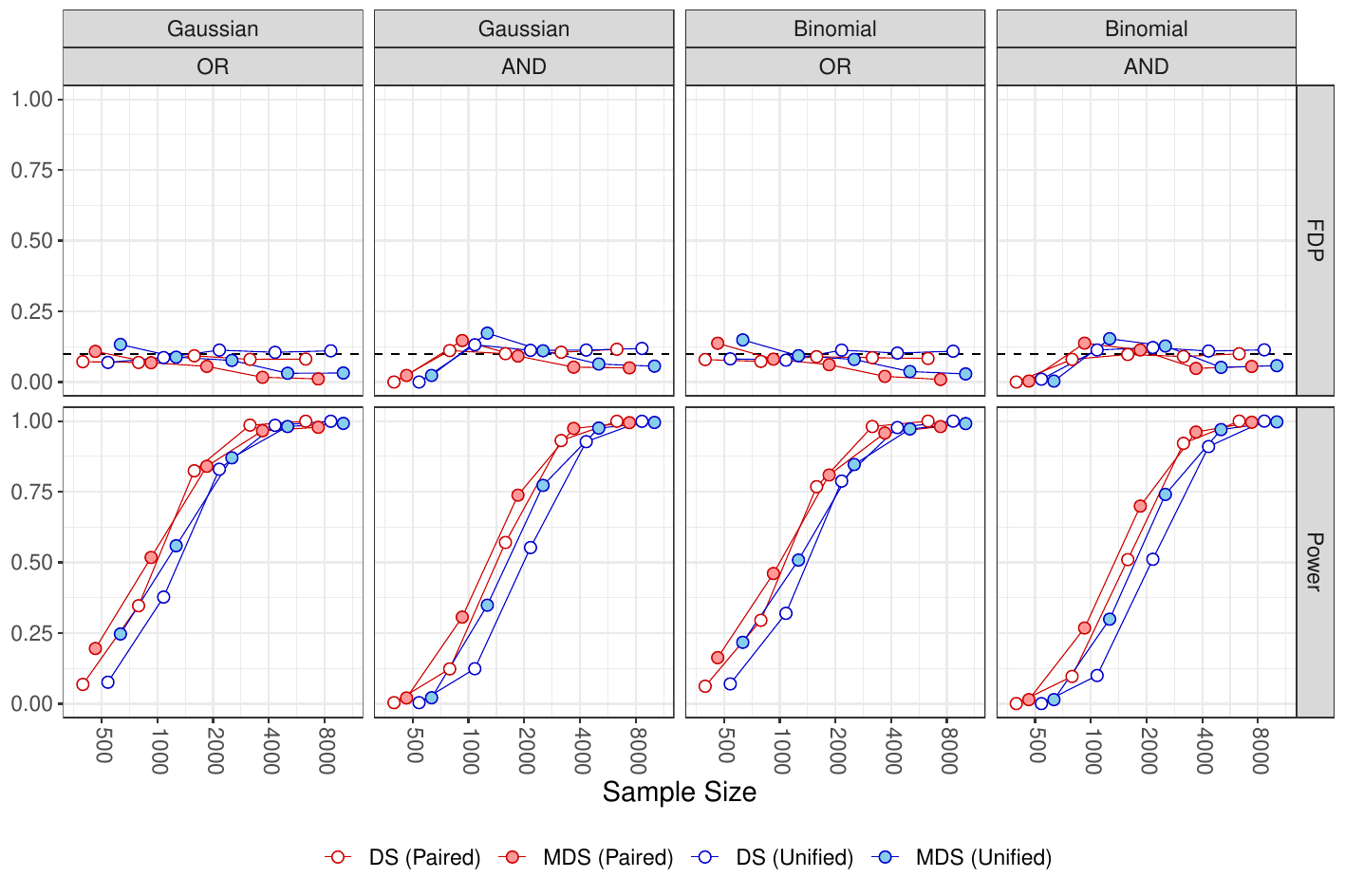}
    \caption{Mean FDP and power of the proposed methods with increasing sample size, using the cross-fitting procedure for parameter estimation. The terms ``OR'' and ``AND'' represent the selection and evaluation criteria, respectively. The outcome distributions are depicted at the top of the figures.}
    \label{fig:exp_asymp_cross}
\end{figure}

Next, we demonstrate that increasing the sample size improves the performance. Figure \ref{fig:exp_asymp_cross} presents the results obtained when parameters were estimated using the cross-fitting procedure. Although FDR control was slightly violated in some instances with $N\le 1000$ in the minimal-set approach, FDR was well-controlled with larger samples. Power increased monotonically as the sample size grew, reaching nearly 1 at $N=8000$. Results from parameter estimations using other methods, such as MLE and the lasso, are provided in Appendix D.

\subsection{FDR, Power, and the Bias of ATE Estimates}\label{sec:comparative}
In this section, we compare the proposed method with the q-values derived from the UIT and the IUT p-values, focusing on FDR control, power, and bias in ATE estimation. The q-values are obtained using BHq \citep{Benjamini1995-jh} and BYq \citep{Benjamini2001-zs}. We used the \texttt{p.adjust} function in the \texttt{stats} package of R in the experiments.

\subsubsection{Low-dimensional Potential Confounders}
In low-dimensional settings ($N=1000, p=90$), we use the MLE, the cross-fitting procedure, and the lasso for parameter estimation. The potential confounders are generated along with two scenarios: one is a multivariate Gaussian distribution with the blockwise Toeplitz correlation structure, and the other is correlated binary variables, which are obtained by thresholding the multivariate Gaussian variables at zero in each dimension. The regression coefficients for both the outcome and the treatment are fixed. We assume that 45 variables are related to either the treatment or the outcome, and 15 of these are related to both the treatment and the outcome. The regression coefficient vectors are designed to have various types of associations with the treatment and the outcome, considering factors such as the absolute value, the sign of the regression coefficients, and the correlation among the potential confounders.

\begin{figure}[ht]
    \centering
    \includegraphics[width=\textwidth]{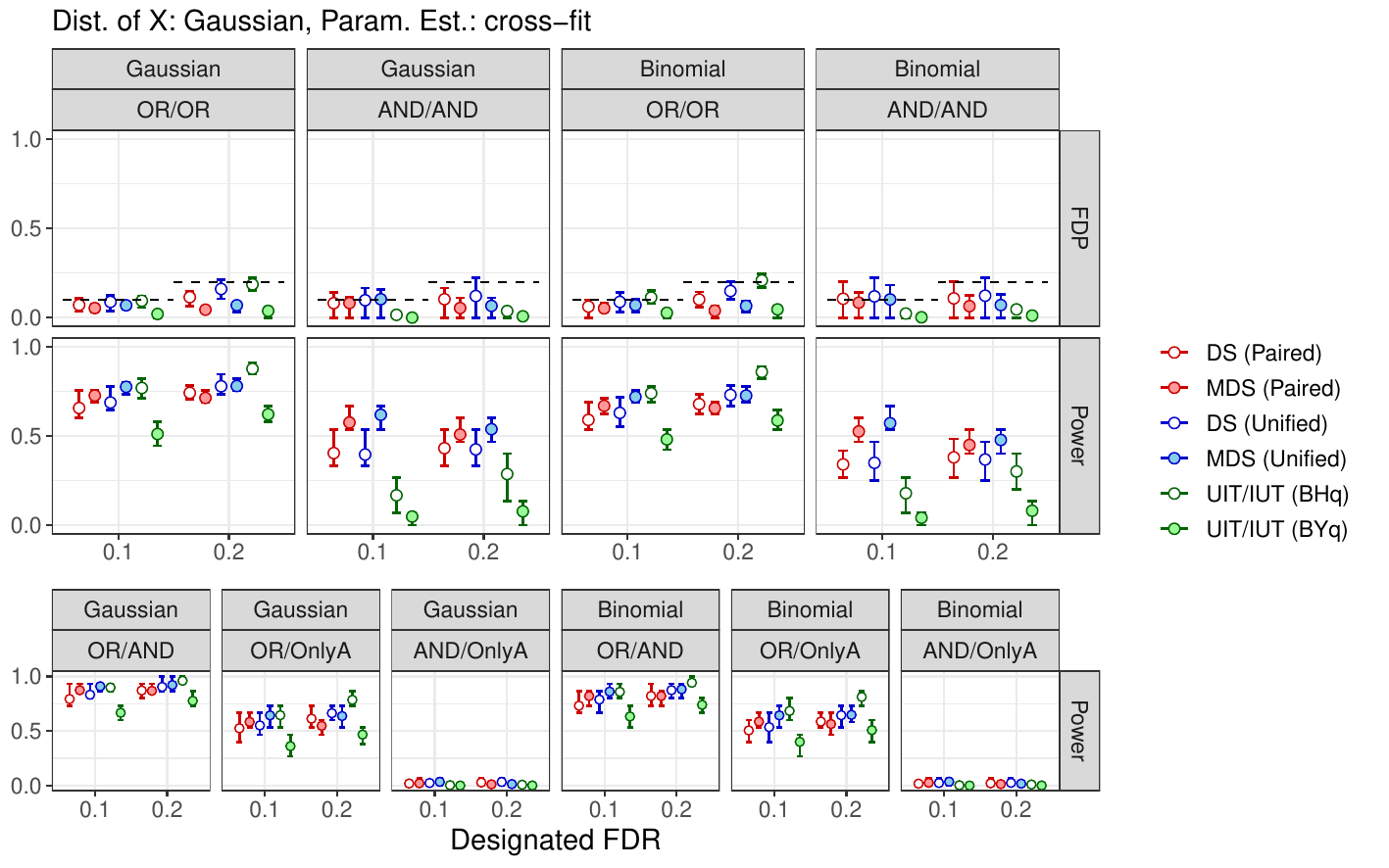}
    \caption{
        FDR and power analysis of the proposed and q-value-based methods in low-dimensional Gaussian settings. Parameter estimates for the mirror statistics were obtained using cross-fitting. The left block presents the scenarios with continuous outcomes, and the right block shows the results for binary outcomes. The notation ``**/**'' indicates the criteria for variable selection and the set of true variables used for evaluation. Each point is a mean value over 100 simulations, and the error bars at each point represent the 1st to 3rd quartiles.}
    \label{fig:exp_comp_low_glm}
\end{figure}

Figure \ref{fig:exp_comp_low_glm} displays the results, in which the regression parameters were obtained using the cross-fitting procedure. Overall, the FDR was effectively controlled. In the union-set approach (OR/OR in the figure), the BHq exhibited the highest power among the six methods, the MDS (Unified) was second, while the BYq was the most conservative. In the minimal-set approach, MDS (Unified) showed superior performance in FDR control and power. MDS outperformed DS with both mirrors, likely due to the stabilization provided by repeating. Furthermore, we assessed the power for the minimal set in the union-set approach (OR/AND) and the power of the variables associated with only the treatment (OR/OnlyA, AND/OnlyA). The variables only associated with the treatment are recommended not to be adjusted for \citep[e.g.,][]{Brookhart2006-zk, Schisterman2009-cw, Patrick2011-ao}. The MDS (Unified) and the BHq demonstrated nearly equivalent performance in detecting the minimal set for continuous and binary outcomes when the designated FDR is 0.1. Notably, the BH q-values for the union-set approach showed high power in detecting 'OnlyA' variables, which is not preferable. The ``OnlyA'' variables were rarely included in the minimal-set approach. These results remained almost consistent across the scenarios using other parameter estimation methods, although the designated FDR was sometimes violated when using MLE for parameter estimation, which was likely because asymptotic approximation did not work well as demonstrated in the previous experiments. For additional results, please refer to Appendix D.

\begin{figure}[ht]
    \centering
    \includegraphics[width=\textwidth]{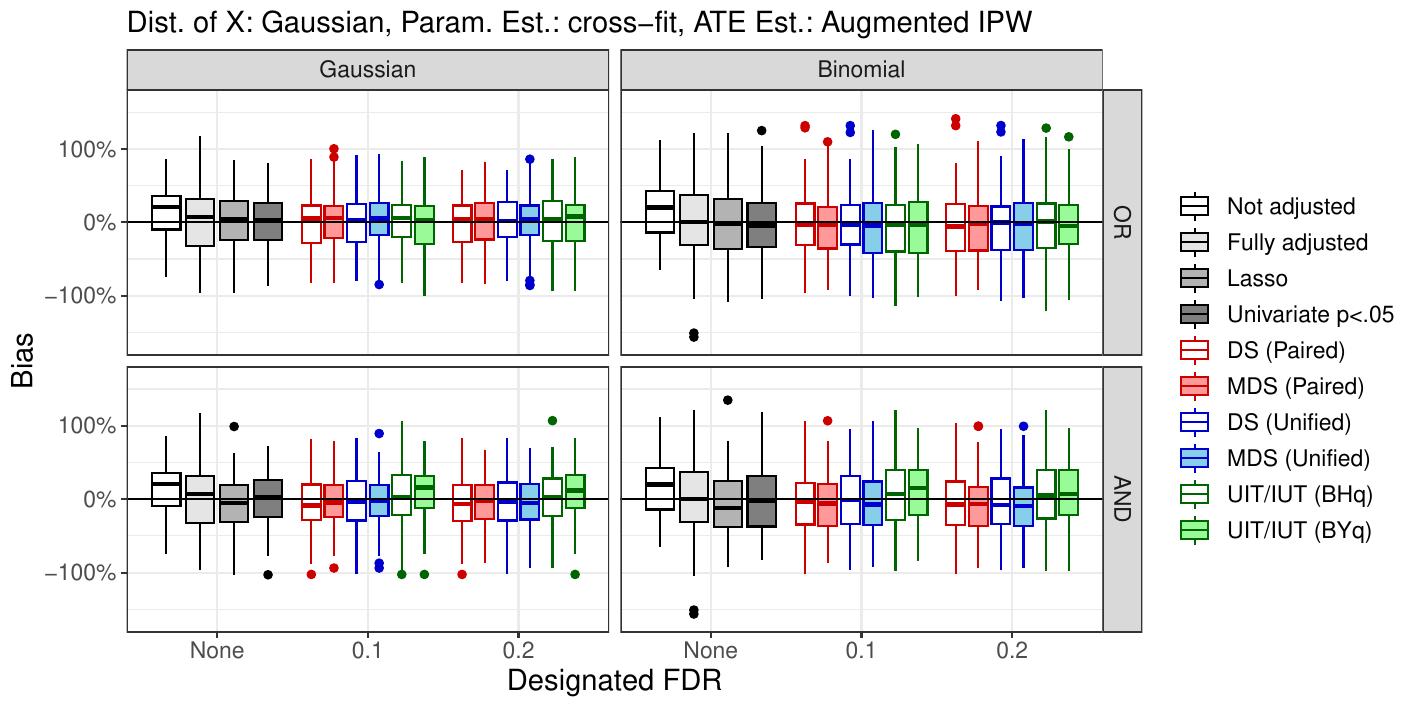}
    \caption{Relative Bias in ATE estimates obtained from 100 simulations. The parameter estimates for the mirror statistics were obtained using the cross-fitting, and the augmented IPW estimators were used to estimate ATE. The distribution of X is Gaussian.}
    \label{fig:exp_ATE_low_glm}
\end{figure}

Using the selected confounders, the ATE was estimated using the Augmented IPW (AIPW) estimator.
Figure \ref{fig:exp_ATE_low_glm} presents the bias of ATE estimates based on selected confounders. For reference, we also performed comparisons between groups without adjusting for variables (Not adjusted), a method that adjusts for all potential confounders (Fully adjusted), a method that applies the lasso regression to both models and selects the union or intersection of the selected sets (Lasso), and a method that selects variables based on the p-values of univariate regression for the outcome and treatment (Univariate p<.05). The unadjusted group difference was biased upward by about 25\%. In the union-set approach, all methods successfully eliminated bias, and no significant difference was observed in the variability of the simulated bias. In the minimal-set approach, while most methods managed to eliminate bias, the BYq method did not, especially when we set the designated FDR to be 0.1; it demonstrated lower power than the others. An examination of the selection proportion for truly relevant variables across 100 simulations revealed that the BYq method selected smaller proportions, even for variables with the largest regression coefficients (see the related figures in Appendix D). In both approaches, selecting confounders resulted in less variability in ATE estimates compared to using all potential confounders. Although the differences between the selection methods were slight, MDS (Unified) often exhibited less variability compared to the other methods.

\subsubsection{High-dimensional Potential Confounders}
In high-dimensional settings ($N=1000$, $p\in\{500, 1000, 1500\}$), we employ the cross-fitting and the lasso for parameter estimation. The UIT and the IUT q-values are obtained in two ways similar to \citep{Xing2023-mp}: (1) the cross-fitting, which selects variables using the lasso based on the first half of the data, and obtains the p-values of the regression coefficients using the other half of the dataset, and finally adjusts them; (2) the marginal q-values obtained by adjusting the p-value of the MLE of the univariate GLM.

\begin{figure}[ht]
    \centering
    \includegraphics[width=\textwidth]{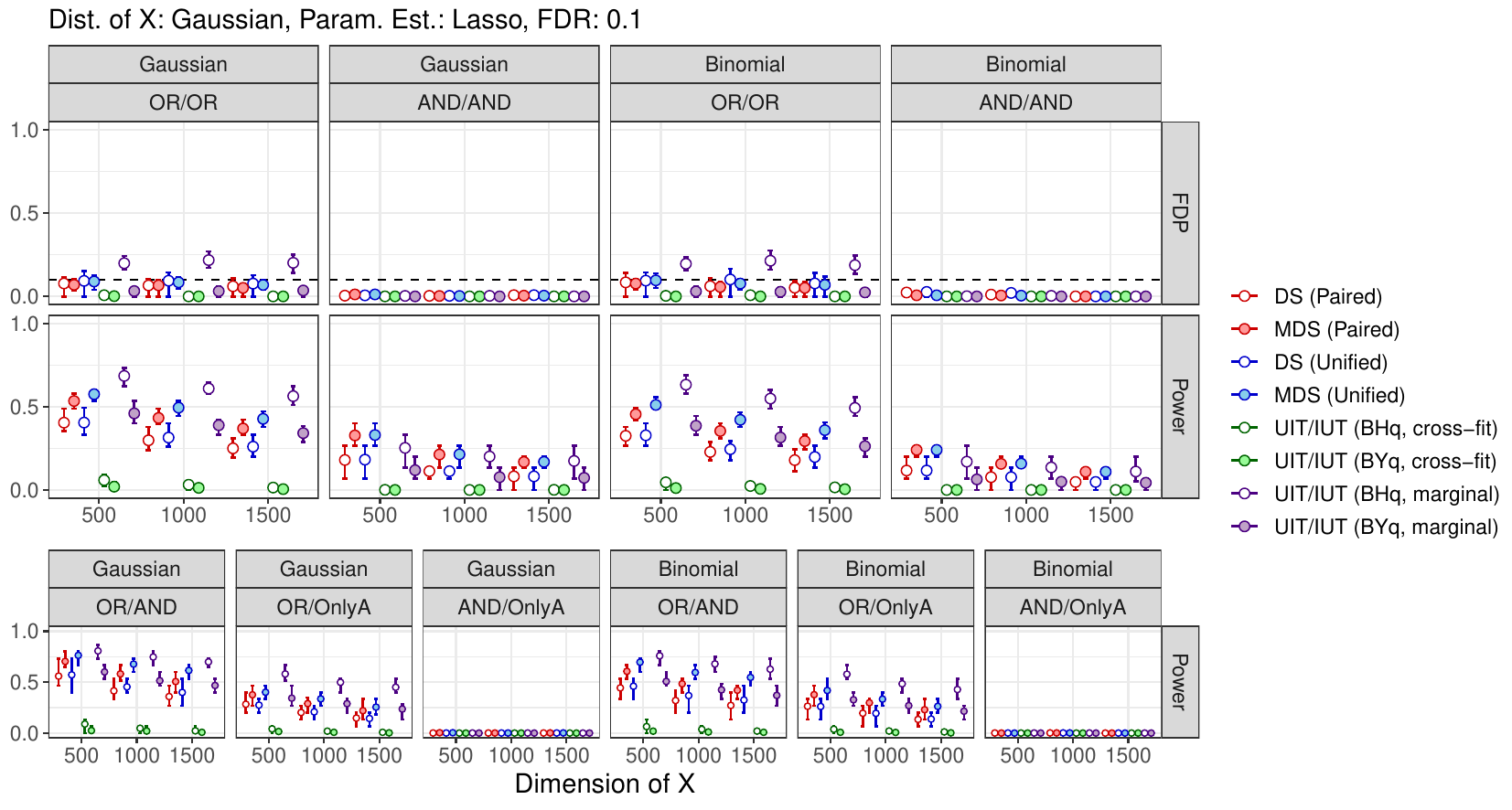}
    \caption{
    FDR and power analysis of the proposed methods and q-value-based methods in high-dimensional settings. Parameter estimates for the mirror statistics were obtained using the lasso. The left block presents the scenarios with continuous outcomes, and the right block shows the results for binary outcomes. The notation ``**/**'' denotes the variable selection criteria and the set of true variables for evaluation. Each point is a mean value over 100 simulations, and the error bars at each point represent the 1st to 3rd quartiles.
    }
    \label{fig:exp_comp_high_lasso}
\end{figure}

Figure \ref{fig:exp_comp_high_lasso} shows the FDR and power analysis when the parameter estimates in the proposed methods were obtained using the lasso. Overall, the proposed methods and the UIT/IUT q-values succeeded in controlling FDR, and an increased dimensionality of $X$ led to reduced power. 
In the union-set approach, MDS was more powerful than DS, and the unified mirrors performed better than the paired mirrors, especially in the 1500-dimensional cases. The cross-fitting UIT/IUT methods were highly conservative. The marginal BHq-values failed to control FDR for both outcomes; this result was consistent with the simulation in \citet{Xing2023-mp}. While the BYq method managed to control FDR, its power was inferior to that of the proposed methods using the lasso, particularly at higher dimensionalities. Besides, it is worth mentioning that marginally selected variables are not always important in multivariate modeling and vice versa. MDS also demonstrated the highest power for the minimal set in the union-set approach (OR/AND).
In the minimal-set approach, all methods conservatively controlled FDR, and MDS exhibited higher power than the other methods. 
The proposed methods worked more efficiently when using the lasso estimates to construct the mirror statistics compared to using the cross-fitting estimates. Please see Appendix D for the results in the other settings.

\begin{figure}[ht]
    \centering
    \includegraphics[width=\textwidth]{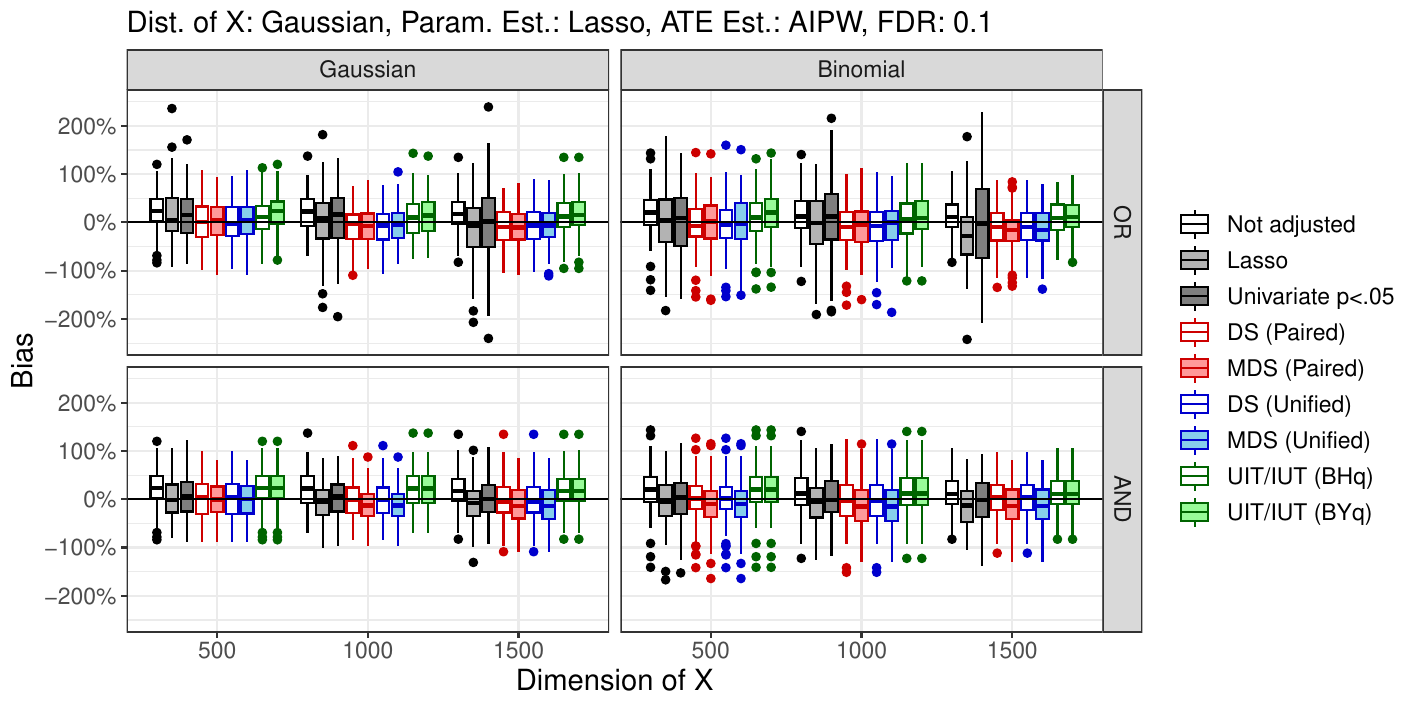}
    \caption{
    Relative Bias in ATE estimates obtained from 100 simulations based on the high-dimensional potential confounders. The parameter estimates for the mirror statistics were obtained using the lasso, and the augmented IPW was used to estimate ATE. The distribution of X is Gaussian.
    }
    \label{fig:exp_ATE_high_lasso1500}
\end{figure}

\begin{figure}[ht]
    \centering
    \includegraphics[width=\textwidth]{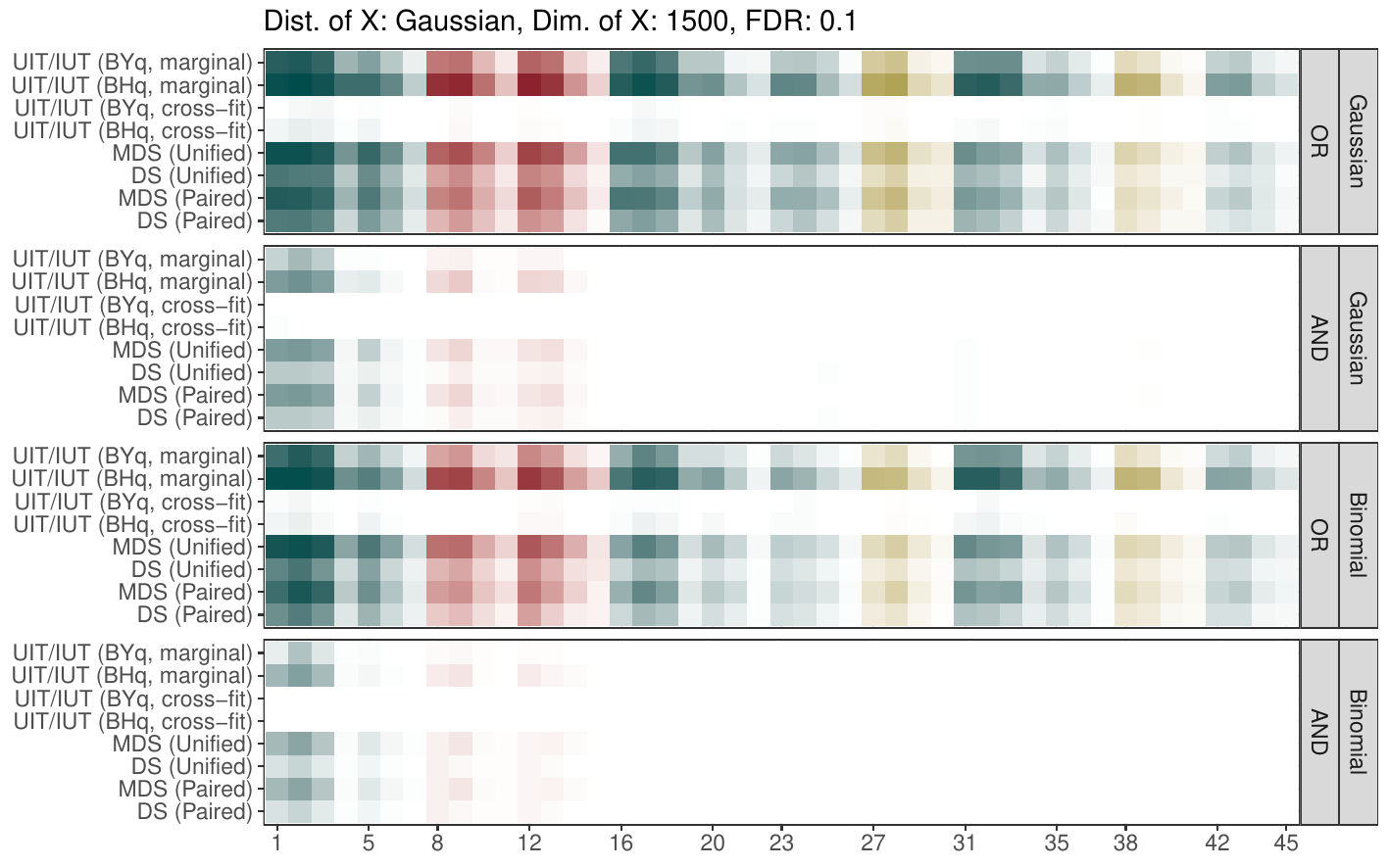}
    \caption{Selected proportion over 100 simulations for each true predictor in high-dimensional setting ($p=1500$). The X-axis is the index of the candidate variables, with indices 1 to 15 being associated with both the outcome and treatment, 16 to 30 with the outcome only, and 31 to 45 with the treatment only. Variables shown with dark blue tiles are positively associated with each corresponding target variable. Variables shown with dark red tiles have opposite signs of regression coefficients for the outcome and treatment, and variables shown with yellow tiles are negatively associated with their corresponding target variables. Variables with indices 1, 5, 16, 20, 31, and 35 have larger absolute regression coefficients, and 4, 7, 11, 15, 19, 22, 26, 30, 37, 41, and 45 have smaller ones. The upper two blocks are for continuous outcomes, while the lower two blocks are for binary outcomes.}
    \label{fig:exp_selected_high_lasso}
\end{figure}

The bias of ATE estimates and selected confounders are also evaluated. Figure \ref{fig:exp_ATE_high_lasso1500} presents the bias of ATE estimates based on the high-dimensional potential confounders following the Gaussian distribution. The unadjusted group difference was biased upward by about 25\%.
In the union-set approach, each method controlling the FDR showed lower variability compared to the lasso or the method based on univariate p-values, especially in the 1500-dimensional setting. The UIT/IUT-based methods using the cross-fitting were biased in almost all settings, probably due to their lower power. The other methods generally succeeded in removing bias. The differences in the variability of the estimates among the FDR-controlling methods were not significant, but, similar to the low-dimensional cases, MDS (Unified/Paired) were relatively stable. As can be seen in Figure \ref{fig:exp_selected_high_lasso}, MDS, especially the unified mirror, detected a significantly higher proportion of positive and strong confounders (indices: 1, 5) than the other methods, as well as a higher proportion of selected confounders (indices: 1, 2, 3) that were correlated with each other and positively associated with the outcome.

% #################################################################################
\section{Application: Right Heart Catheterization}
Right heart catheterization (RHC) is an invasive diagnostic procedure used to assess conditions related to the heart and the pulmonary artery. 
The primary purpose of RHC is to measure the pressure in the right side of the heart and the pulmonary artery. These measurements are crucial for evaluating the heart's function and diagnosing conditions such as pulmonary hypertension, heart failure, and congenital heart disease. Despite its benefits, RHC carries risks such as bleeding, infection, and, in rare cases, damage to the heart or blood vessels. 
For further reading, the American Heart Association provides a comprehensive guideline for heart catheterization \citep{AHA2023}.

\citet{Connors1996-hp} conducted a prospective cohort study that examined the risks and benefits of RHC in critically ill patients admitted to the intensive care unit (ICU). The study involved 5,735 critically ill patients requiring ICU care, dividing them into groups with and without RHC, and compared the 30-day mortality rate. The results indicated that the group of RHC had higher mortality rates, costs, and ICU durations than the control group, triggering discussions about the potential risks of RHC and appropriate patient selection. 

In this section, we reanalyze the dataset of \citet{Connors1996-hp} obtained from \url{http://hbiostat.org/data}, courtesy of the Vanderbilt University Department of Biostatistics. The RHC dataset consists of the outcome variables and 53 covariates, which is detailed in Appendix C. The outcome is death within 30 days after admission, and the treatment is RHC. For simplicity, the secondary disease category, one of the activity indices (\texttt{adld3p}), and urine output were excluded due to the large number of missing values (4,535, 4,296, and 3,028, respectively), and the diagnoses of trauma and orthopedic were also excluded due to the small number of patients with these diagnoses (52 and 7). The risk difference was estimated using the AIPW. The propensity score was obtained using the logistic regression with selected covariates. The confounder selection was conducted in the same ways in the numerical experiments: no covariates (unadjusted), all covariates, marginal BH/BY q-values, and the paired mirrors and the unified mirrors.

\begin{figure}[ht]
    \centering
    \includegraphics[width=\textwidth]{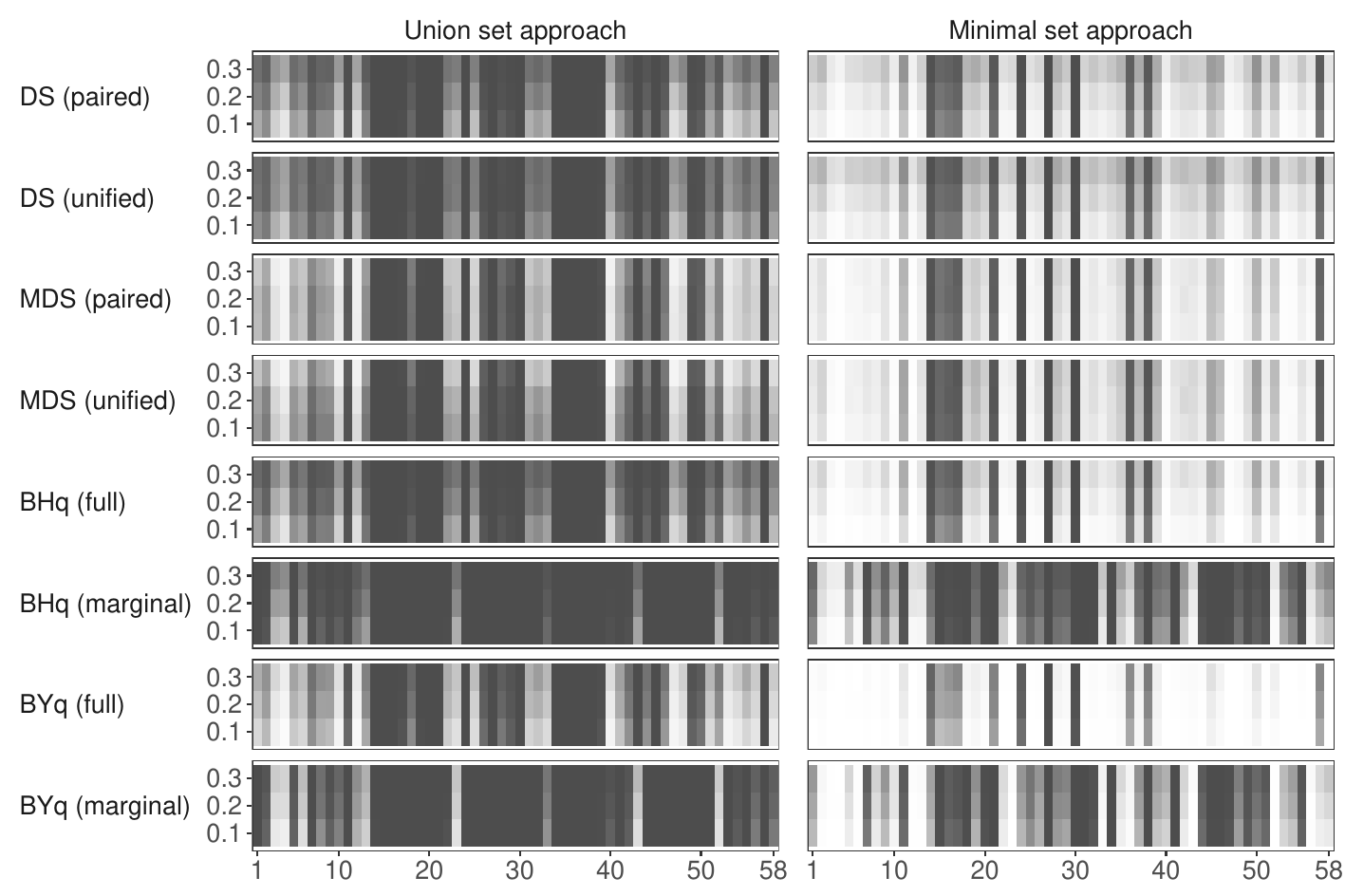}
    \caption{Selection probability based on the 1,000 bootstrap iterations. The left block shows the results in the union-set approach, and the right one shows those in the minimal-set approach. The horizontal axis represents the variable index, and the values 0.1, 0.2, and 0.3 on the vertical axis for each method indicate the designated FDR.}
    \label{fig:rhc_selected}
\end{figure}

Figure \ref{fig:rhc_selected} shows the selection proportion over the 1,000 bootstrap iterations. 
Overall, the methods based on the union-set approach selected more variables. Specifically, the marginal BHq and BYq selected most of the potential confounders. Selection based on the minimal-set approach was much sparser, and the marginal q-values selected more variables than the other multivariable methods. The primary and secondary disease (index = 14 to 17), neurological and hematological diagnosis at admission (21, 24), do-not-resuscitate on the first day (27), SUPPORT score (30) \citep{Knaus1991-ae}, PaO2/FIO2 ratio (36), partial pressure of arterial carbon dioxide (38), and transfer from another hospital (57) were frequently selected in the minimal-set approach. All of these were chosen in the union-set approach. Please refer to Appendix C to see the other variables.

\begin{figure}[ht]
    \centering
    \includegraphics[width=\textwidth]{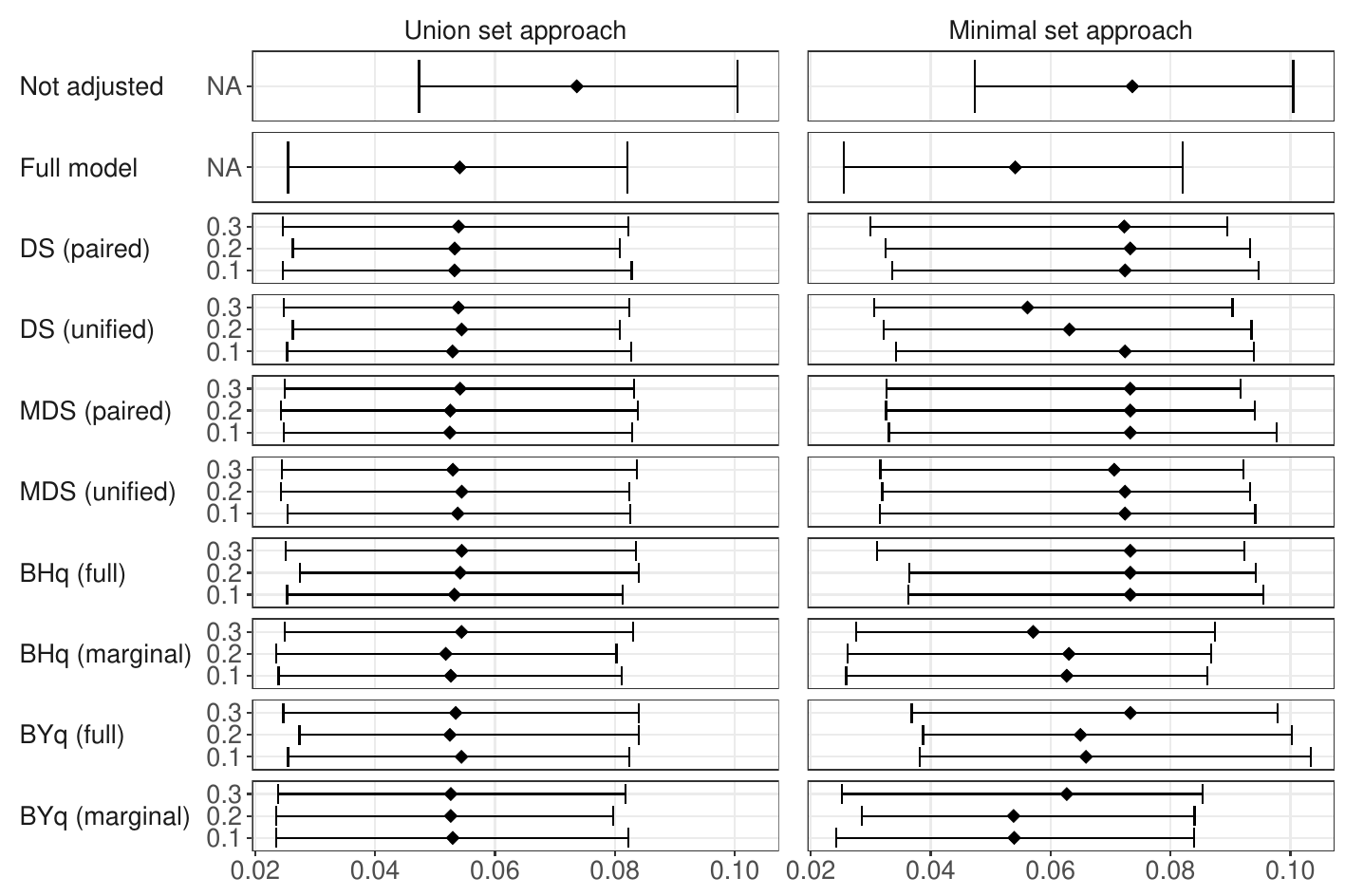}
    \caption{The augmented IPW estimates of the risk difference with 95\% bootstrap CIs. }
    \label{fig:rhc_OR}
\end{figure}

Figure \ref{fig:rhc_OR} shows the estimates of the risk difference with bootstrap 95\% CIs. The intervals were constructed using the percentile method with 1,000 bootstrap iterations. 
The selection based on the union-set approach, regardless of which methods were used, yielded similar estimates to the estimate of the full model. On the contrary, the estimates based on the minimal-set approach were relatively biased to the full-model estimate. The marginal BHq and BYq estimates had relatively small biases, probably because they selected a large number of variables regardless of false discoveries and thus succeeded in incorporating a large number of variables to be adjusted for. In the RHC example, the entire set of potential confounders was carefully selected by the medical experts, so it is believed that a stringent variable selection policy such as the minimal-set approach was not considered to be suitable.

% #################################################################################
\section{Discussion}
We defined the false discovery rate for variable selection approaches in confounder selection: the union-set and the minimal-set approaches. Our proposed methods, the paired and the unified mirrors, are designed to control FDR under both approaches, assuming some symmetries in the distribution of the estimates of the regression coefficients of each potential confounder. Our methods are p-value-free, similar to the original mirror statistic, and can be combined with a variety of estimation methods, including data-adaptive methods such as lasso regression at least formally. Exhaustive numerical experiments have demonstrated that our method effectively controls the FDR corresponding to the confounder selection criteria. Furthermore, these methods have been shown to have higher power compared to the p-value-based methods aimed at controlling FDR for confounder selection, particularly in high-dimensional settings.

However, several challenges persist. Firstly, we have shown the ability to control FDR with finite sample sizes only through numerical experiments. Extending these methods to control FDR with finite samples in confounder selection criteria, similar to the Model-X knockoff framework \citep{Candes2016-mi}, remains an open problem. Secondly, although the MDS demonstrated stabilization in variable selection and enhanced power through repeated estimations by data splitting, it is computationally intensive, particularly with large datasets. Lastly, this study focused on a simple confounding scenario; in real-world applications, the true causal structure often remains unknown, and there is no assurance that conventional criteria will select the correct variable set for identifying causal effects --- a challenge that is not unique to this study but common across other data-driven confounder selection methods. Determining when various methods, including the proposed ones, effectively adjust for confounding --- and when they might fail --- under more realistic situations is an exceedingly important practical question.

%%%%%%%%%%%%%%%%%%%%%%%%%%%%%%%%%%%%%%%%%%%%%%%%%%%%%%%%%%%%%%%%%%%%%%%%%%%%%%%%%%%%%%%%%%%%%%%%%%%%%%%%%%%%%%%%%%%%%%%%%%%%

\section*{Software}
The R source code for numerical experiments and real data analysis is available on the GitHub repository (\url{https://github.com/kazuharu-harada-stat/FDR-control-for-confounder-selection/tree/main}).

\section*{Supplementary Materials}
Supplementary material is available online.

\section*{Acknowledgements}
Kazuharu Harada is partially supported by JSPS KAKENHI Grant Number 22K21286.
Masataka Taguri is partially supported by JSPS KAKENHI Grant Number 24K14862.

%%%%%%%%%%%%%%%%%%%%%%%%%%%%%%%%%%%%%%%%%%%%%%%%%%%%%%%%%%%%%%%%%%%%%%%%%%%%%%%%%%%%%%%%%%%%%%%%%%%%%%%%%%%%%%%%%%%%%%%%%%%%

\bibliographystyle{plainnat}
\bibliography{main}

\clearpage
\appendix
% \section*{Appendix}

% \renewcommand{\thepage}{A\arabic{page}} % ページ番号に"A"を追加
% \setcounter{page}{1} % ページ番号をリセット

\setcounter{figure}{0}
\setcounter{table}{0}
\setcounter{equation}{0}
\renewcommand{\thesection}{Appendix \Alph{section}}
\renewcommand{\thefigure}{A\arabic{figure}}
\renewcommand{\thetable}{A\arabic{table}}
\renewcommand{\theequation}{A\arabic{equation}}
\renewcommand{\thetheorem}{A\arabic{theorem}}
\renewcommand{\thelemma}{A\arabic{lemma}}

\section{Proofs}\label{app:proofs}
\subsection{Proof of Proposition \ref{prp:approx_OR}}
\begin{proof}
    For the variables with $j\notin S_{\mathrm{OR}}$, the joint distribution of $(M^{Y}_j, M^{A}_j)$ are symmetric with respect to the origin $(0,0)$ since $M^{Y}_j$ and $M^{A}_j$ are marginally symmetric about 0 and are independent. 
    For $j\notin S_{\mathrm{OR}}$ and any $t>0$, $\mathbb{P}(M^Y_j > t \vee M^A_j > t) = 1 - \mathbb{P}(M^Y_j \le t \wedge M^A_j \le t) = 1 - \mathbb{P}(M^Y_j \le t)\cdot\mathbb{P}(M^A_j \le t) = 1 - \mathbb{P}(M^Y_j \ge -t)\cdot\mathbb{P}(M^A_j \ge -t) = \mathbb{P}(M^Y_j < -t \vee M^A_j < t)$.
    Then, the proposition holds.
\end{proof}

\subsection{Proof of Proposition \ref{prp:approx_AND}}
\begin{proof}
    The null set $\overline{S_{\mathrm{AND}}}$ is partitioned into three subsets:
    \begin{gather*}
        \overline{S_{\mathrm{AND}}} 
        = \underbrace{(S_{Y}\cap \overline{S_{A}})}_{R_1} 
        \cup \underbrace{(\overline{S_{Y}}\cap S_{A})}_{R_2}  
        \cup \underbrace{(\overline{S_{Y}}\cap \overline{S_{A}})}_{R_3}.
    \end{gather*}
    Now let $\hat{S}_{\mathrm{AND}} = \{j:M^{Y}_j > t \wedge M^{A}_j > t\}$, and for $j\notin S_{\mathrm{AND}}$, the selected probability is decomposed into
    \begin{align*}
        &~ \mathbb{P}(j\in \hat{S}_{\mathrm{AND}}\cap \overline{S_{\mathrm{AND}}}) \\
        =&~ \mathbb{P}(j\in \hat{S}_{\mathrm{AND}} \cap R_1)
        + \mathbb{P}(j\in \hat{S}_{\mathrm{AND}} \cap \in R_2)
        + \mathbb{P}(j\in \hat{S}_{\mathrm{AND}} \cap \in R_3).
    \end{align*}
    By the symmetry of the mirror statistics, these components can be rewritten as
    \begin{align*}
        \mathbb{P}(j\in \hat{S}_{\mathrm{AND}} \cap R_1) 
            =&~ \mathbb{P}(j\in \hat{S}_{1} \cap R_1) \\
        \mathbb{P}(j\in \hat{S}_{\mathrm{AND}} \cap R_2) 
            =&~ \mathbb{P}(j\in \hat{S}_{2} \cap R_2) \\
        \mathbb{P}(j\in \hat{S}_{\mathrm{AND}} \cap R_3) 
            =&~ \mathbb{P}(j\in \hat{S}_{3} \cap R_3) 
    \end{align*}
    On the other hand, the following approximation holds:
    \begin{align*}
        &~ \mathbb{P}(j\in \hat{S}_{1} \cap \overline{S_{\mathrm{AND}}}) \\
        =&~ \mathbb{P}(j\in \hat{S}_{1} \cap R_1)
        + \mathbb{P}(j\in \hat{S}_{1} \cap R_2)
        + \mathbb{P}(j\in \hat{S}_{1} \cap R_3) \\
        \approx&~ \mathbb{P}(j\in \hat{S}_{1} \cap R_1) +
        \mathbb{P}(j\in \hat{S}_{1} \cap R_3),
    \end{align*}
    since $M_j^A$ is supposed to be large for $j \in R_2$ by the second property of the mirror statistic. Similarly, the following approximations hold:
    \begin{align*}
        \mathbb{P}(j\in \hat{S}_{2} \cap \overline{S_{\mathrm{AND}}}) &\approx
        \mathbb{P}(j\in \hat{S}_{2} \cap R_2) +
        \mathbb{P}(j\in \hat{S}_{2} \cap R_3) \\
        \mathbb{P}(j\in \hat{S}_{3} \cap \overline{S_{\mathrm{AND}}}) &\approx
        \mathbb{P}(j\in \hat{S}_{3} \cap R_3).
    \end{align*}
    Then, we have 
    \begin{align*}
        & \mathbb{P}(j\in \hat{S}_{\mathrm{AND}} \cap \overline{S_{\mathrm{AND}}})\\
        =&~ \mathbb{P}(j\in \hat{S}_{\mathrm{AND}} \cap R_1)
        + \mathbb{P}(j\in \hat{S}_{\mathrm{AND}} \cap R_2)
        + \mathbb{P}(j\in \hat{S}_{\mathrm{AND}} \cap R_3) \\
        =&~ \mathbb{P}(j\in \hat{S}_{1} \cap R_1)
        + \mathbb{P}(j\in \hat{S}_{2} \cap R_2)
        + \mathbb{P}(j\in \hat{S}_{3} \cap R_3) \\
        \approx&~ \mathbb{P}(j\in \hat{S}_{1}\cap\overline{S_{\mathrm{AND}}}) + \mathbb{P}(j\in \hat{S}_{2}\cap\overline{S_{\mathrm{AND}}}) - \mathbb{P}(j\in \hat{S}_{3}\cap\overline{S_{\mathrm{AND}}}),
    \end{align*}
    which implies that the proposition holds.
\end{proof}

\subsection{Proof of Proposition \ref{prp:sym_unif_AND}}
\begin{proof}
    Let $f_3\left(\theta^{(1)}_j, \theta^{(2)}_j\right) = \sin2\theta^{(1)}_j\cdot\sin2\theta^{(2)}_j$. The null set $\overline{S_{\mathrm{AND}}}$ is partitioned into three subsets: $S_{Y}\cap \overline{S_{A}}$, $\overline{S_{Y}}\cap S_{A}$, and $\overline{S_{Y}}\cap \overline{S_{A}}$.
    When $j \in S_{Y}\cap \overline{S_{A}}$, let $\mathbf{T}_j^{(m)}$ is drawn from an y-axis symmetric distribution $F$ with location $(0, c)$, assuming $c>0$ without loss of generality. We have $\mathbb{P}(\theta_j^{(m)} \in (0,\pi/2]) = \mathbb{P}(\theta_j^{(m)} \in (\pi/2,\pi])$, which implies $\mathbb{P}\left(\sin{2\theta_j^{(m)}} \ge 0\right) = \mathbb{P}\left(\sin{2\theta_j^{(m)}} < 0\right) = 1/2$. Since $\mathbf{T}_j^{(1)}$ and $\mathbf{T}_j^{(2)}$ are independent, $\mathbb{P}\left(f_3\left(\theta^{(1)}_j, \theta^{(2)}_j\right) \ge 0\right) = \mathbb{P}\left(f_3\left(\theta^{(1)}_j, \theta^{(2)}_j\right) < 0\right) = 1/2$.
    The cases of $\overline{S_{Y}}\cap S_{A}$ and $\overline{S_{Y}}\cap \overline{S_{A}}$ can be shown in a similar manner to case of $j \in S_{Y}\cap \overline{S_{A}}$.
\end{proof}

\clearpage
\section{Simulation Settings}\label{app:settings}
\subsection{Choise of Functional Forms and Asymptotic Properties}
The data generation model used for the selection of functional forms and the evaluation of asymptotic properties in Section 5.1 of the main text is as follows:
\begin{itemize}
    \item The sample size is set to $\{500, 1000, 2000, 4000, 8000\}$.
    \item The candidate variables $X$ are generated from a multivariate normal distribution with $p=100$. The mean vector is zero, and the covariance matrix is a block diagonal matrix, with each block being a 5-dimensional Toeplitz matrix. These blocks have 1s on the diagonal and $\rho^{|k-j|}$ as the off-diagonal components, where $k$ and $j$ are the indices within each block. 
    \item The treatment $A$ is binary. We generate $A$ from a Bernoulli distribution with 
    \begin{gather*}
        \mathbb{P}(A=1 \mid X) = g_A^{-1}(X^T{\bs\alpha}),
    \end{gather*}
    where $g_A$ is the logit link function. 
    \item For the continuous outcome, we generate $Y$ as 
    \begin{gather*}
        Y = \tau A + X^T{\bs\beta} + \varepsilon,
    \end{gather*}
    where $\varepsilon\sim N(0,1)$. For the binary outcome, we draw $Y$ from a Bernoulli distribution with 
    \begin{gather*}
        \mathbb{P}(Y=1 \mid A, X) = g_Y^{-1}(\tau A + X^T{\bs\beta}),
    \end{gather*}
    where $g_Y$ is the logit link function.
    \item The true relevance sets $S_Y$ and $S_Y$ are randomly selected 30 out of 100 variables so that $|S_Y\cap S_A| = 15$, respectively.
    \item When $Y$ is continuous, $\beta_j$ is set to $\pm 0.08$ with equal probability for $j\in S_Y$. When $Y$ is binary, $\beta_j$ is set to be $\pm 0.16$ with equal probability for $j\in S_Y$. For $j\in S_A$, $\alpha_j$ is set to be $\pm 0.16$ with equal probability. For non-relevant variables, the corresponding coefficients are set to zero.
\end{itemize}

\subsection{FDR, Power, and the Bias of ATE Estimates}
In the comparative experiments of Section 5, the data generation model similar to the previous section is used, although there are some different settings. In the low-dimensional setting, the data generation model differs from the previous section in the following points:
\begin{itemize}
    \item The sample size is set to 1000.
    \item The dimension of $X$ is set to 90. 
    \item When $X$ is Gaussian, it is the same as in Section 5.1. When $X$ is binary, we generate multi-dimensional binary variables with correlations by binarizing the generated Gaussian variables at zero in each dimension.
    \item The sets of truly relevant variables, as well as the signs and magnitudes of the regression coefficients, are not randomized but fixed. The fixed values are summarized in the table below. This approach is chosen because, when set randomly, the bias in estimating the ATE using simple group differences would vary across samples, rendering it unsuitable for bias evaluation in ATE estimation.
    \item Since the entire $X$ is 90-dimensional, there are not enough correlated blocks to realize the structure in Table A. Then, in the low-dimensional experiments, after generating $X$ of 150 dimensions once, $X$ is generated by picking up 45 dimensions of relevant variables and 15 dimensions of null variables.
\end{itemize}

In high-dimensional settings, the data generation model is similar to those in the low-dimensional cases except for the following points:
\begin{itemize}
    \item The sample size is set to 1000.
    \item The dimension of $X$ is set to $\{500, 1000, 1500\}$.
\end{itemize}
Note that the term ``high-dimensional'' often means situations with $N \le p$, while we treat the setting $N=1000, p=500$ as high-dimensional. This is because the proposed methods split the dataset and can use only half $N$, which is high-dimensional in that the ordinary estimation methods such as MLE and OLS do not work.

\begin{table}[ht]
    \small
    \centering
    \caption{Setting of the true coefficients in the comparative experiments. Variables identified by Block no. and Index in blocks have signs corresponding to the relevance in the table for each model of $Y$ and $A$. The strength of the regression coefficients is set to 0.1 for continuous responses and 0.2 for binary outcome responses, multiplied by the magnitude for each variable (variables with a blank magnitude are multiplied by 1). For binary X, these are further doubled. Variables within the same blocks are correlated, as indicated by the ``Correlated'' column.}
    \label{tab:true_coef} \vspace{1mm}
    \begin{tabular}{ccccccc} 
    \hline
Index & Relevance to Y & Relevance to A & Block no. & Index in blocks & Magnitude & Correlated \\ \hline
1 & $+$& $+$& 1 & 1 & $\times$1.25 & \checkmark \\
2 & $+$& $+$& 1 & 2 &  & \checkmark \\
3 & $+$& $+$& 1 & 3 &  & \checkmark \\
4 & $+$& $+$& 1 & 4 & $\times$0.75 & \checkmark \\
5 & $+$& $+$& 2 & 1 & $\times$1.25 &  \\
6 & $+$& $+$& 3 & 1 &  &  \\
7 & $+$& $+$& 4 & 1 & $\times$0.75 &  \\ \hline
8 & $+$& $-$& 5 & 1 &  & \checkmark \\
9 & $+$& $-$& 5 & 2 &  & \checkmark \\
10 & $+$& $-$& 6 & 1 &  &  \\
11 & $+$& $-$& 7 & 1 & $\times$0.75 &  \\ \hline
12 & $-$& $+$& 8 & 1 &  & \checkmark \\
13 & $-$& $+$& 8 & 2 &  & \checkmark \\
14 & $-$& $+$& 9 & 1 &  &  \\
15 & $-$& $+$& 10 & 1 & $\times$0.75 &  \\ \hline
16 & $+$& $0$& 11 & 1 & $\times$1.25 & \checkmark \\
17 & $+$& $0$& 11 & 2 &  & \checkmark \\
18 & $+$& $0$& 11 & 3 &  & \checkmark \\
19 & $+$& $0$& 11 & 4 & $\times$0.75 & \checkmark \\
20 & $+$& $0$& 12 & 1 & $\times$1.25 &  \\
21 & $+$& $0$& 13 & 1 &  &  \\
22 & $+$& $0$& 14 & 1 & $\times$0.75 &  \\ \hline
23 & $+$& $0$& 15 & 1 &  & \checkmark \\
24 & $+$& $0$& 15 & 2 &  & \checkmark \\
25 & $+$& $0$& 16 & 1 &  &  \\
26 & $+$& $0$& 17 & 1 & $\times$0.75 &  \\ \hline
27 & $-$& $0$& 18 & 1 &  & \checkmark \\
28 & $-$& $0$& 18 & 2 &  & \checkmark \\
29 & $-$& $0$& 19 & 1 &  &  \\
30 & $-$& $0$& 20 & 1 & $\times$0.75 &  \\ \hline
31 & $0$& $+$& 21 & 1 & $\times$1.25 & \checkmark \\
32 & $0$& $+$& 21 & 2 &  & \checkmark \\
33 & $0$& $+$& 21 & 3 &  & \checkmark \\
34 & $0$& $+$& 21 & 4 & $\times$0.75 & \checkmark \\
35 & $0$& $+$& 22 & 1 & $\times$1.25 &  \\
36 & $0$& $+$& 23 & 1 &  &  \\
37 & $0$& $+$& 24 & 1 & $\times$0.75 &  \\ \hline
38 & $0$& $-$& 25 & 1 &  & \checkmark \\
39 & $0$& $-$& 25 & 2 &  & \checkmark \\
40 & $0$& $-$& 26 & 1 &  &  \\
41 & $0$& $-$& 27 & 1 & $\times$0.75 &  \\ \hline
42 & $0$& $+$& 28 & 1 &  & \checkmark \\
43 & $0$& $+$& 28 & 2 &  & \checkmark \\
44 & $0$& $+$& 29 & 1 &  &  \\
45 & $0$& $+$& 30 & 1 & $\times$0.75 &  \\
\hline
    \end{tabular}
\end{table}

\subsection{Parameter Scaling}
When we use paired mirrors, we combine two different mirror statistics, so proper scaling is required to select variables based on a common threshold $t$ for both the outcome and the treatment. Similarly, it is better to scale the regression coefficients before constructing the unified mirrors. When using the GLM or cross-fitting, we standardize the estimated regression coefficients using their standard errors and construct the mirror statistics using the standardized estimates. When using the lasso for parameter estimation, standardization is done in an \textit{ad hoc} manner, using estimates with GLM-like standard error estimates. Using the debiased lasso might seem like a sensible idea, but as is well-known, the approximation of the precision matrix is computationally heavy. Therefore, we determined that it is not suitable for the proposed method of repeatedly selecting variables in MDS.

\clearpage
\newgeometry{left=2.5cm,right=1cm}
\begin{landscape}
\section{Definitions of the variables in the RHC datasets}
The definition of variables is based on \url{https://hbiostat.org/data/repo/rhc} (Accessed: 15 May, 2024) and \url{https://hbiostat.org/data/repo/crhc} (Accessed: 15 May, 2024).
\vspace{-3mm}

\begin{table}[ht]
    \small
    \centering
    \caption{Variables in RHC dataset}\vspace{2mm}
    \begin{tabular}{cp{5cm}p{8cm}p{5cm}}
\hline
No. & Original Variable Names & Definition & Note \\
\hline
1 & age & Age &  \\
2 & sex & Sex (Female) & Ref. class is Male \\
3 & race & Race (Black) & Ref. class is White \\
4 & race & Race (Other) & Ref. class is White \\
5 & edu & Years of education &  \\
6 & income & Income (\$11k-\$25k) & Ref. class is Under \$11k \\
7 & income & Income (\$25k-\$50k) & Ref. class is Under \$11k \\
8 & income & Income (> \$50k) & Ref. class is Under \$11k \\
9 & insurance & Insurance (Medicare) & Ref. class is Private \\
10 & insurance & Insurance (Private \& Medicare) & Ref. class is Private \\
11 & insurance & Insurance (Medicaid) & Ref. class is Private \\
12 & insurance & Insurance (Medicare \& Medicaid) & Ref. class is Private \\
13 & insurance & Insurance (No Insurance) & Ref. class is Private \\
14 & cat1 & Primary disease (ARF) & Ref. class is Others \\
15 & cat1 & Primary disease (MOSF) & Ref. class is Others \\
16 & cat1 & Primary disease (CHF) & Ref. class is Others \\
17 & cat2 & Secondary disease (Other) & Ref. class is None \\
18 & cat2 & Secondary disease (MOSF) & Ref. class is None \\
19 & resp & Admission Diagnosis (Respiratory) & Yes/No \\
20 & card & Admission Diagnosis (Cardiovascular) & Yes/No \\
21 & neuro & Admission Diagnosis (Neurological) & Yes/No \\
22 & renal & Admission Diagnosis (Renal) & Yes/No \\
23 & meta & Admission Diagnosis (Metabolic) & Yes/No \\
24 & hema & Admission Diagnosis (Hematologic) & Yes/No \\
25 & seps & Admission Diagnosis (Sepsis) & Yes/No \\
26 & das2d3pc & DASI ( Duke Activity Status Index) &  \\
27 & dnr1Yes & DNR status on day1 & Yes/No \\
\hline \\
    \end{tabular}
    \label{app:rhc_varaibles}
\end{table}

\addtocounter{table}{-1}
\begin{table}[ht]
    \small
    \centering
    \caption{Variables in RHC dataset (continued)}\vspace{2mm}
    \begin{tabular}{cp{5cm}p{8cm}p{5cm}}
\hline
No. & Original Variable Names & Definition & Note \\
\hline
28 & cancer & Cancer (Yes) & Ref. class is No \\
29 & cancer & Cancer (Metastatic) & Ref. class is No \\
30 & surv2md1 & Support model estimate of the prob. of surviving 2 months &  \\
31 & aps1 & APACHE score &  \\
32 & scoma1 & Glasgow Coma Score &  \\
33 & temp1 & Temperature &  \\
34 & meanbp1 & Mean blood pressure &  \\
35 & resp1 & Respiratory rate &  \\
36 & hrt1 & Heart rate &  \\
37 & pafi1 & PaO2/FIO2 ratio &  \\
38 & paco21 & PaCo2 &  \\
39 & ph1 & Ph &  \\
40 & wblc1 & WBC &  \\
41 & hema1 & Hematocrit &  \\
42 & sod1 & Sodium &  \\
43 & pot1 & Potassium &  \\
44 & crea1 & Creatinine &  \\
45 & bili1 & Billirubin &  \\
46 & alb1 & Albumin &  \\
47 & cardiohx1 & Cardio & Yes/No \\
48 & chfhx1 & Comorbidities (cardiovascular) & Yes/No \\
49 & dementhx1 & Comorbidities (dementia) & Yes/No \\
50 & psychhx1 & Comorbidities (psychiatric) & Yes/No \\
51 & chrpulhx1 & Comorbidities (chronic pulmonary) & Yes/No \\
52 & renalhx1 & Comorbidities (renal) & Yes/No \\
53 & liverhx1 & Comorbidities (liver) & Yes/No \\
54 & gibledhx1 & Comorbidities (Upper GI bleeding) & Yes/No \\
55 & malighx1 & Comorbidities (malignancies) & Yes/No \\
56 & immunhx1 & Comorbidities (immunology) & Yes/No \\
57 & transhx1 & Comorbidities (transfer from another hosp.) & Yes/No \\
58 & amihx1 & Comorbidities (myocardial infarction) & Yes/No \\
\hline \\
    \end{tabular}
\end{table}

\end{landscape}

\restoregeometry

\clearpage
\section{Other Simulation Results}\label{app:other_results}
\subsection{Choice of Functional Forms and Large-sample Behavior}
In addition to the following figures, a supplementary spreadsheet is provided in the Web Appendix.

\begin{figure}[ht]
    \centering
    \includegraphics[width=\textwidth]{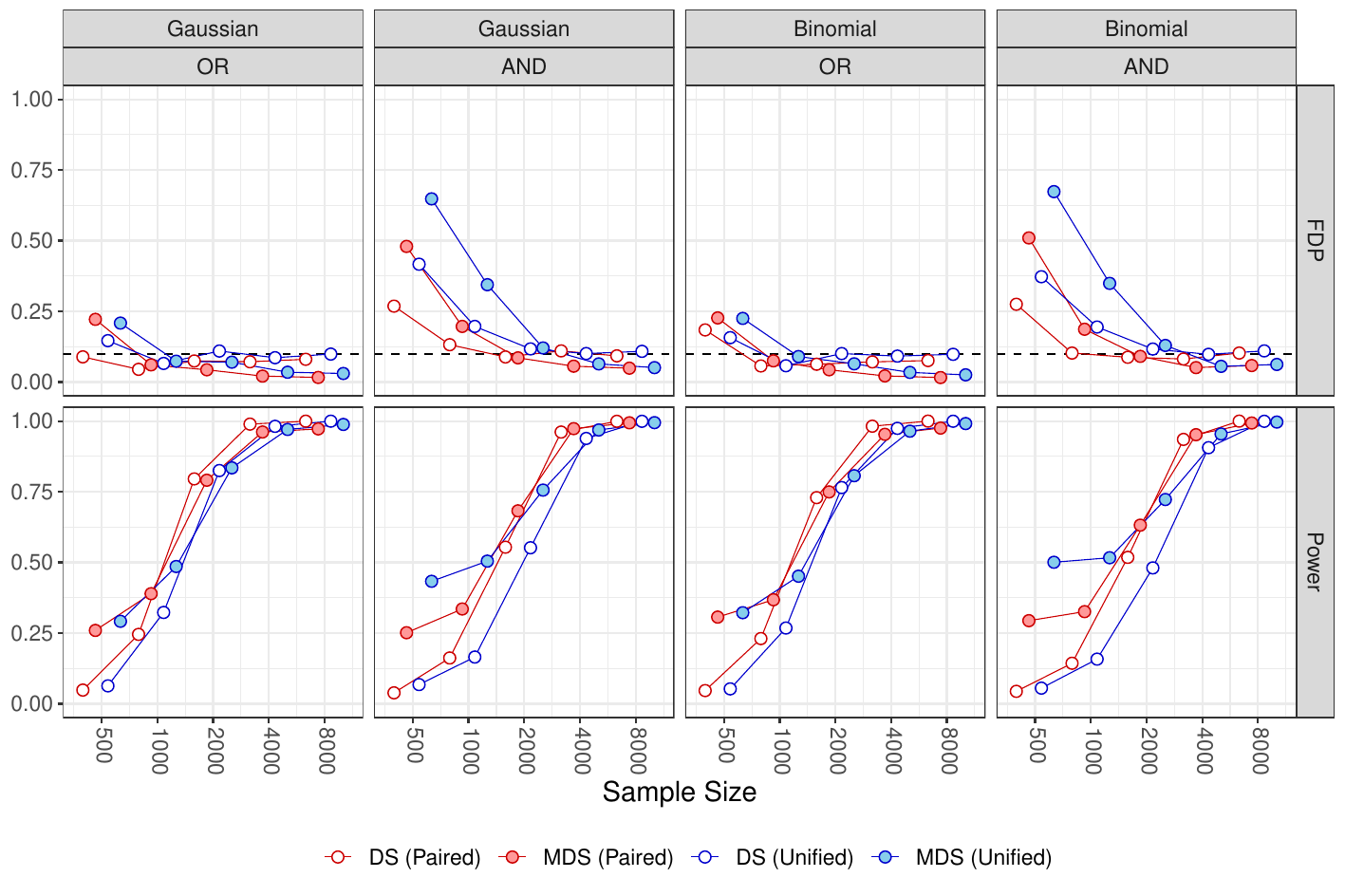}
    \caption{Mean FDP and power of the proposed methods with increasing sample size, using MLE for parameter estimation. The terms ``OR'' and ``AND'' represent the selection and evaluation criteria, respectively. The outcome distributions are depicted at the top of the figures.}
\end{figure}

\begin{figure}[ht]
    \centering
    \includegraphics[width=\textwidth]{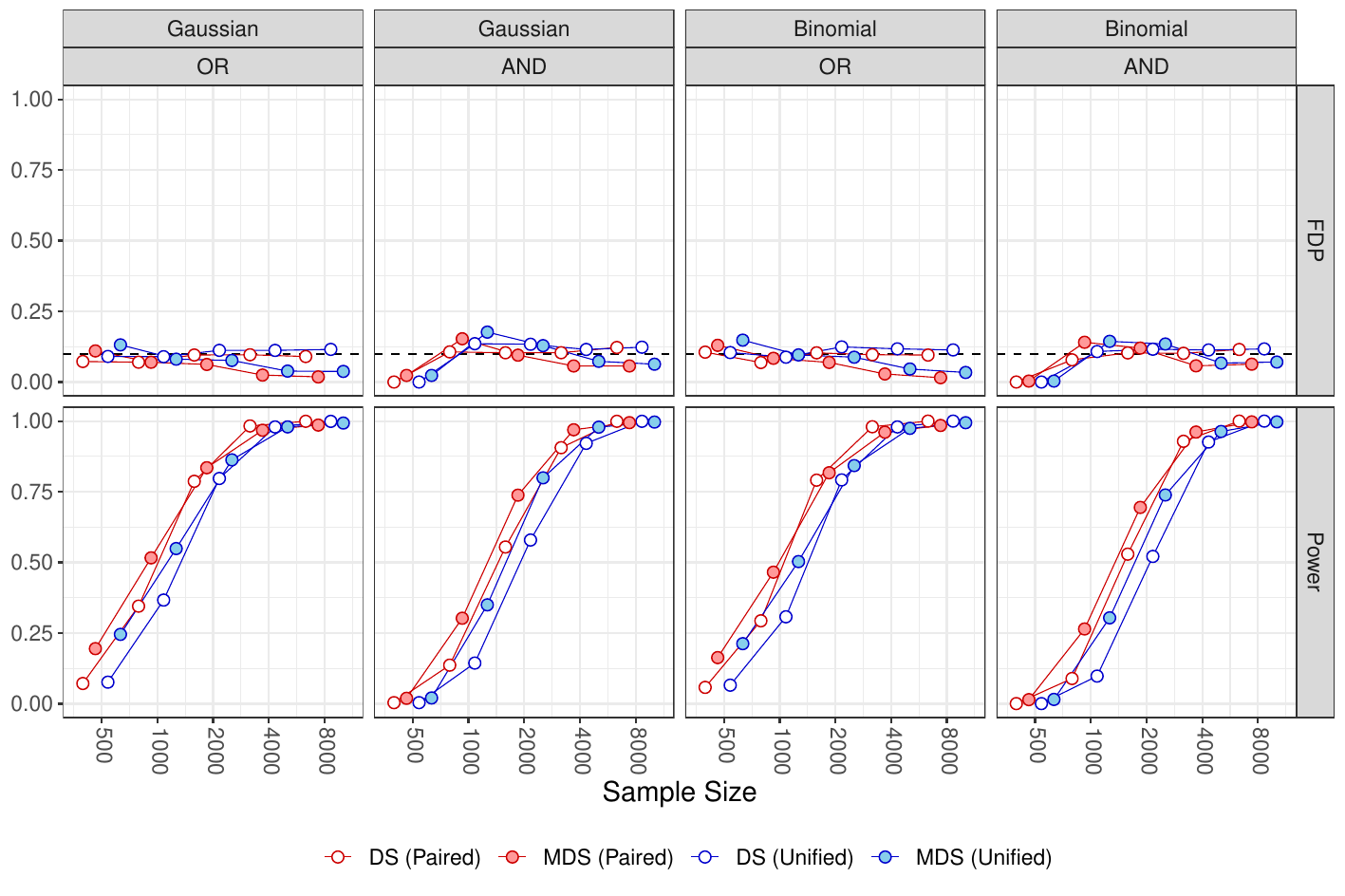}
    \caption{Mean FDP and power of the proposed methods with increasing sample size, using the lasso for parameter estimation. The terms ``OR'' and ``AND'' represent the selection and evaluation criteria, respectively. The outcome distributions are depicted at the top of the figures.}
\end{figure}

\clearpage
\subsection{Low-dimensional Potential Confounders}

\begin{figure}[ht]
    \centering
    \includegraphics[width=\textwidth]{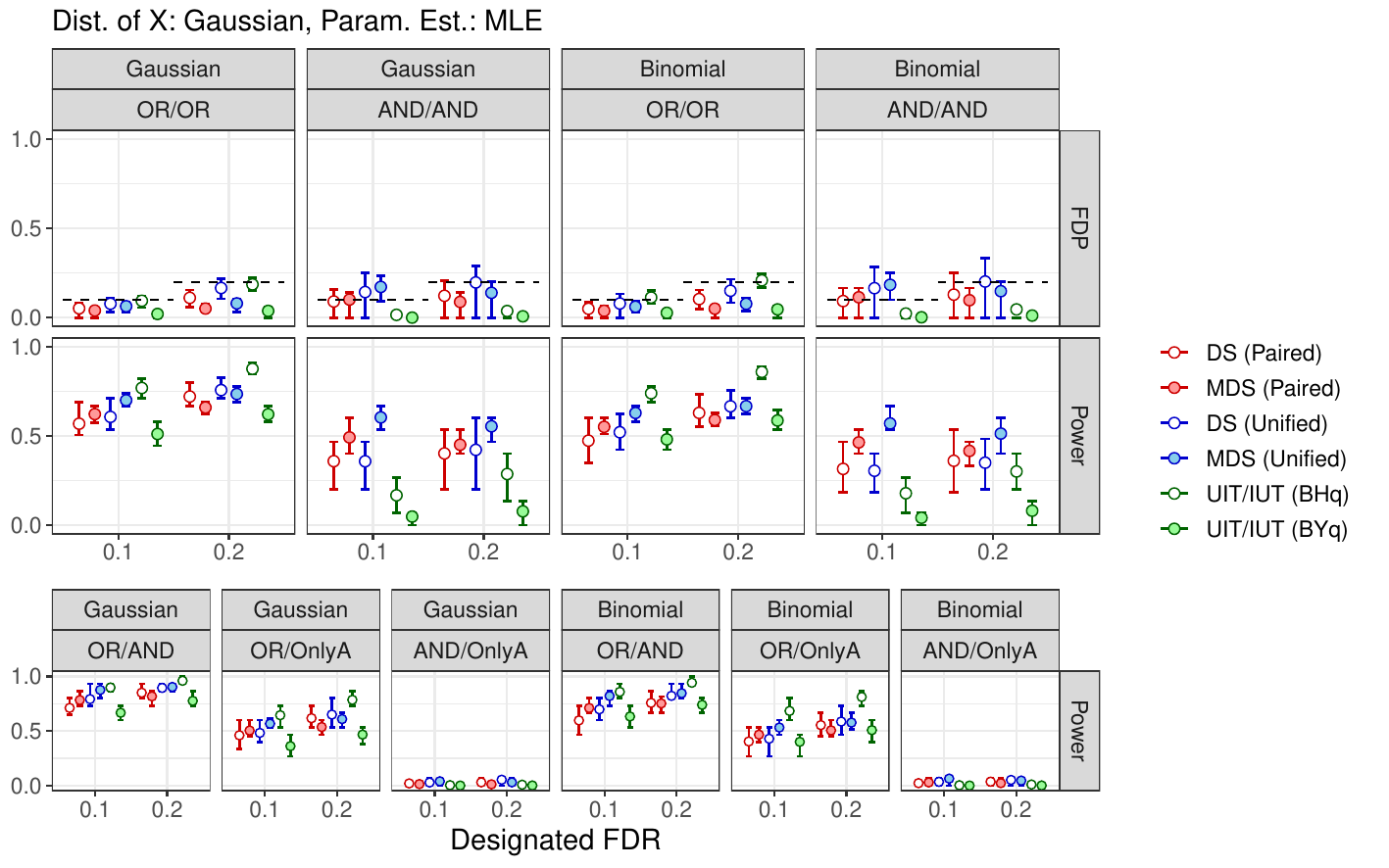}
    \caption{
        FDR and power analysis of the proposed and q-value-based methods in low-dimensional Gaussian settings. Parameter estimates for the mirror statistics were obtained using GLM. The left block presents the scenarios with continuous outcomes, and the right block shows the results for binary outcomes. The notation ``**/**'' indicates the criteria for variable selection and the set of true variables used for evaluation. Each point is a mean value over 100 simulations, and the error bars at each point represent the 1st to 3rd quartiles.}
\end{figure}

\begin{figure}[ht]
    \centering
    \includegraphics[width=\textwidth]{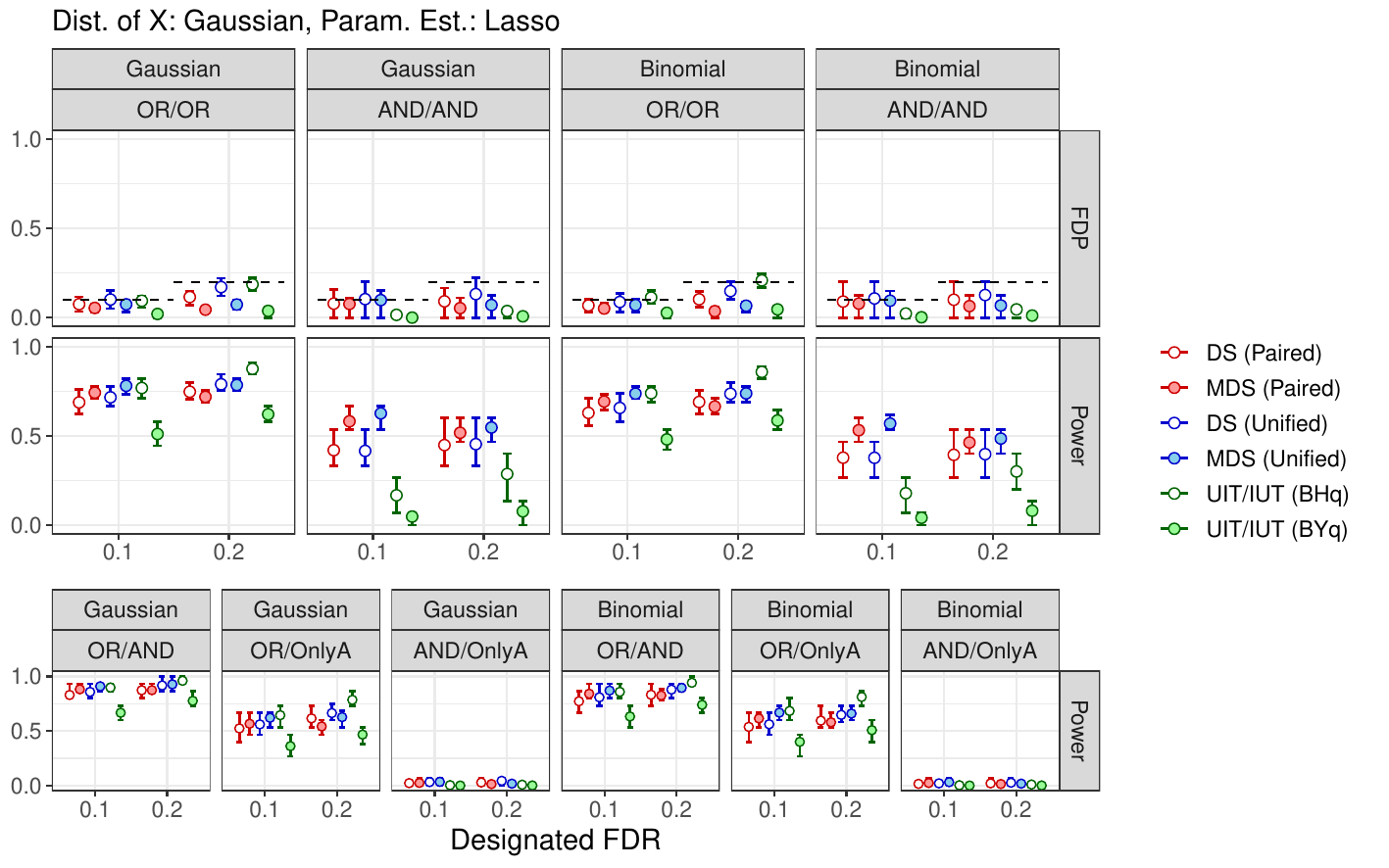}
    \caption{
        FDR and power analysis of the proposed and q-value-based methods in low-dimensional Gaussian settings. Parameter estimates for the mirror statistics were obtained using the lasso. The left block presents the scenarios with continuous outcomes, and the right block shows the results for binary outcomes. The notation ``**/**'' indicates the criteria for variable selection and the set of true variables used for evaluation. Each point is a mean value over 100 simulations, and the error bars at each point represent the 1st to 3rd quartiles.}
\end{figure}

\begin{figure}[ht]
    \centering
    \includegraphics[width=\textwidth]{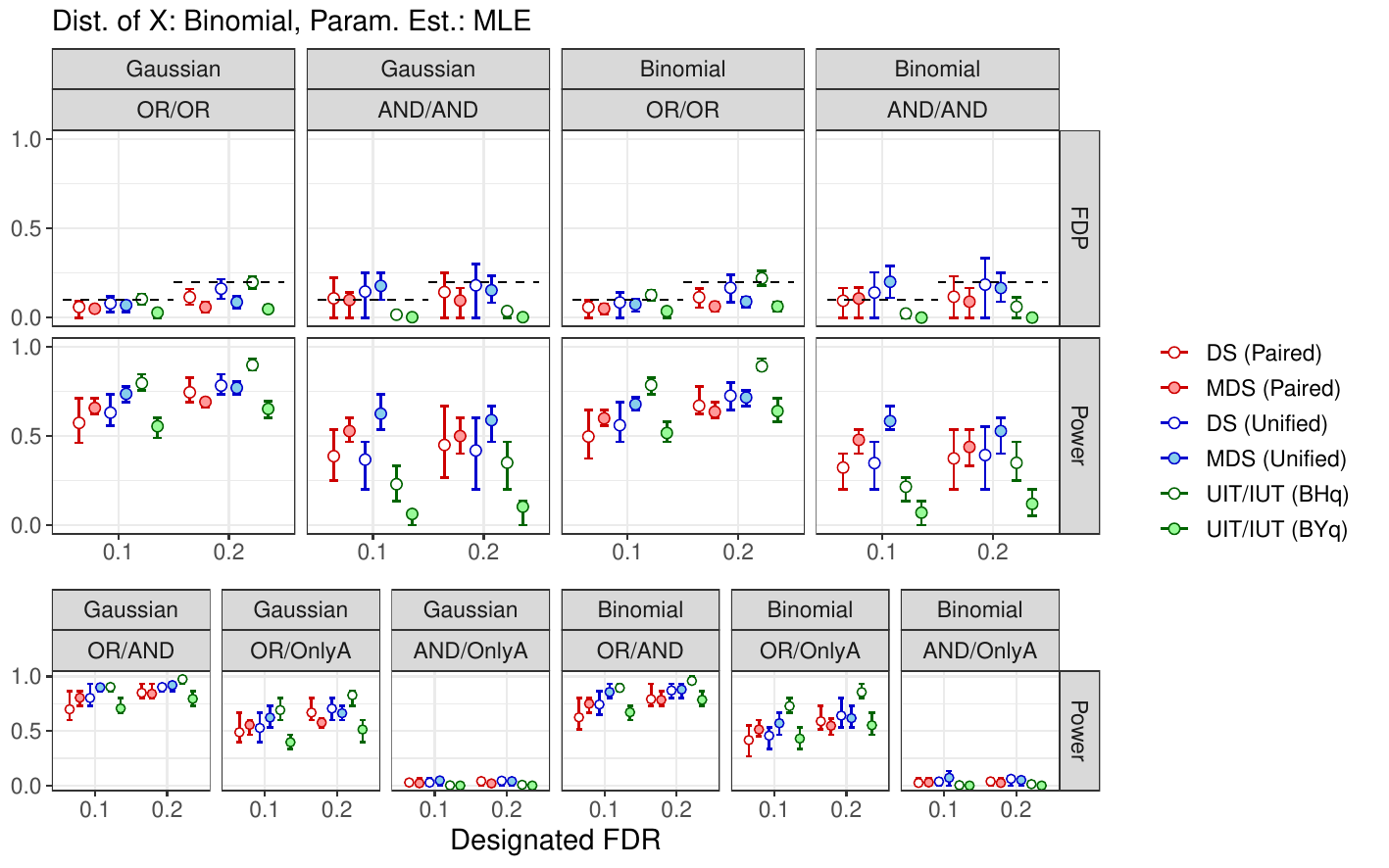}
    \caption{
        FDR and power analysis of the proposed and q-value-based methods in low-dimensional binomial settings. Parameter estimates for the mirror statistics were obtained using GLM. The left block presents the scenarios with continuous outcomes, and the right block shows the results for binary outcomes. The notation ``**/**'' indicates the criteria for variable selection and the set of true variables used for evaluation. Each point is a mean value over 100 simulations, and the error bars at each point represent the 1st to 3rd quartiles.}
\end{figure}

\begin{figure}[ht]
    \centering
    \includegraphics[width=\textwidth]{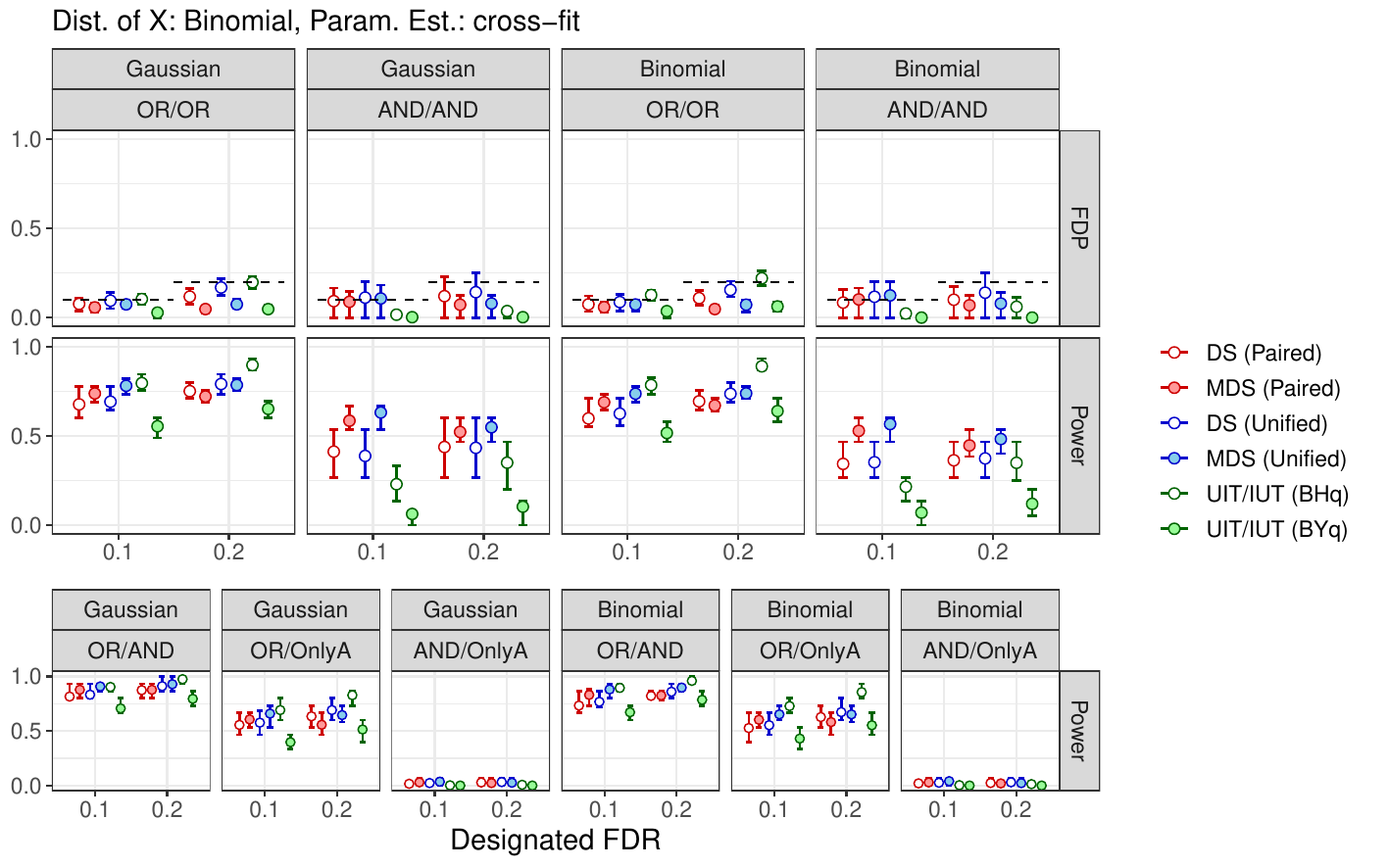}
    \caption{
        FDR and power analysis of the proposed and q-value-based methods in low-dimensional binomial settings. Parameter estimates for the mirror statistics were obtained using cross-fitting. The left block presents the scenarios with continuous outcomes, and the right block shows the results for binary outcomes. The notation ``**/**'' indicates the criteria for variable selection and the set of true variables used for evaluation. Each point is a mean value over 100 simulations, and the error bars at each point represent the 1st to 3rd quartiles.}
\end{figure}

\begin{figure}[ht]
    \centering
    \includegraphics[width=\textwidth]{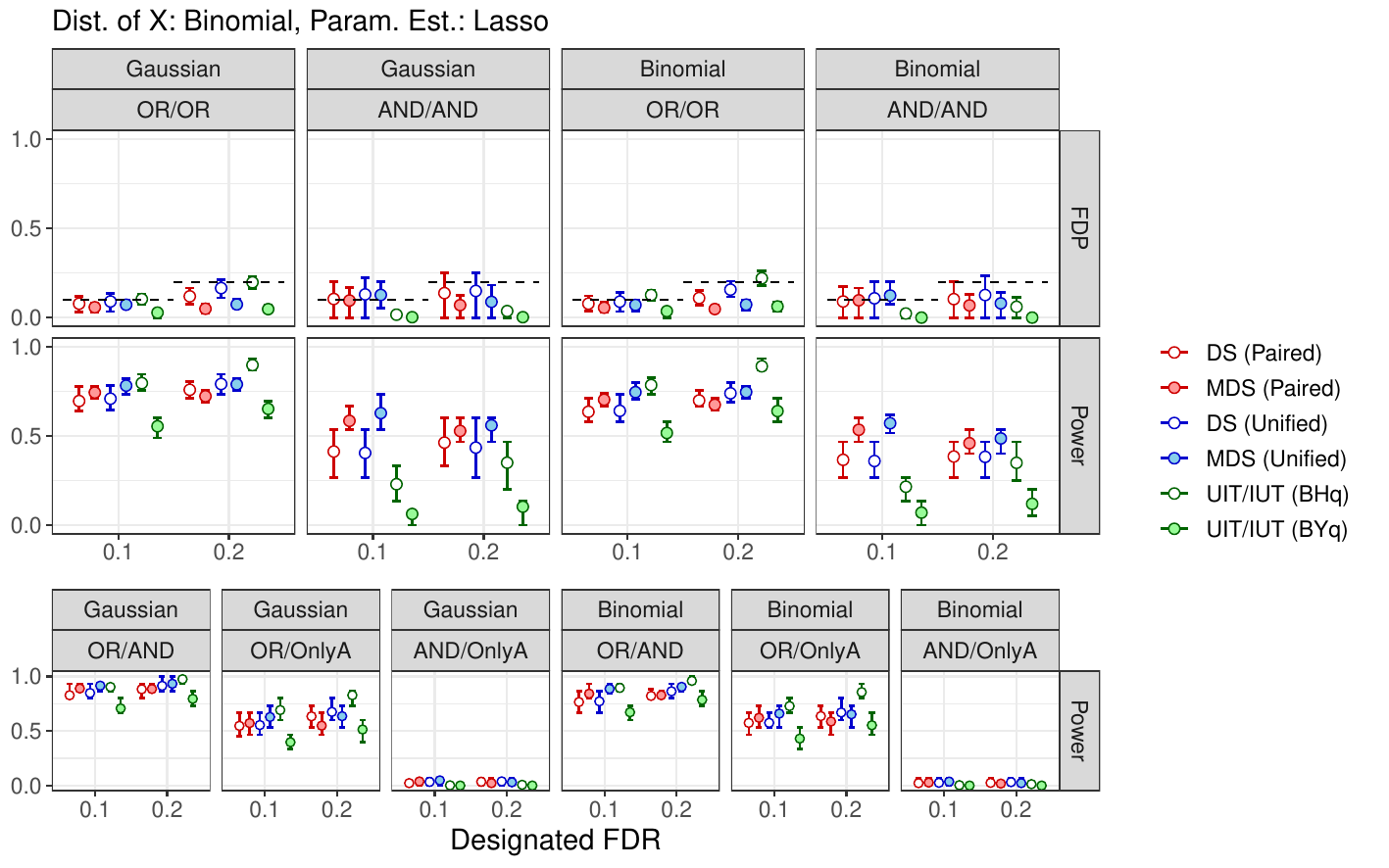}
    \caption{
        FDR and power analysis of the proposed and q-value-based methods in low-dimensional binomial settings. Parameter estimates for the mirror statistics were obtained using the lasso. The left block presents the scenarios with continuous outcomes, and the right block shows the results for binary outcomes. The notation ``**/**'' indicates the criteria for variable selection and the set of true variables used for evaluation. Each point is a mean value over 100 simulations, and the error bars at each point represent the 1st to 3rd quartiles.}
\end{figure}

\begin{figure}[ht]
    \centering
    \includegraphics[width=\textwidth]{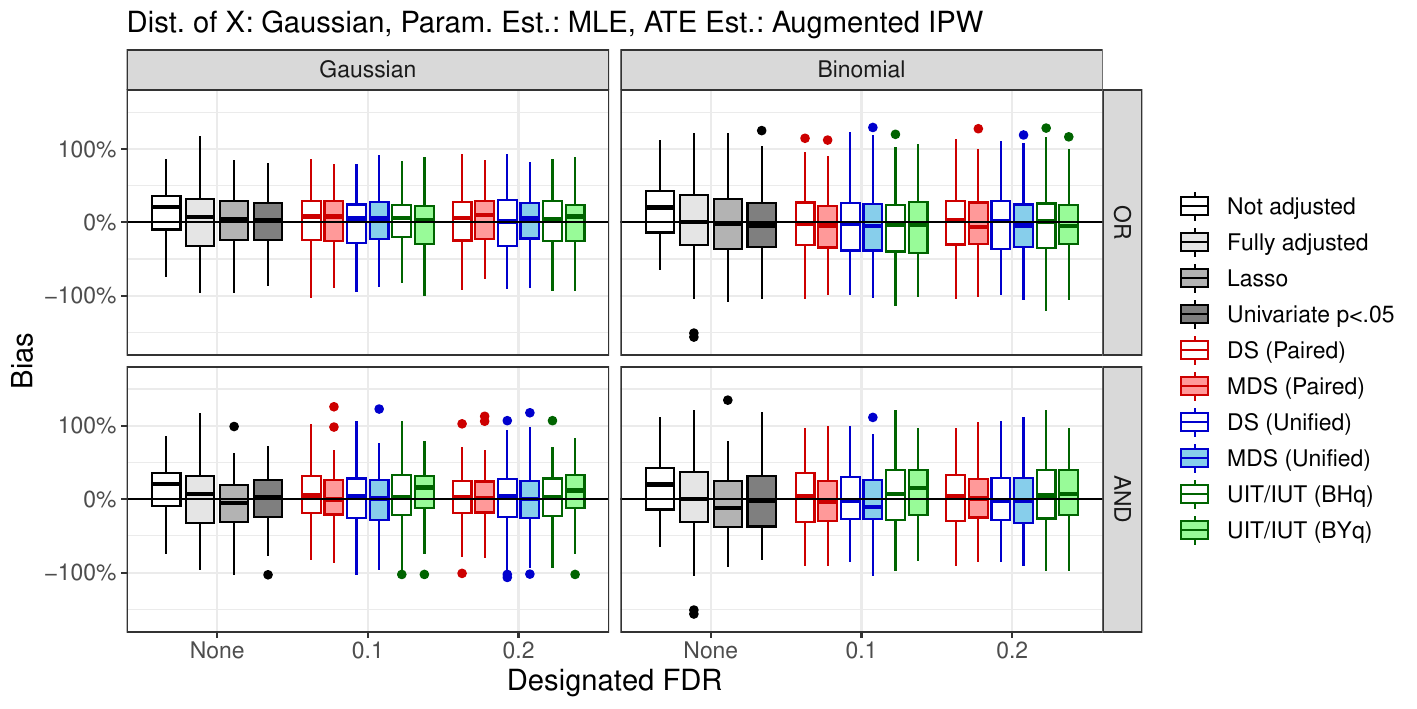}
    \caption{Relative Bias in ATE estimates obtained from 100 simulations. The parameter estimates for the mirror statistics were obtained using GLM, and the augmented IPW estimators were used to estimate ATE. The distribution of X is Gaussian.}
\end{figure}

\begin{figure}[ht]
    \centering
    \includegraphics[width=\textwidth]{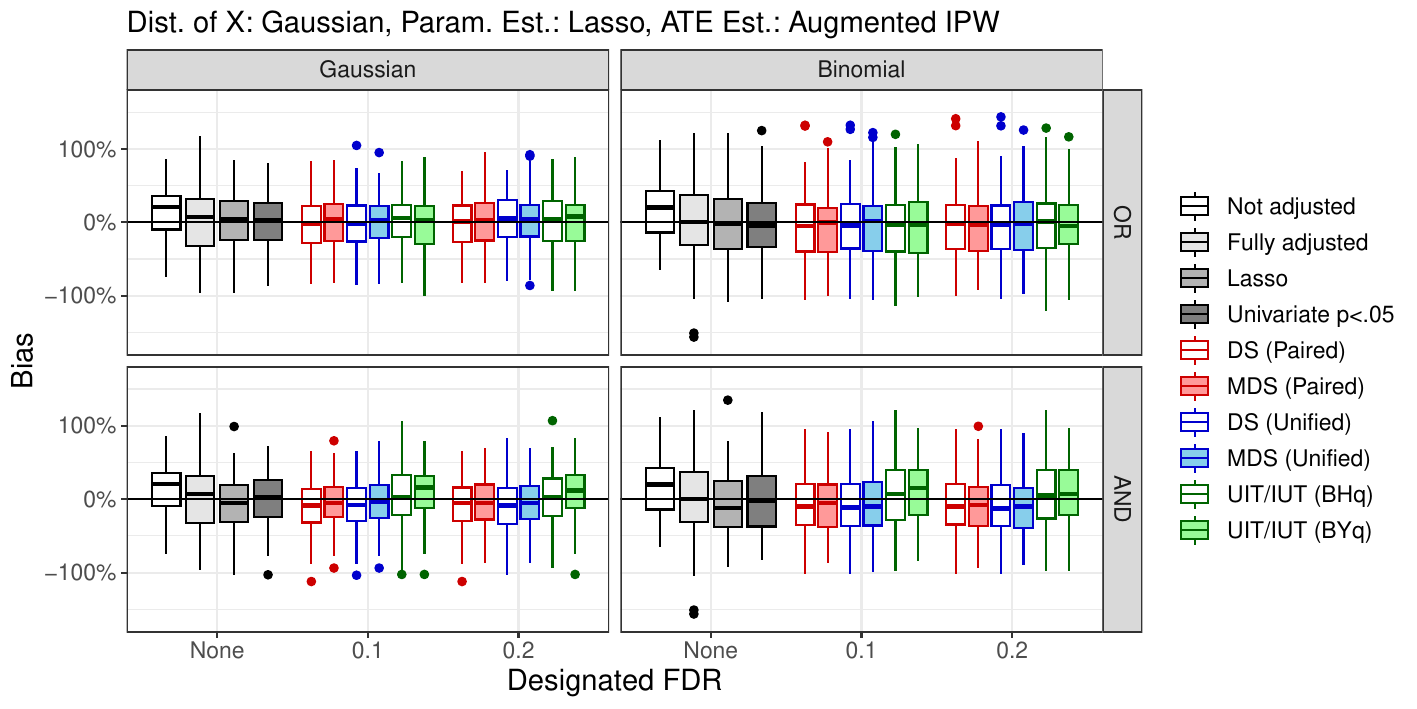}
    \caption{Relative Bias in ATE estimates obtained from 100 simulations. The parameter estimates for the mirror statistics were obtained using the lasso, and the augmented IPW estimators were used to estimate ATE. The distribution of X is Gaussian.}
\end{figure}

\begin{figure}[ht]
    \centering
    \includegraphics[width=\textwidth]{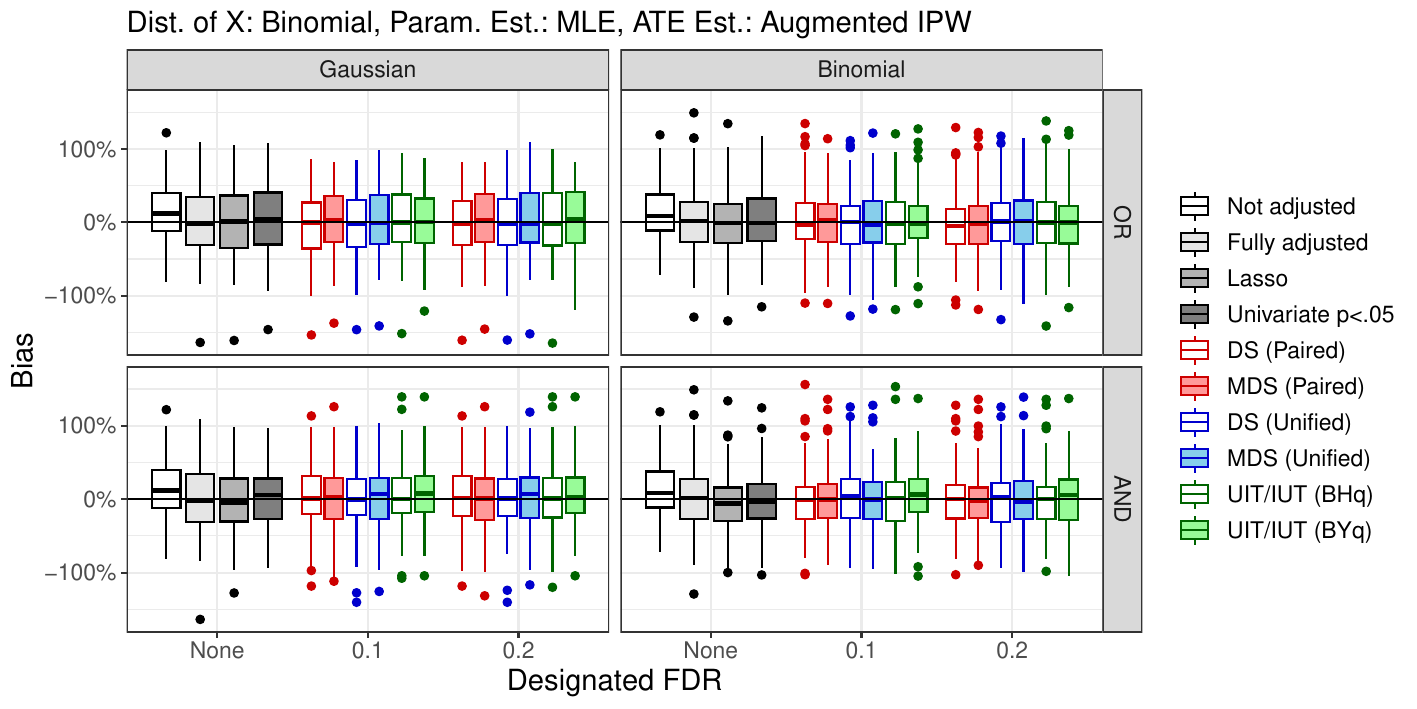}
    \caption{Relative Bias in ATE estimates obtained from 100 simulations. The parameter estimates for the mirror statistics were obtained using GLM, and the augmented IPW estimators were used to estimate ATE. The distribution of X is binomial.}
\end{figure}

\begin{figure}[ht]
    \centering
    \includegraphics[width=\textwidth]{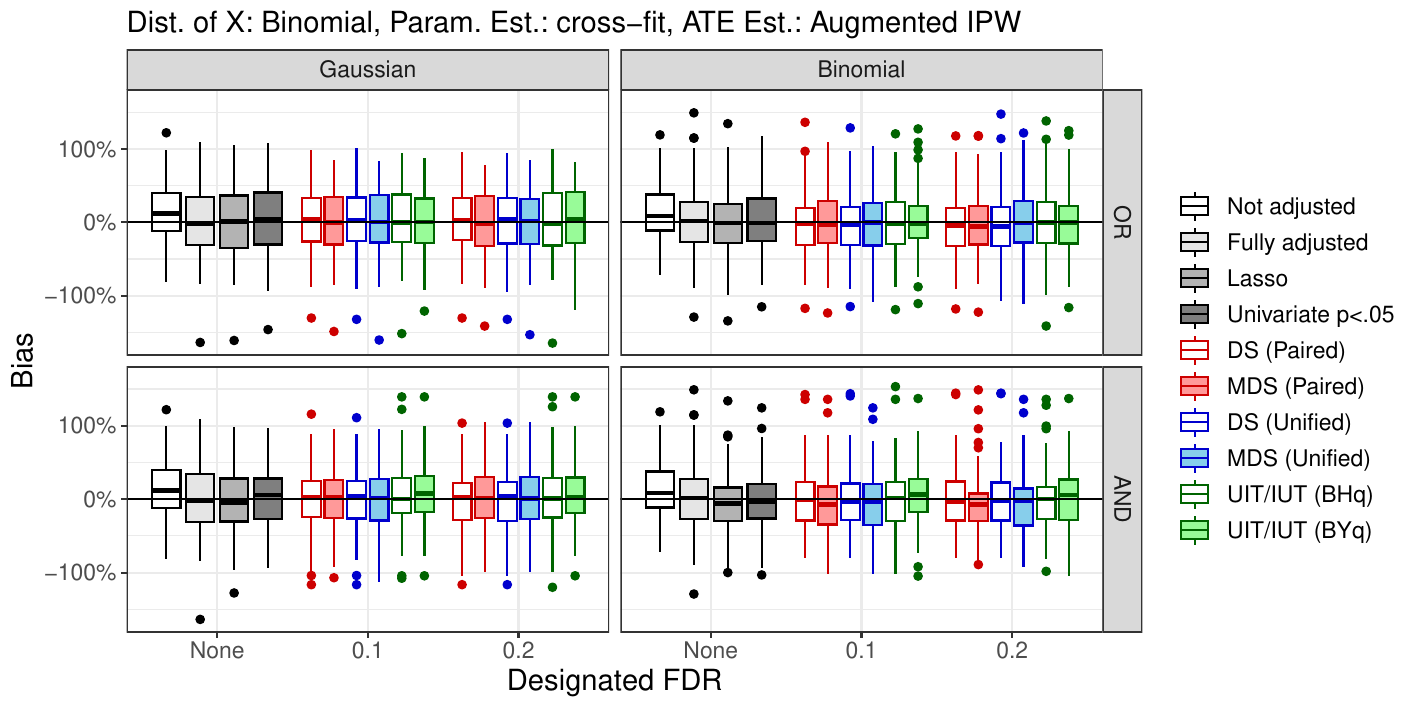}
    \caption{Relative Bias in ATE estimates obtained from 100 simulations. The parameter estimates for the mirror statistics were obtained using the cross-fitting, and the augmented IPW estimators were used to estimate ATE. The distribution of X is binomial.}
\end{figure}

\begin{figure}[ht]
    \centering
    \includegraphics[width=\textwidth]{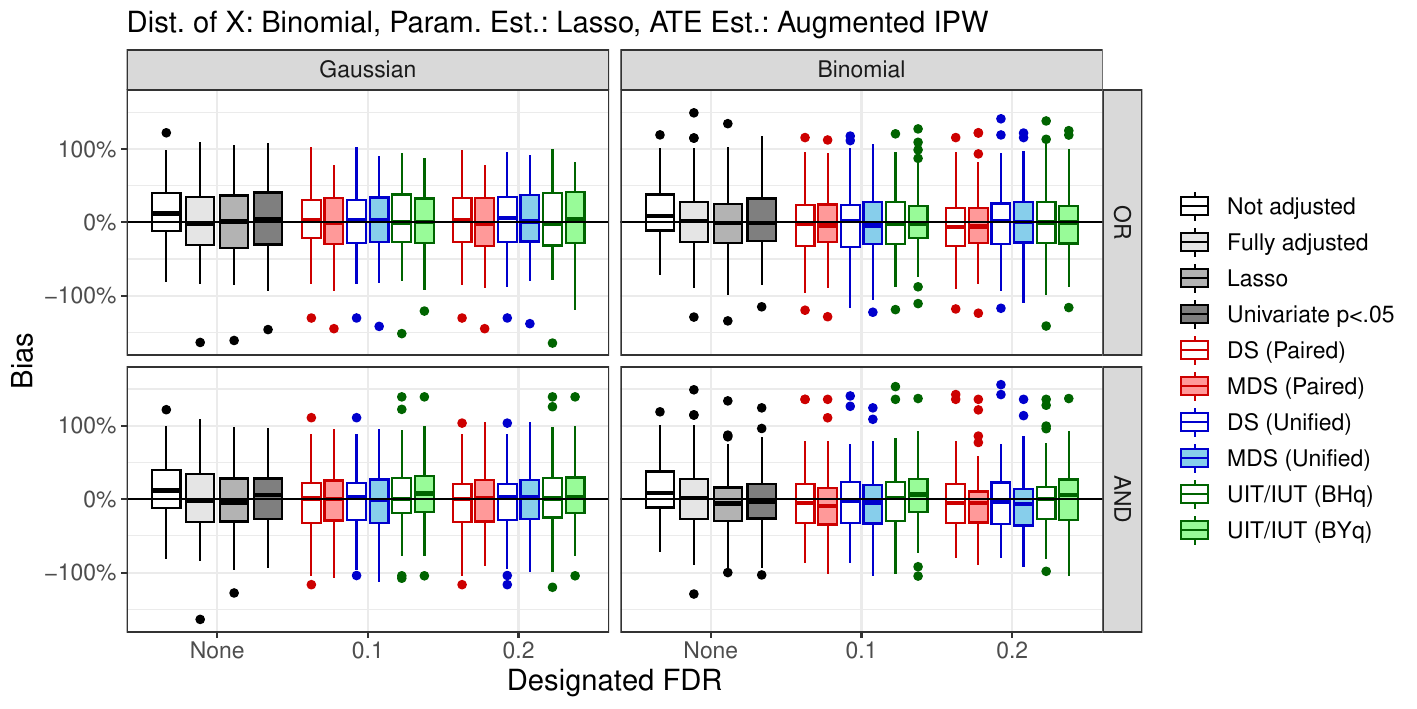}
    \caption{Relative Bias in ATE estimates obtained from 100 simulations. The parameter estimates for the mirror statistics were obtained using the lasso, and the augmented IPW estimators were used to estimate ATE. The distribution of X is binomial.}
\end{figure}

\clearpage
Below, we show the selection rates over 100 simulations for 45 variables related to the outcome or treatment by each method under several conditions. The X-axis is the index of the potential confounders, with indices 1 to 15 being associated with both the outcome and treatment, 16 to 30 with the outcome only, and 31 to 45 with the treatment only. Variables shown with dark blue tiles are positively associated with each corresponding target variable. Variables shown with dark red tiles have opposite signs of regression coefficients for the outcome and treatment, and variables shown with yellow tiles are negatively associated with their corresponding target variables. Variables with indices 1, 5, 16, 20, 31, and 35 have larger absolute regression coefficients, and 4, 7, 11, 15, 19, 22, 26, 30, 37, 41, and 45 have smaller ones. The upper two blocks are for continuous outcomes, while the lower two blocks are for binary outcomes.

\begin{figure}[ht]
    \centering
    \includegraphics[width=\textwidth]{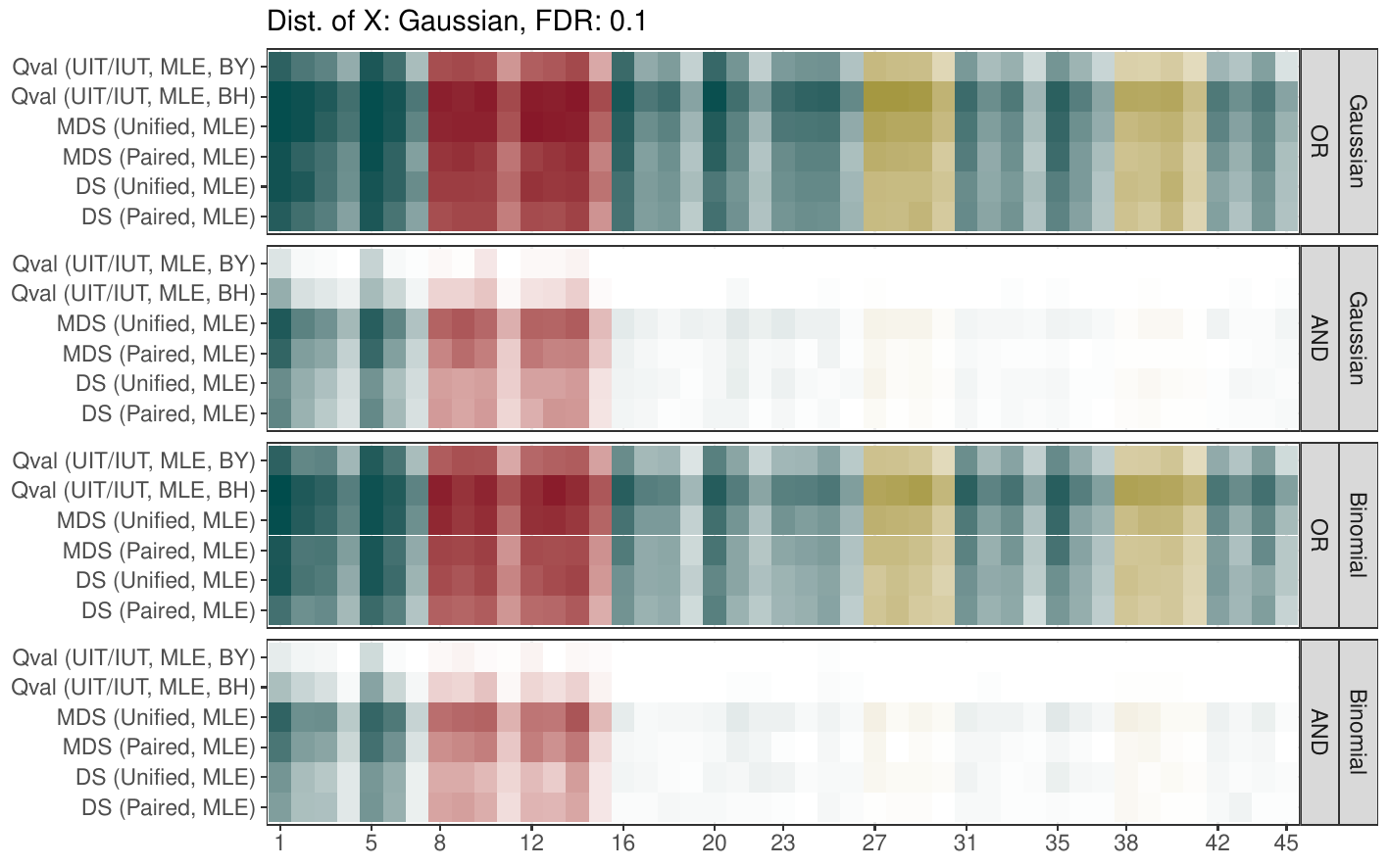}
    \caption{Selection rates over 100 simulations for each true predictor in the low-dimensional Gaussian setting. The parameters were estimated using the MLE for the mirror statistics, and the designated FDR is 0.1.}
\end{figure}

\begin{figure}[ht]
    \centering
    \includegraphics[width=\textwidth]{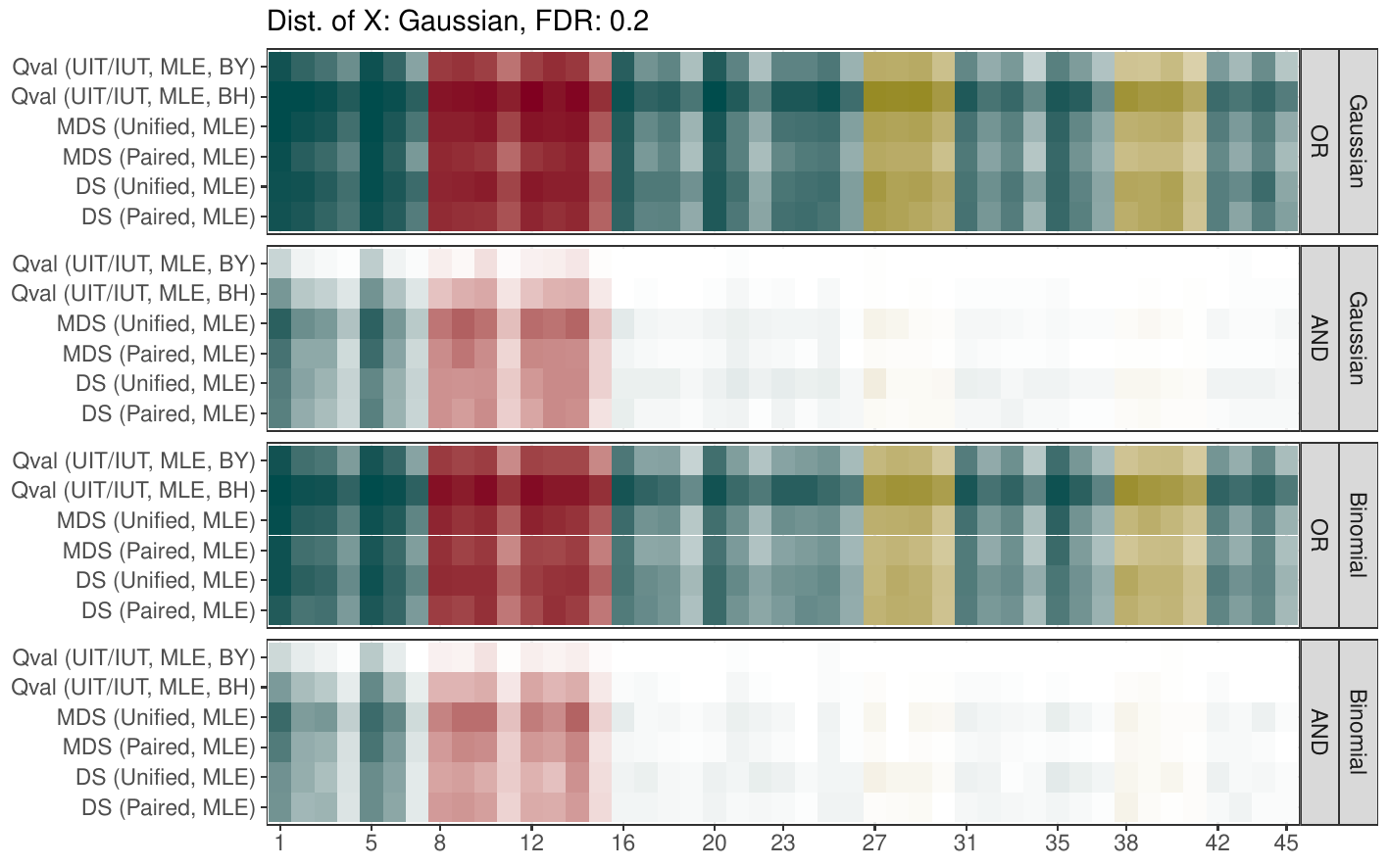}
    \caption{Selection rates over 100 simulations for each true predictor in the low-dimensional Gaussian setting. The parameters were estimated using the MLE for the mirror statistics, and the designated FDR is 0.2.}
\end{figure}

\begin{figure}[ht]
    \centering
    \includegraphics[width=\textwidth]{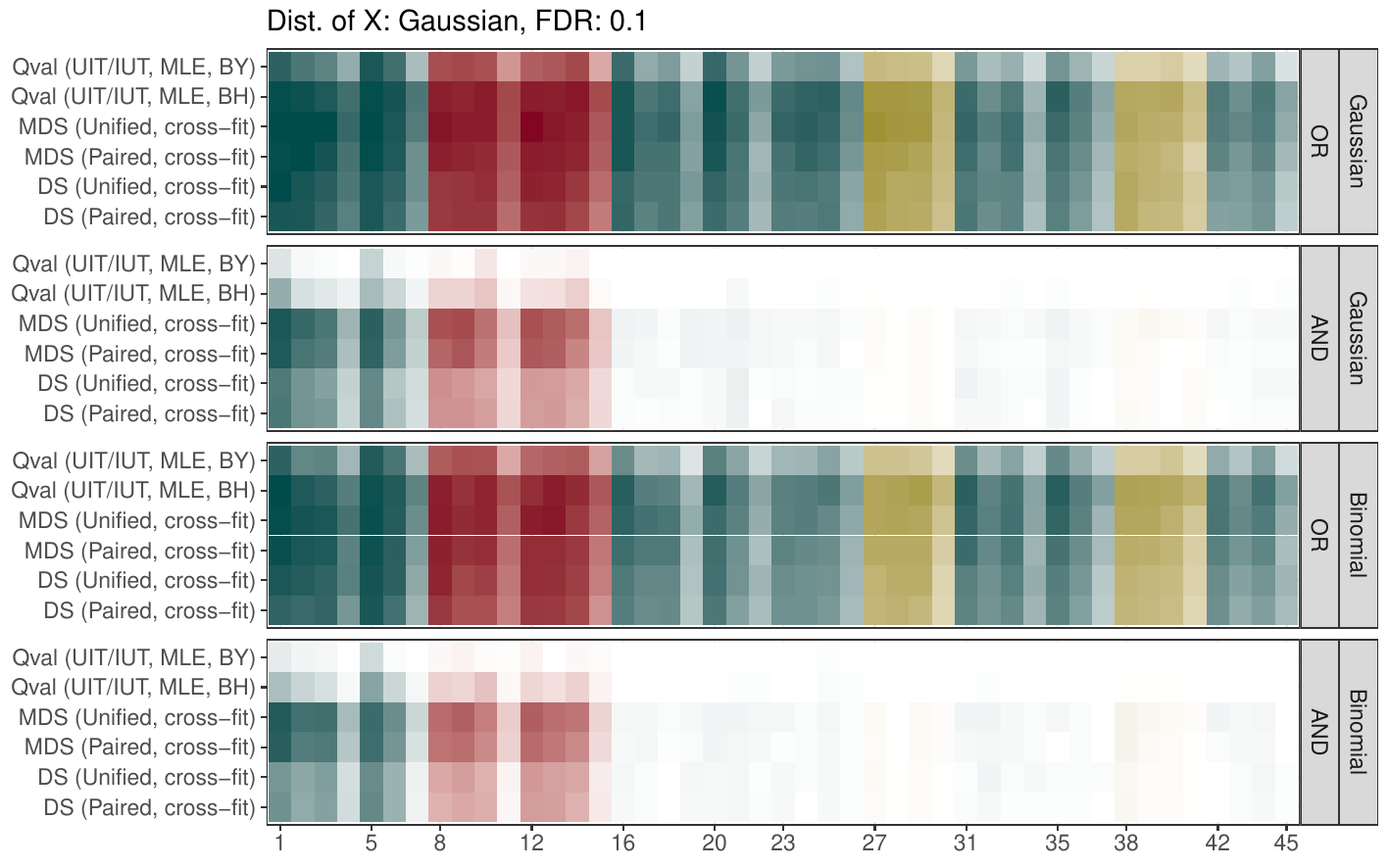}
    \caption{Selection rates over 100 simulations for each true predictor in the low-dimensional Gaussian setting. The parameters were estimated using the cross-fitting for the mirror statistics, and the designated FDR is 0.1.}
\end{figure}

\begin{figure}[ht]
    \centering
    \includegraphics[width=\textwidth]{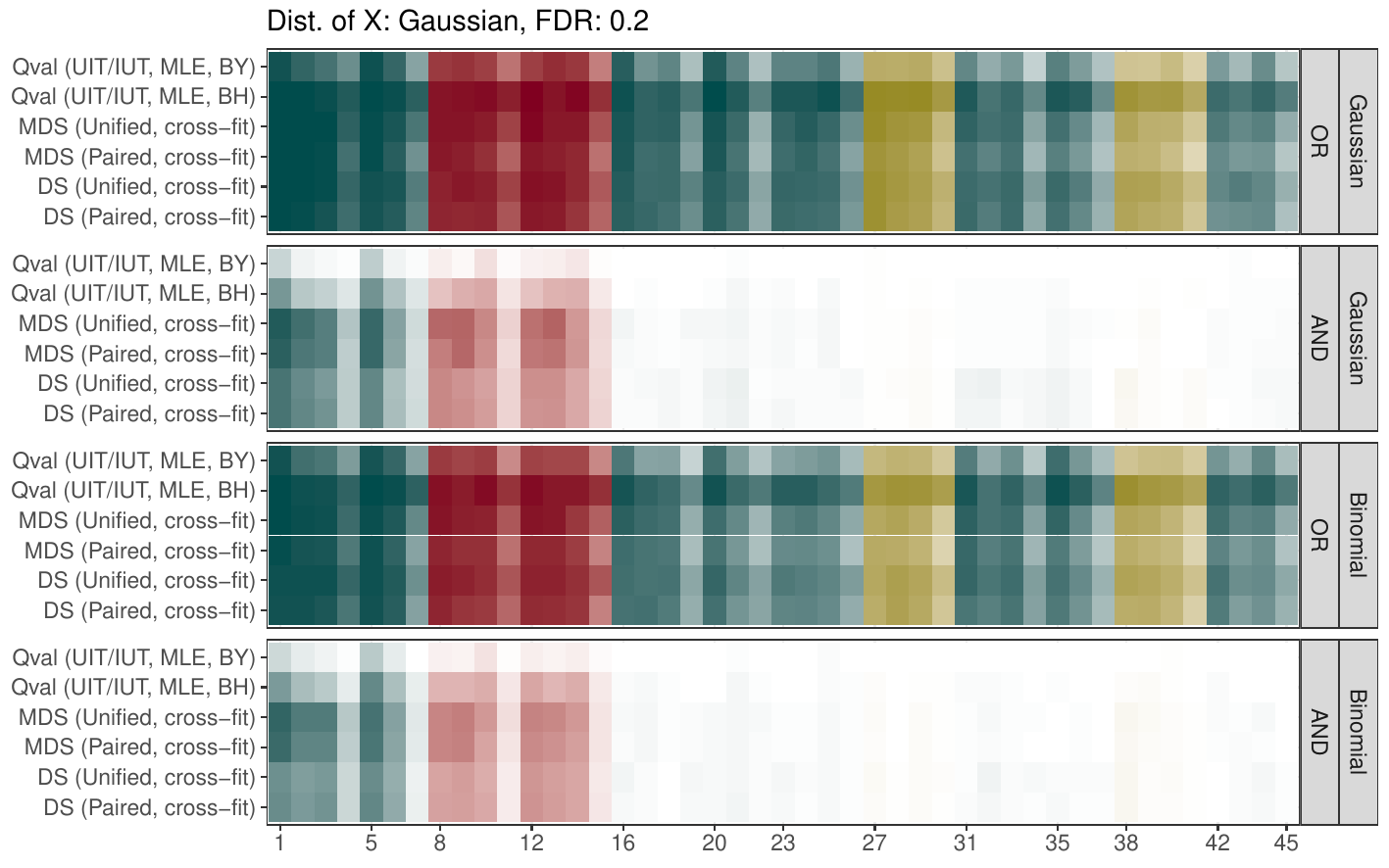}
    \caption{Selection rates over 100 simulations for each true predictor in the low-dimensional Gaussian setting. The parameters were estimated using the cross-fitting for the mirror statistics, and the designated FDR is 0.2.}
\end{figure}

\begin{figure}[ht]
    \centering
    \includegraphics[width=\textwidth]{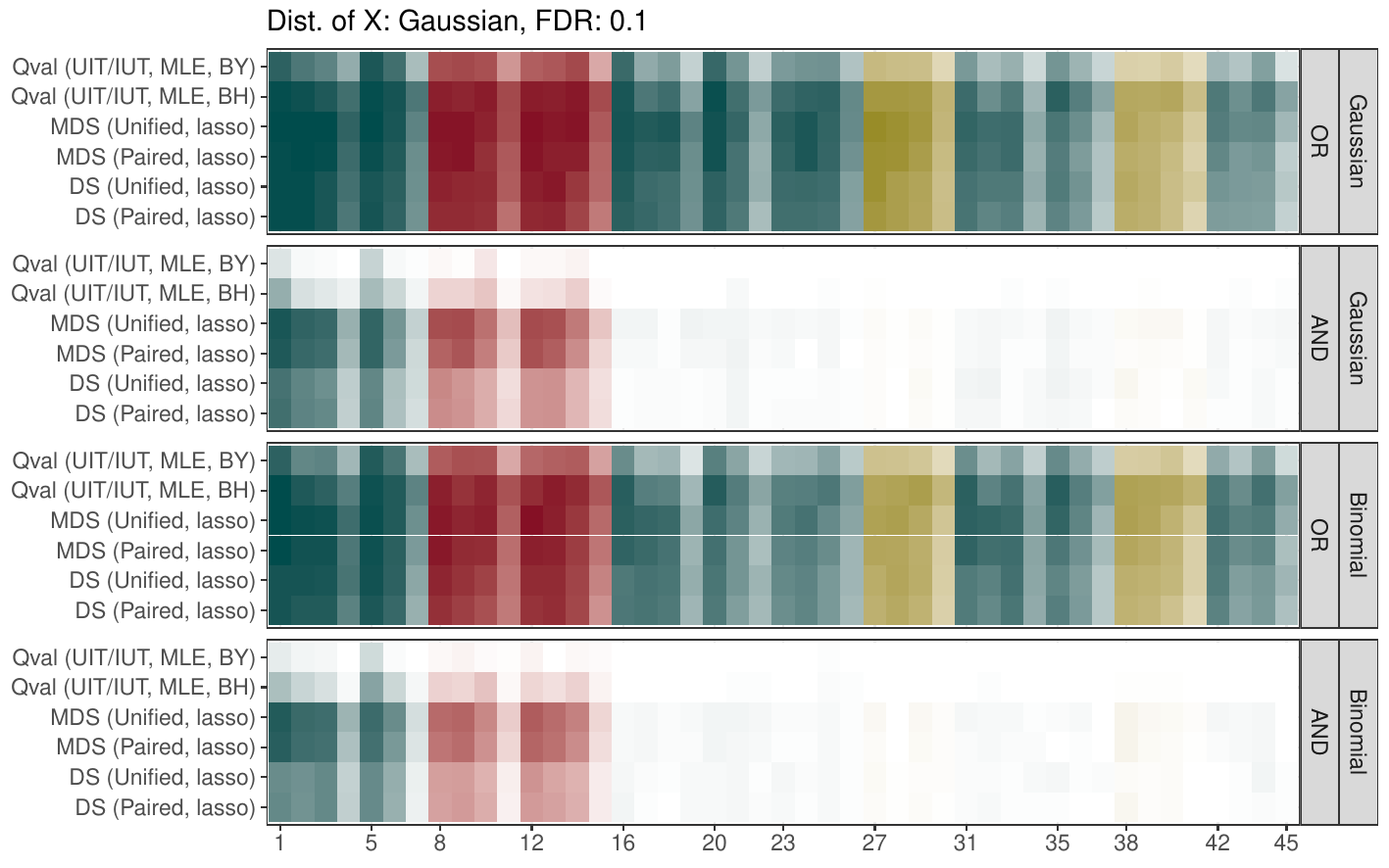}
    \caption{Selection rates over 100 simulations for each true predictor in the low-dimensional Gaussian setting. The parameters were estimated using the lasso for the mirror statistics, and the designated FDR is 0.1.}
\end{figure}

\begin{figure}[ht]
    \centering
    \includegraphics[width=\textwidth]{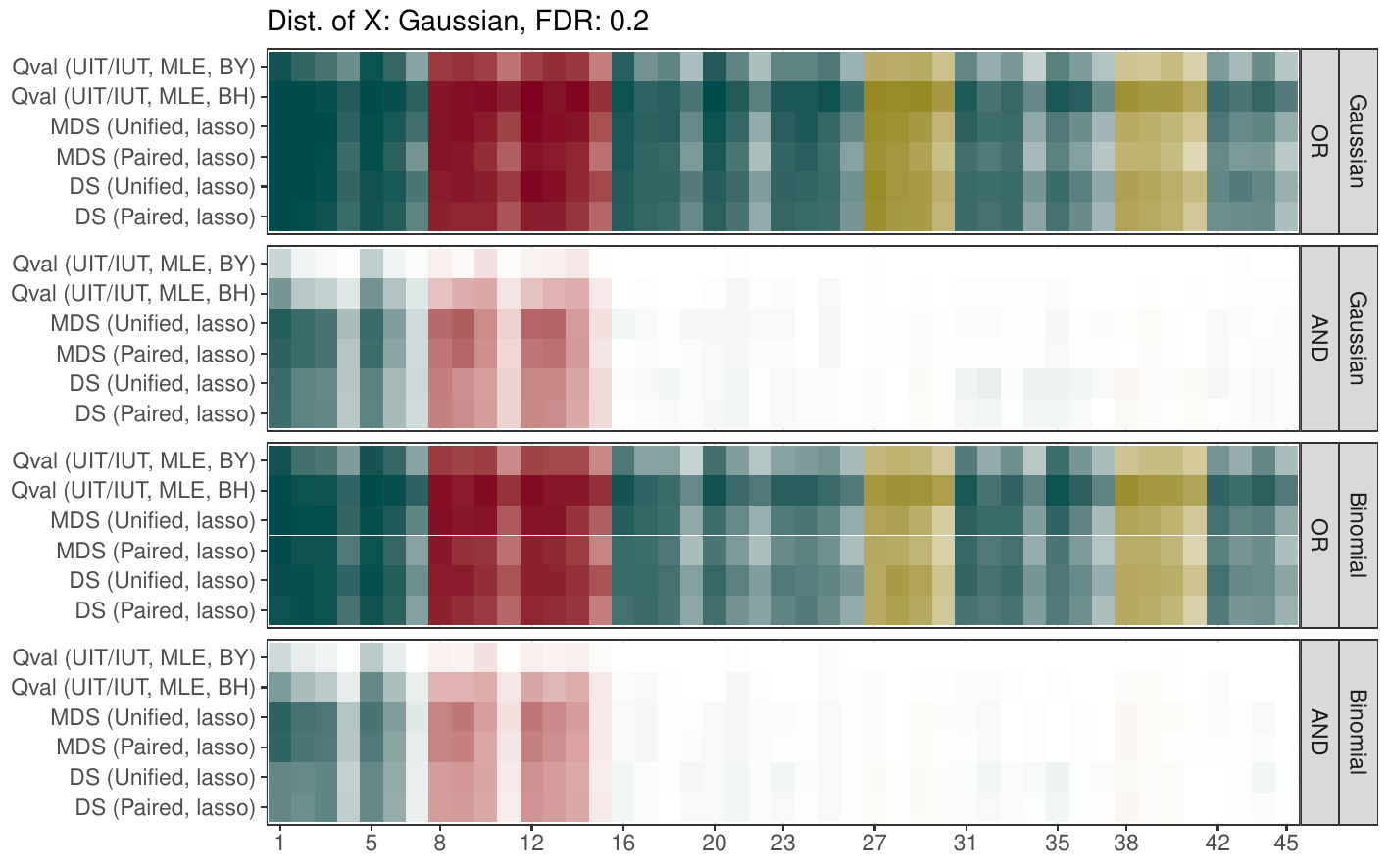}
    \caption{Selection rates over 100 simulations for each true predictor in the low-dimensional Gaussian setting. The parameters were estimated using the lasso for the mirror statistics, and the designated FDR is 0.2.}
\end{figure}

\begin{figure}[ht]
    \centering
    \includegraphics[width=\textwidth]{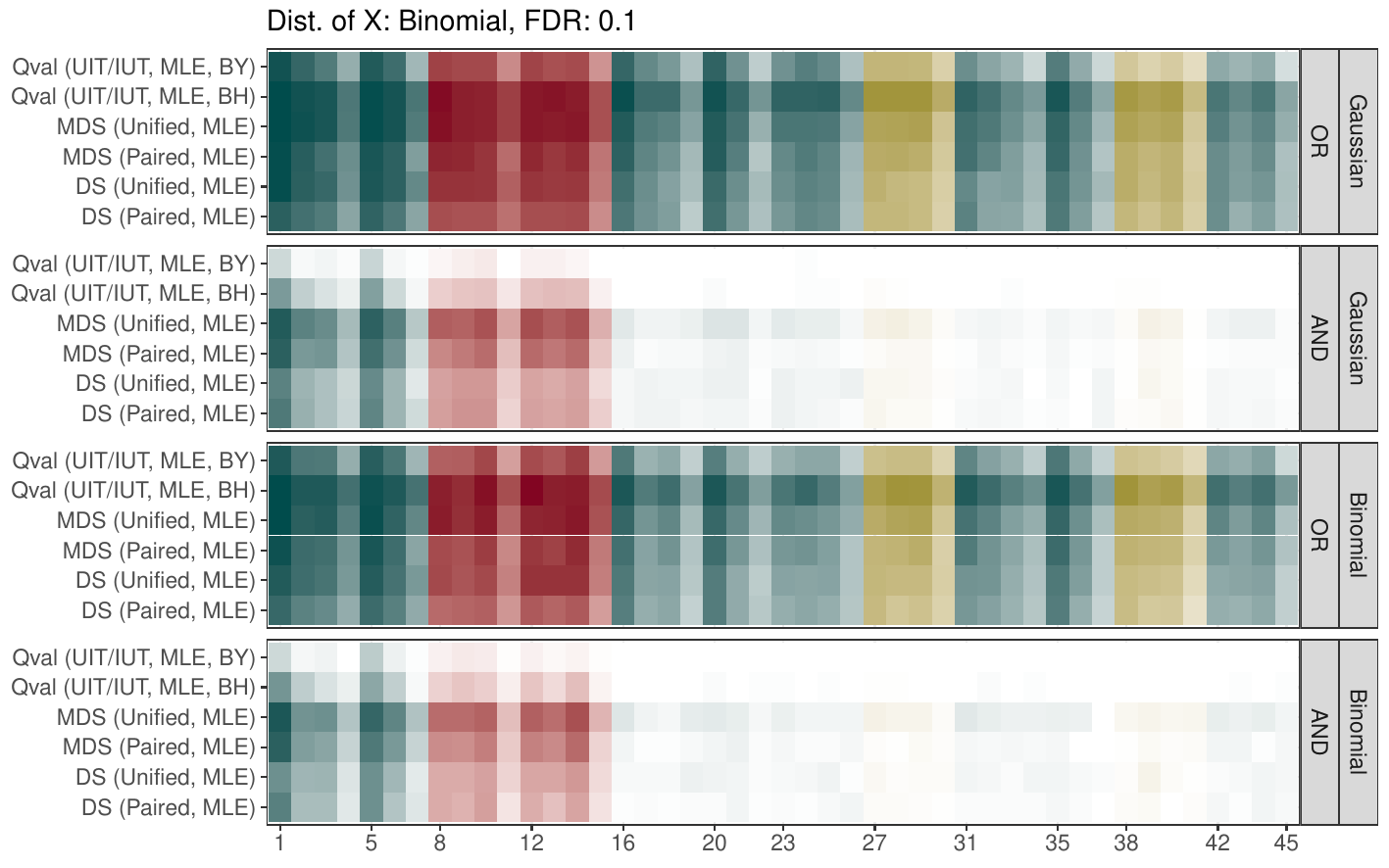}
    \caption{Selection rates over 100 simulations for each true predictor in the low-dimensional binomial setting. The parameters were estimated using the MLE for the mirror statistics, and the designated FDR is 0.1.}
\end{figure}

\begin{figure}[ht]
    \centering
    \includegraphics[width=\textwidth]{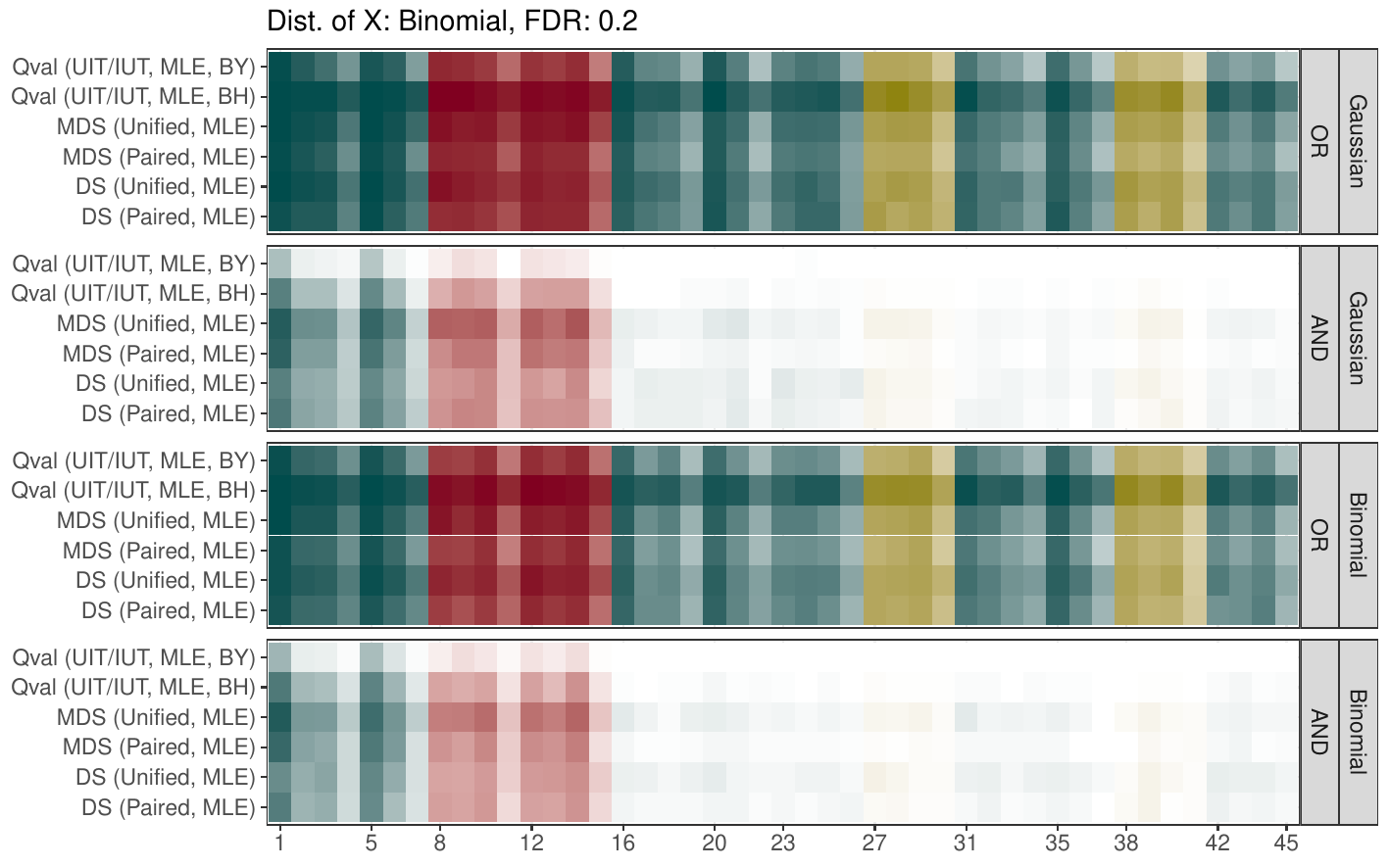}
    \caption{Selection rates over 100 simulations for each true predictor in the low-dimensional binomial setting. The parameters were estimated using the MLE for the mirror statistics, and the designated FDR is 0.2.}
\end{figure}

\begin{figure}[ht]
    \centering
    \includegraphics[width=\textwidth]{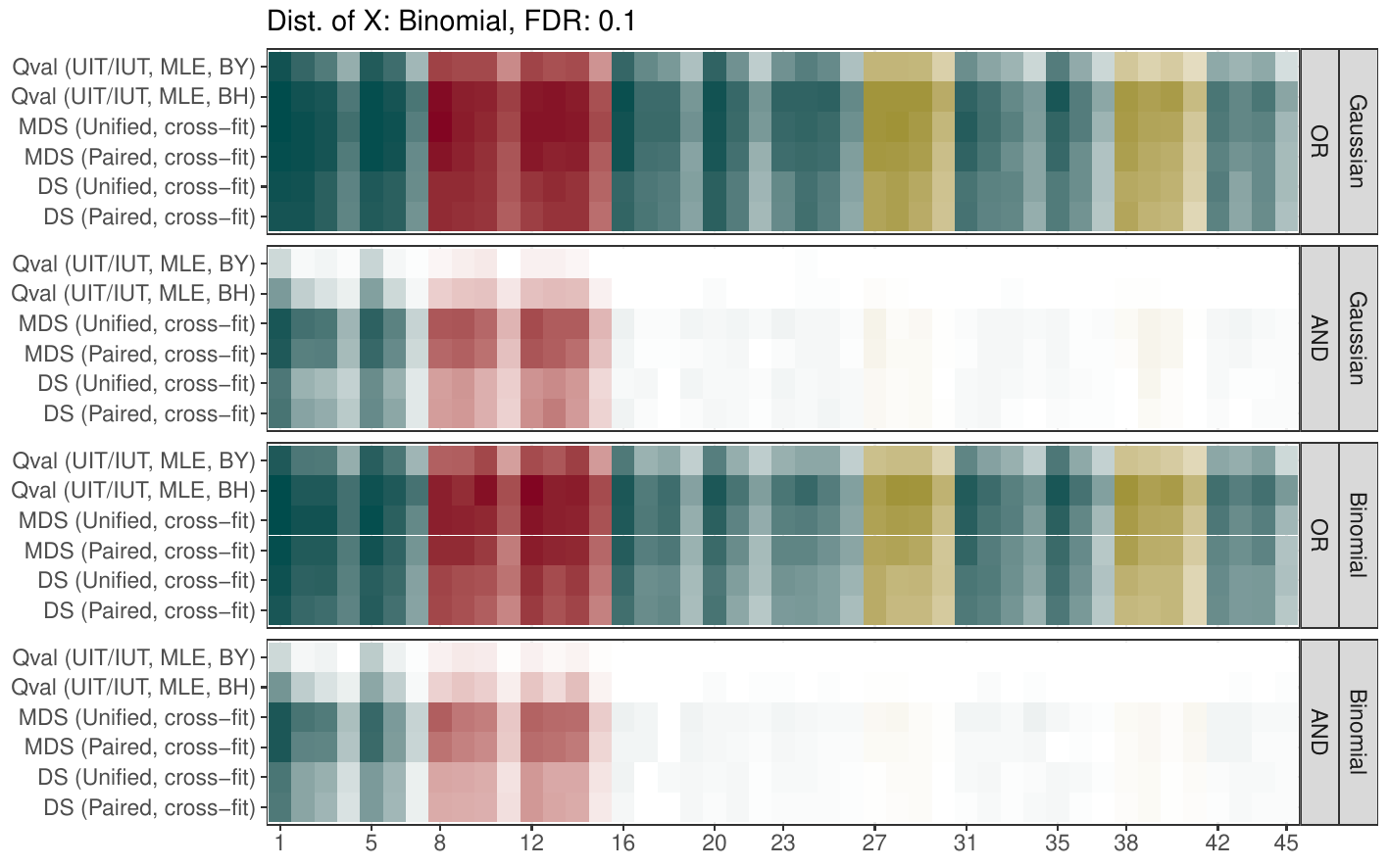}
    \caption{Selection rates over 100 simulations for each true predictor in the low-dimensional binomial setting. The parameters were estimated using the cross-fitting for the mirror statistics, and the designated FDR is 0.1.}
\end{figure}

\begin{figure}[ht]
    \centering
    \includegraphics[width=\textwidth]{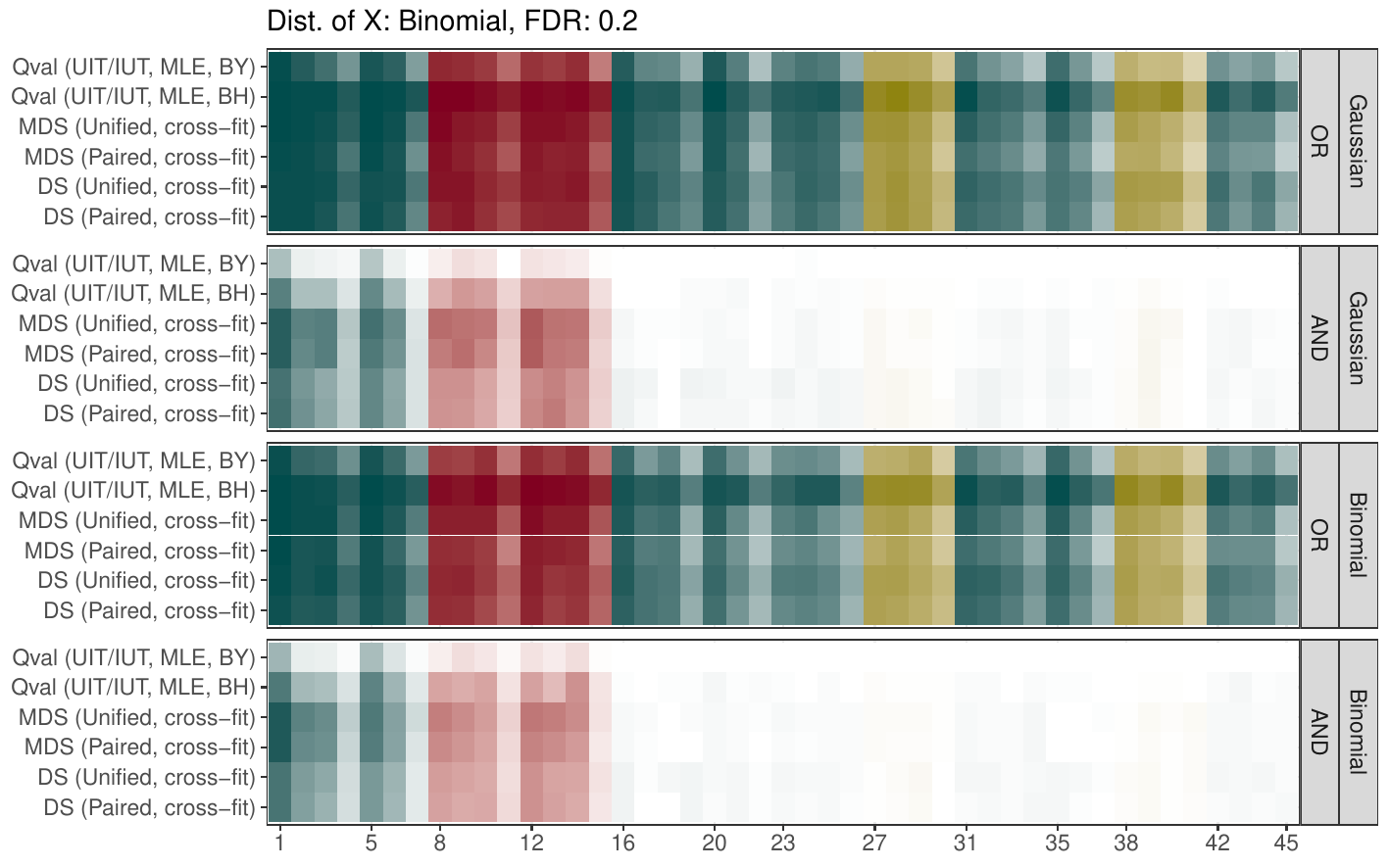}
    \caption{Selection rates over 100 simulations for each true predictor in the low-dimensional binomial setting. The parameters were estimated using the cross-fitting for the mirror statistics, and the designated FDR is 0.2.}
\end{figure}

\begin{figure}[ht]
    \centering
    \includegraphics[width=\textwidth]{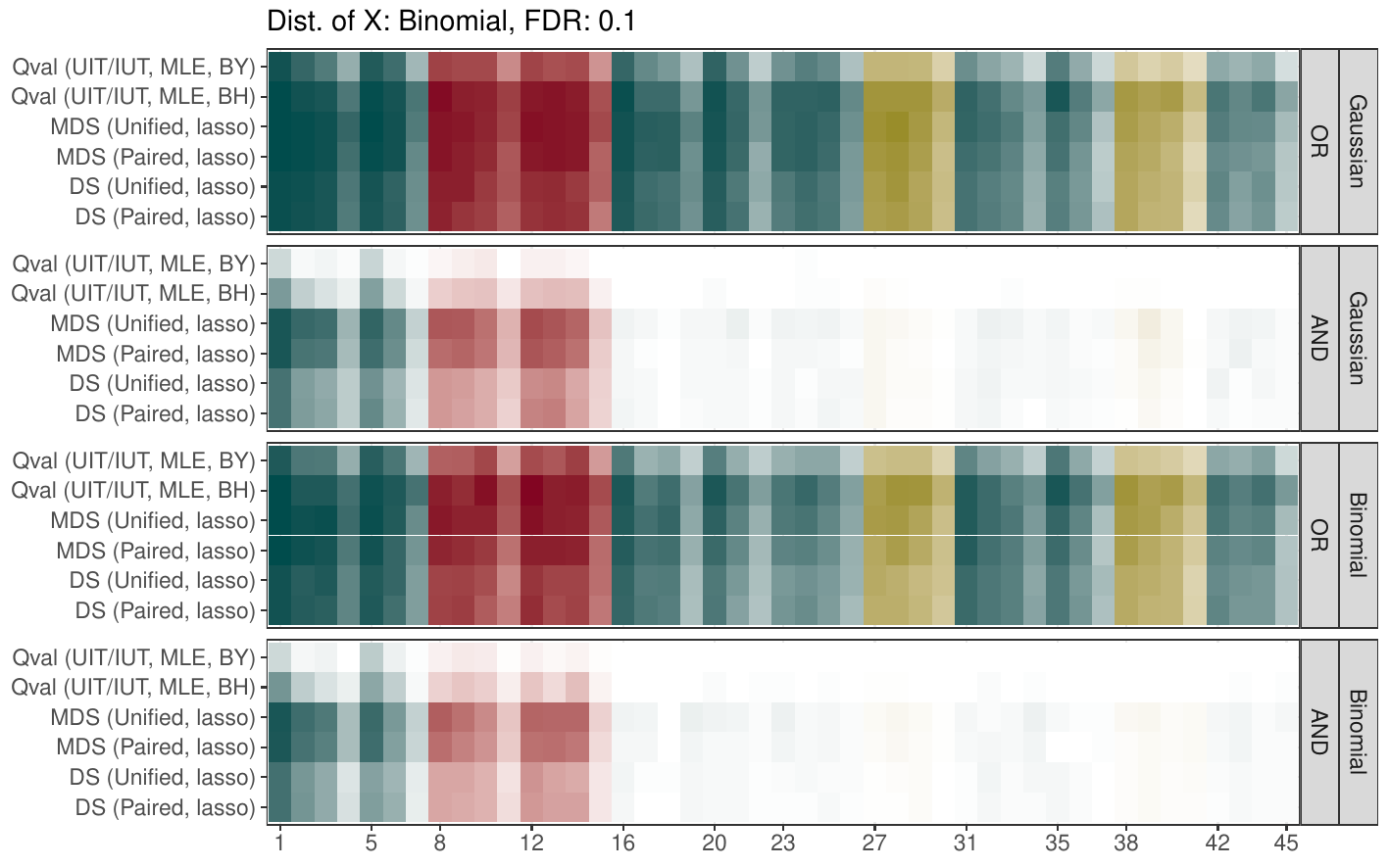}
    \caption{Selection rates over 100 simulations for each true predictor in the low-dimensional binomial setting. The parameters were estimated using the lasso for the mirror statistics, and the designated FDR is 0.1.}
\end{figure}

\begin{figure}[ht]
    \centering
    \includegraphics[width=\textwidth]{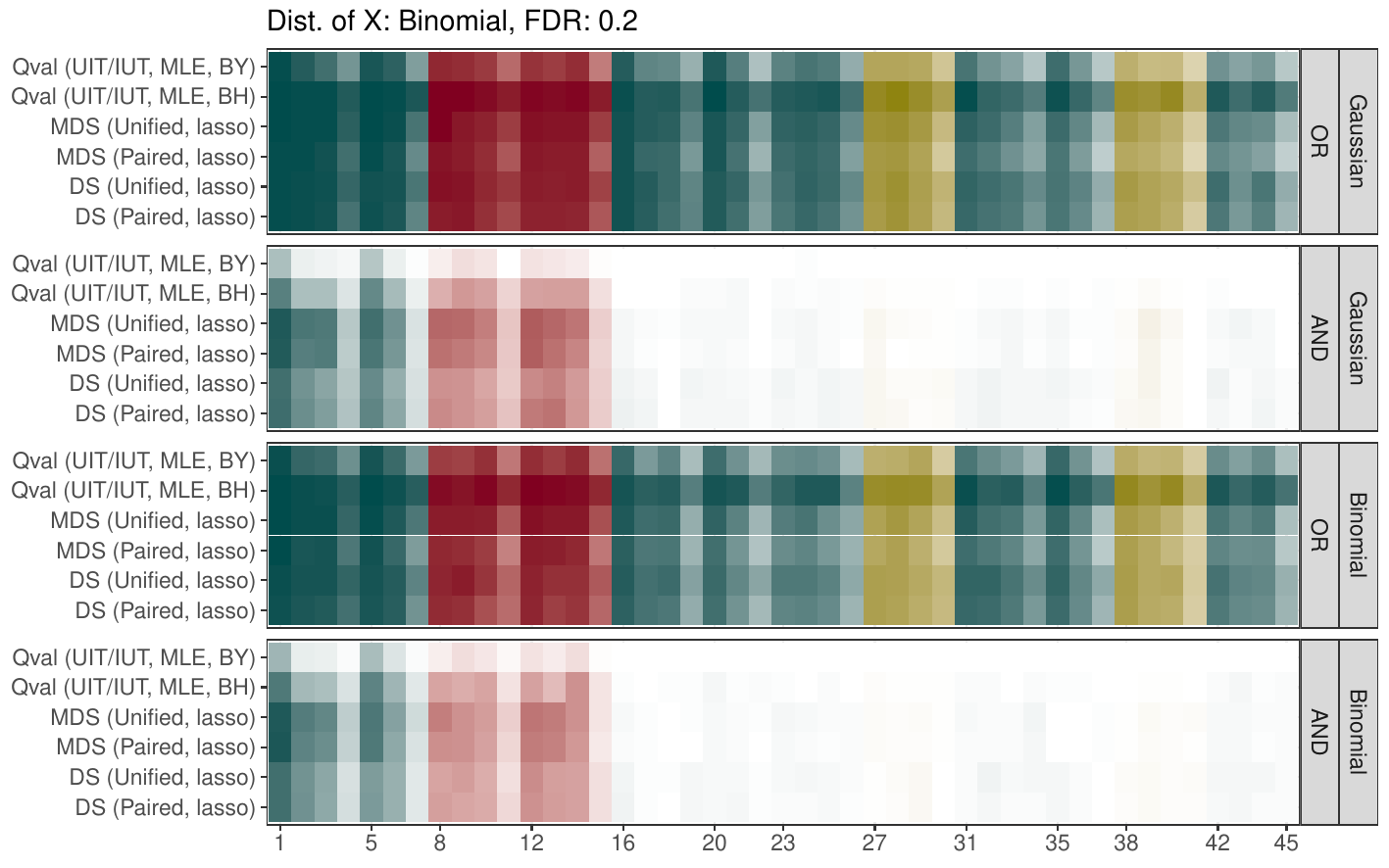}
    \caption{Selection rates over 100 simulations for each true predictor in the low-dimensional binomial setting. The parameters were estimated using the lasso for the mirror statistics, and the designated FDR is 0.2.}
\end{figure}

\clearpage
\subsection{High-dimensional Potential Confounders}

\begin{figure}[ht]
    \centering
    \includegraphics[width=\textwidth]{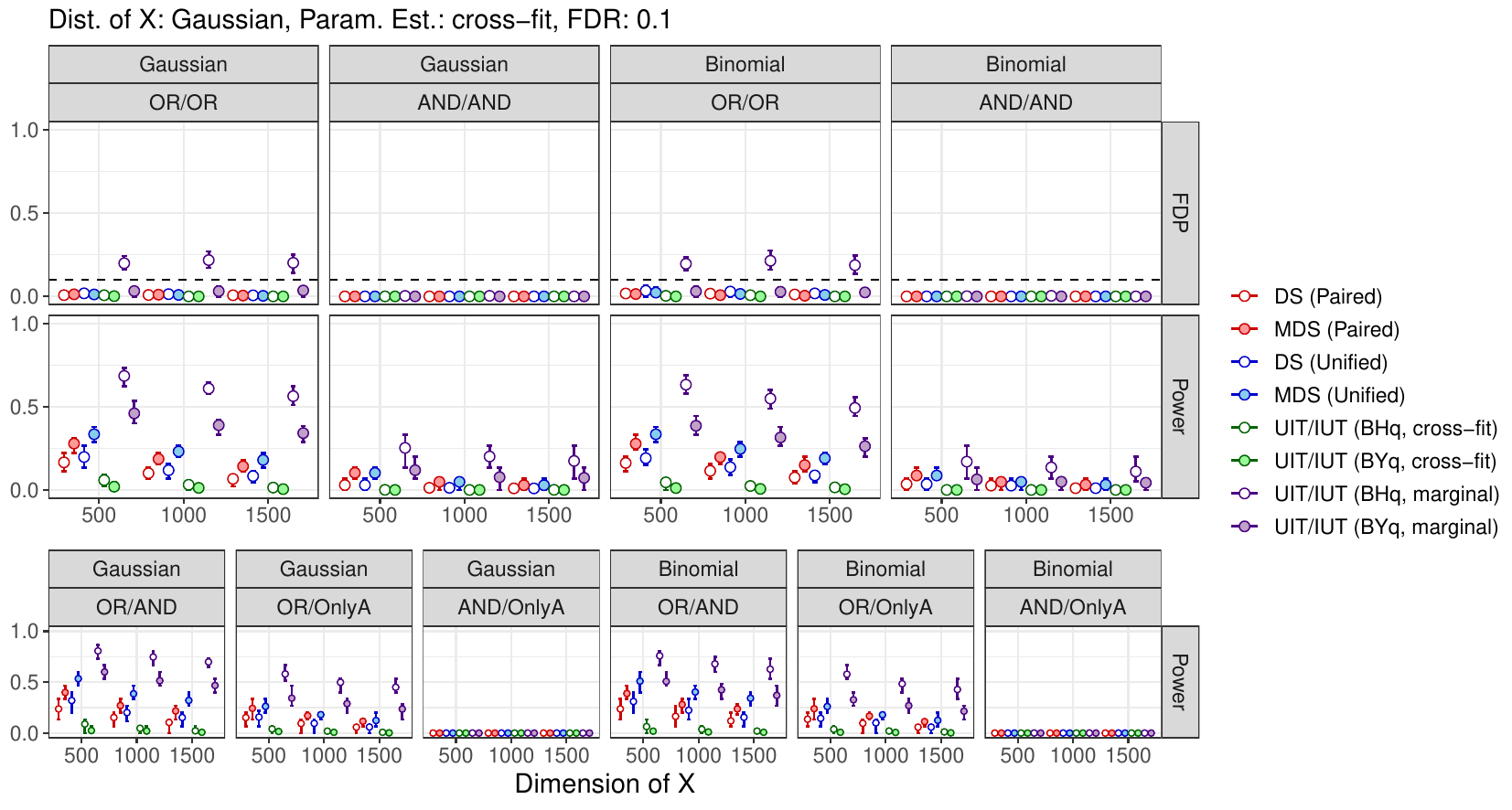}
    \caption{
    FDR and power analysis of the proposed and q-value-based methods in high-dimensional Gaussian settings. Parameter estimates for the mirror statistics were obtained using the cross-fitting. The left block presents the scenarios with continuous outcomes, and the right block shows the results for binary outcomes. The notation ``**/**'' denotes the variable selection criteria and the set of true variables for evaluation. Each point is a mean value over 100 simulations, and the error bars at each point represent the 1st to 3rd quartiles.
    }
\end{figure}

\begin{figure}[ht]
    \centering
    \includegraphics[width=\textwidth]{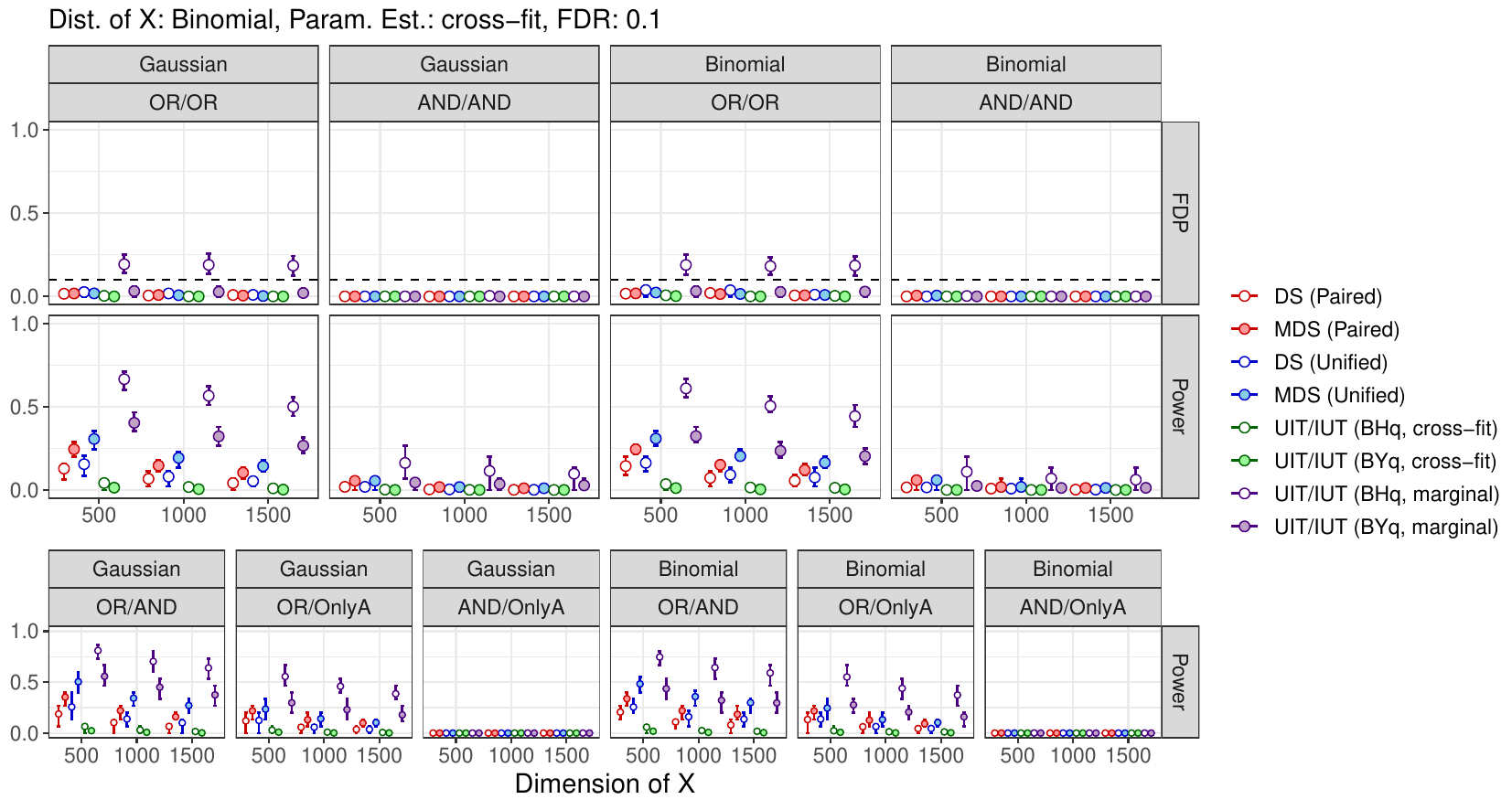}
    \caption{
    FDR and power analysis of the proposed and q-value-based methods in high-dimensional binomial settings. Parameter estimates for the mirror statistics were obtained using the cross-fitting. The left block presents the scenarios with continuous outcomes, and the right block shows the results for binary outcomes. The notation ``**/**'' denotes the variable selection criteria and the set of true variables for evaluation. Each point is a mean value over 100 simulations, and the error bars at each point represent the 1st to 3rd quartiles.
    }
\end{figure}

\begin{figure}[ht]
    \centering
    \includegraphics[width=\textwidth]{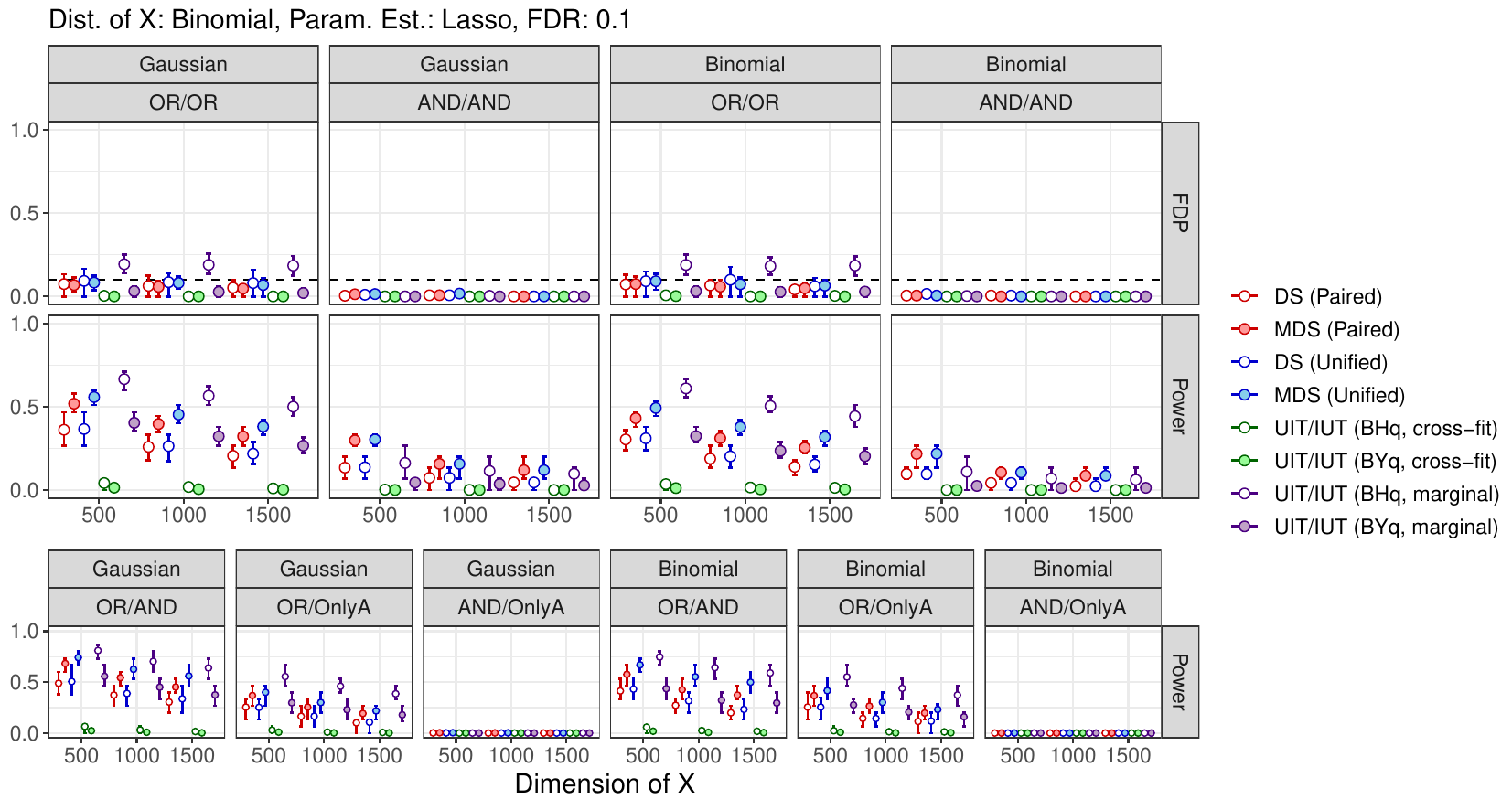}
    \caption{
    FDR and power analysis of the proposed and q-value-based methods in high-dimensional binomial settings. Parameter estimates for the mirror statistics were obtained using the lasso. The left block presents the scenarios with continuous outcomes, and the right block shows the results for binary outcomes. The notation ``**/**'' denotes the variable selection criteria and the set of true variables for evaluation. Each point is a mean value over 100 simulations, and the error bars at each point represent the 1st to 3rd quartiles.
    }
\end{figure}

\begin{figure}[ht]
    \centering
    \includegraphics[width=\textwidth]{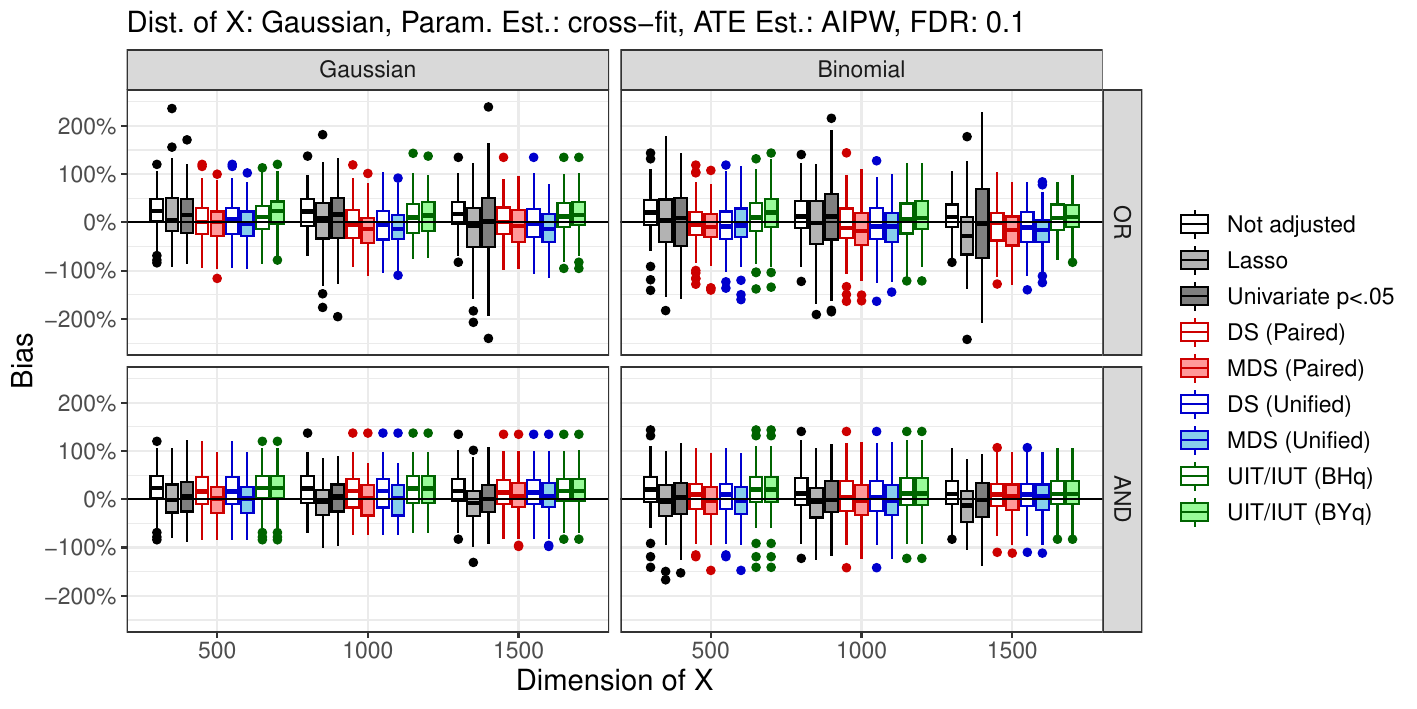}
    \caption{
    Relative Bias in ATE estimates obtained from 100 simulations based on the high-dimensional potential confounders. The parameter estimates for the mirror statistics were obtained using cross-fitting, and the augmented IPW was used to estimate ATE. The distribution of X is Gaussian.
    }
\end{figure}

\begin{figure}[ht]
    \centering
    \includegraphics[width=\textwidth]{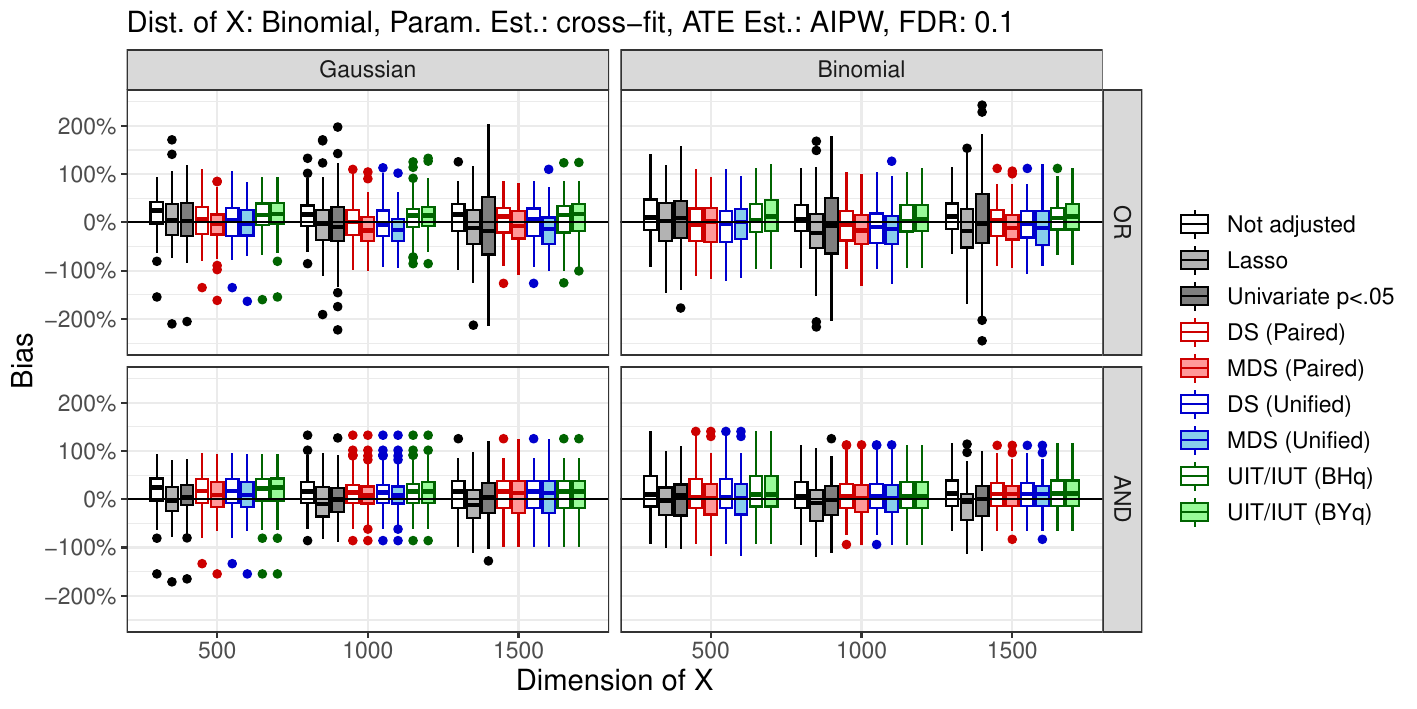}
    \caption{
    Relative Bias in ATE estimates obtained from 100 simulations based on the high-dimensional potential confounders. The parameter estimates for the mirror statistics were obtained using cross-fitting, and the augmented IPW was used to estimate ATE. The distribution of X is binomial.
    }
\end{figure}

\begin{figure}[ht]
    \centering
    \includegraphics[width=\textwidth]{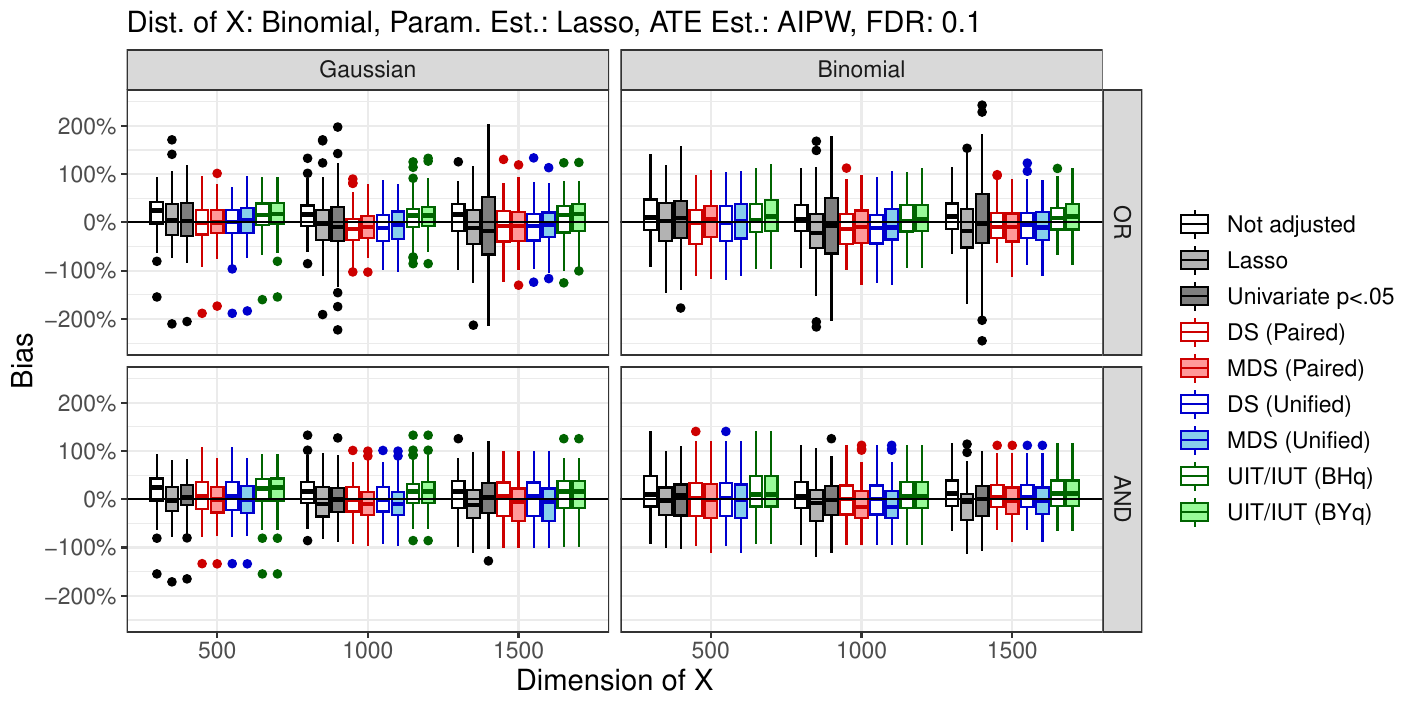}
    \caption{
    Relative Bias in ATE estimates obtained from 100 simulations based on the high-dimensional potential confounders. The parameter estimates for the mirror statistics were obtained using the lasso, and the augmented IPW was used to estimate ATE. The distribution of X is binomial.
    }
\end{figure}

\clearpage
Below, we show the selection rates over 100 simulations for 45 variables related to the outcome or treatment by each method under several conditions. The X-axis is the index of the potential confounders, with indices 1 to 15 being associated with both the outcome and treatment, 16 to 30 with the outcome only, and 31 to 45 with the treatment only. Variables shown with dark blue tiles are positively associated with each corresponding target variable. Variables shown with dark red tiles have opposite signs of regression coefficients for the outcome and treatment, and variables shown with yellow tiles are negatively associated with their corresponding target variables. Variables with indices 1, 5, 16, 20, 31, and 35 have larger absolute regression coefficients, and 4, 7, 11, 15, 19, 22, 26, 30, 37, 41, and 45 have smaller ones. The upper two blocks are for continuous outcomes, while the lower two blocks are for binary outcomes.

\begin{figure}[ht]
    \centering
    \includegraphics[width=\textwidth]{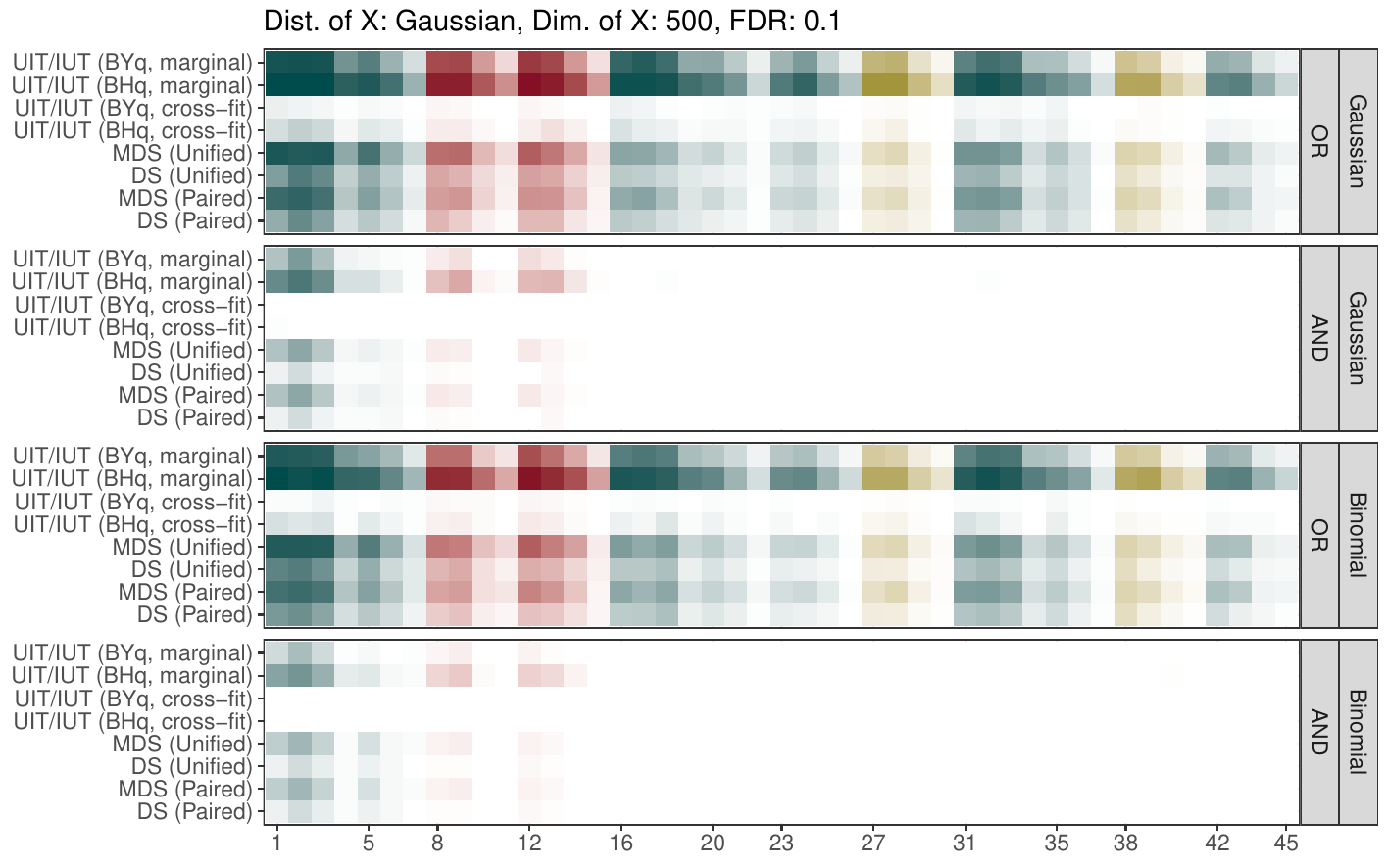}
    \caption{Selection rates over 100 simulations for each true predictor in the high-dimensional Gaussian setting ($p=500$). The parameters were estimated using cross-fitting for the mirror statistics, and the designated FDR is 0.1.}
\end{figure}

\begin{figure}[ht]
    \centering
    \includegraphics[width=\textwidth]{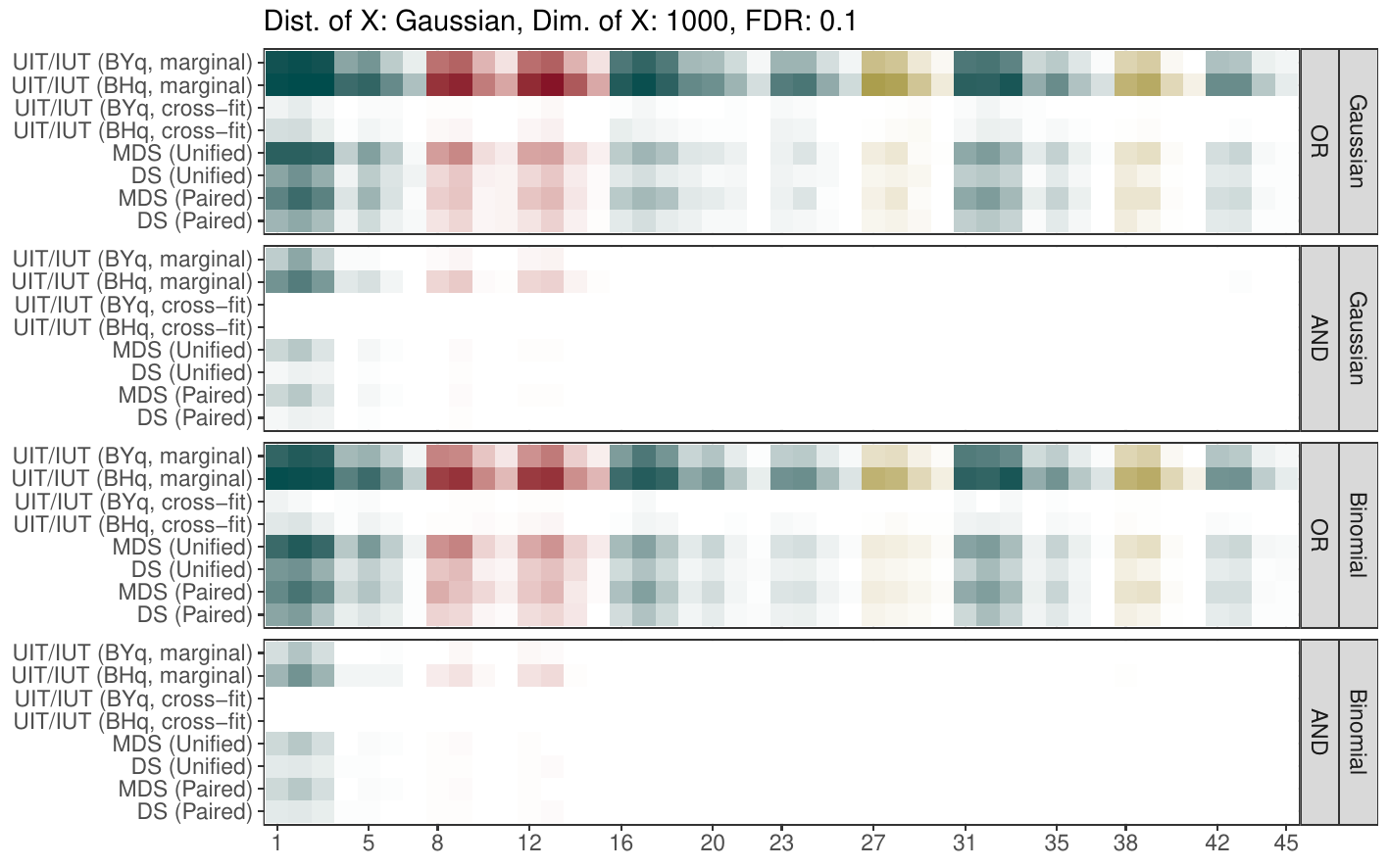}
    \caption{Selection rates over 100 simulations for each true predictor in the high-dimensional Gaussian setting ($p=1000$). The parameters were estimated using cross-fitting for the mirror statistics, and the designated FDR is 0.1.}
\end{figure}

\begin{figure}[ht]
    \centering
    \includegraphics[width=\textwidth]{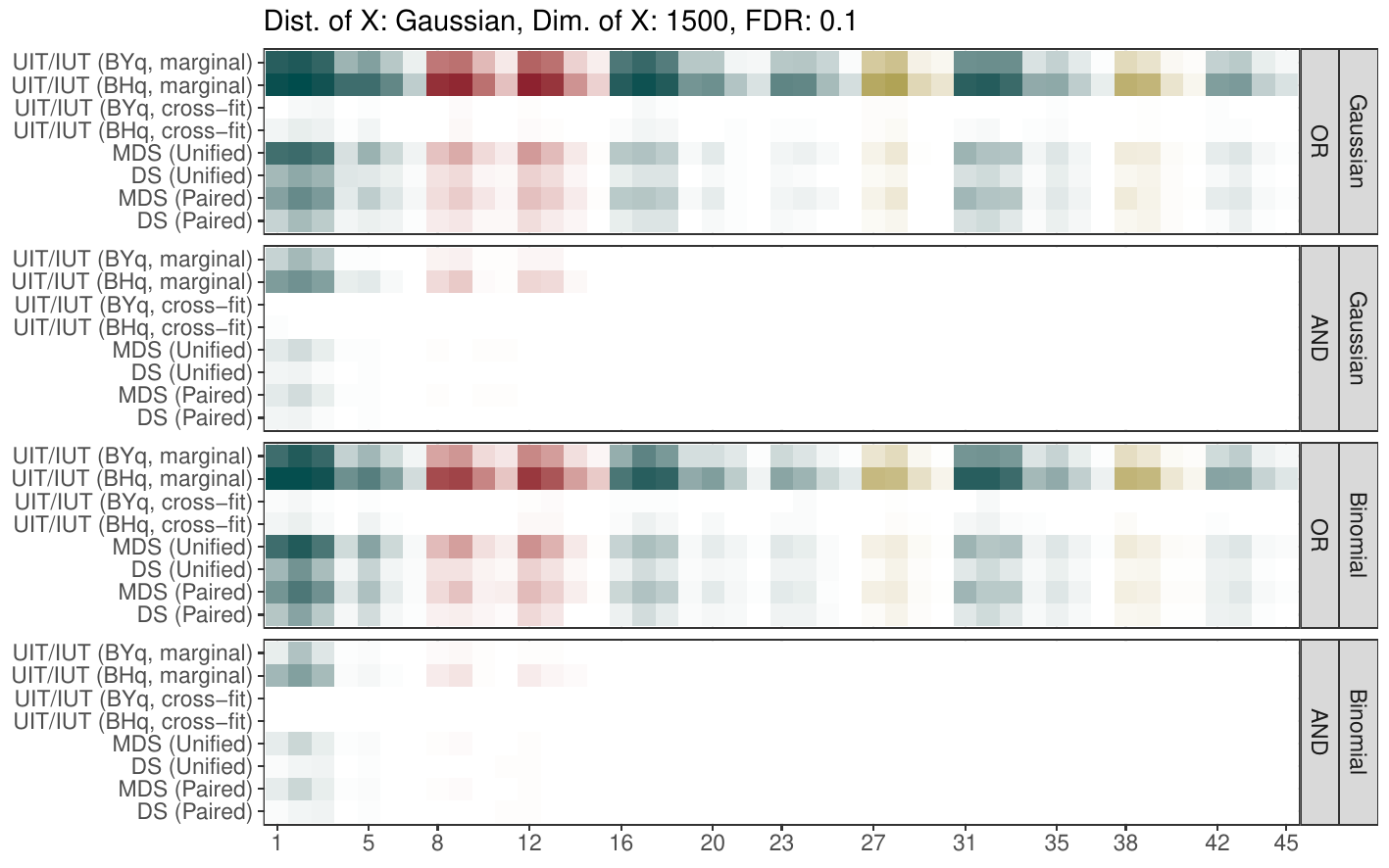}
    \caption{Selection rates over 100 simulations for each true predictor in the high-dimensional Gaussian setting ($p=1500$). The parameters were estimated using cross-fitting for the mirror statistics, and the designated FDR is 0.1.}
\end{figure}

\begin{figure}[ht]
    \centering
    \includegraphics[width=\textwidth]{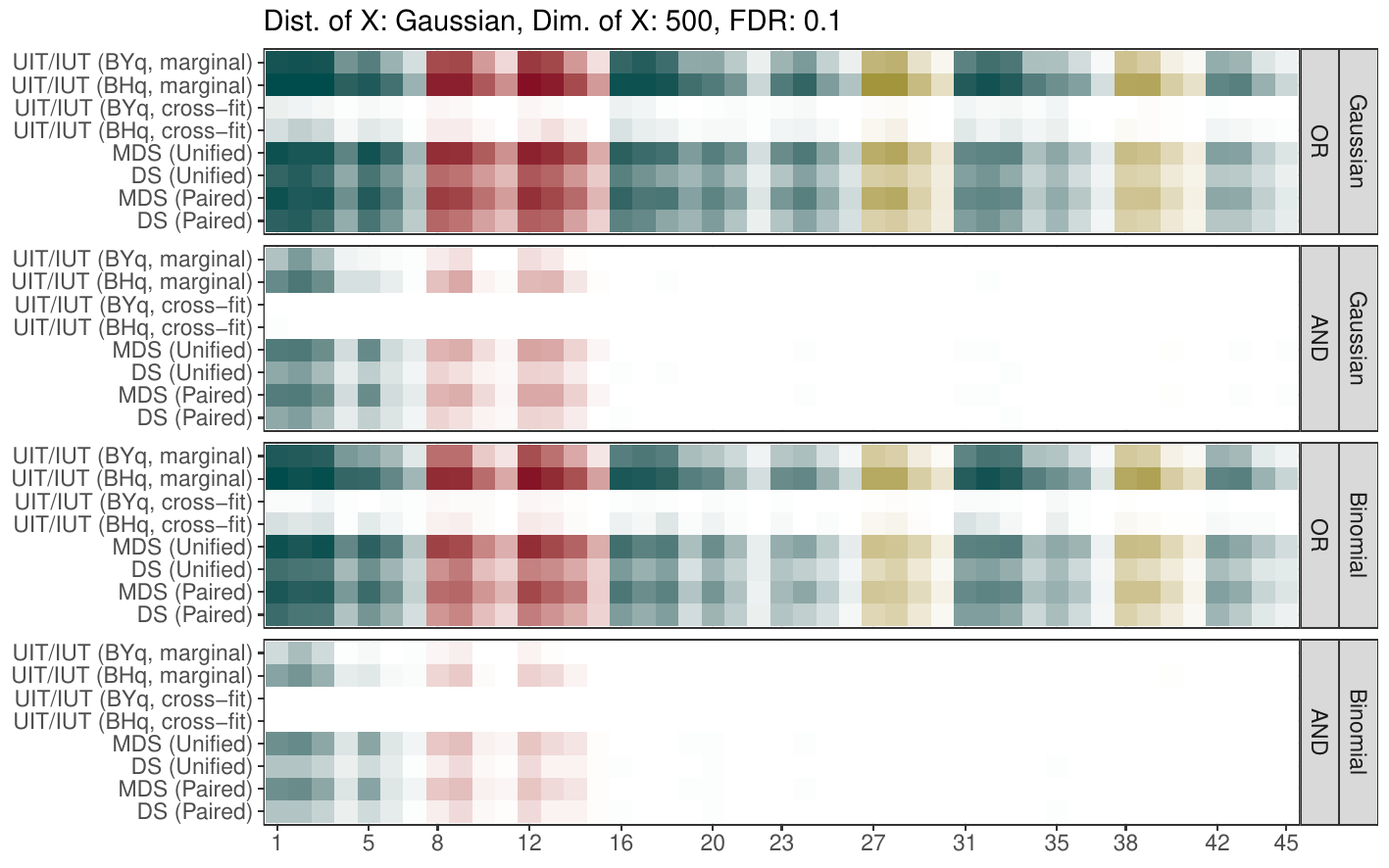}
    \caption{Selection rates over 100 simulations for each true predictor in the high-dimensional Gaussian setting ($p=500$). The parameters were estimated using the lasso for the mirror statistics, and the designated FDR is 0.1.}
\end{figure}

\begin{figure}[ht]
    \centering
    \includegraphics[width=\textwidth]{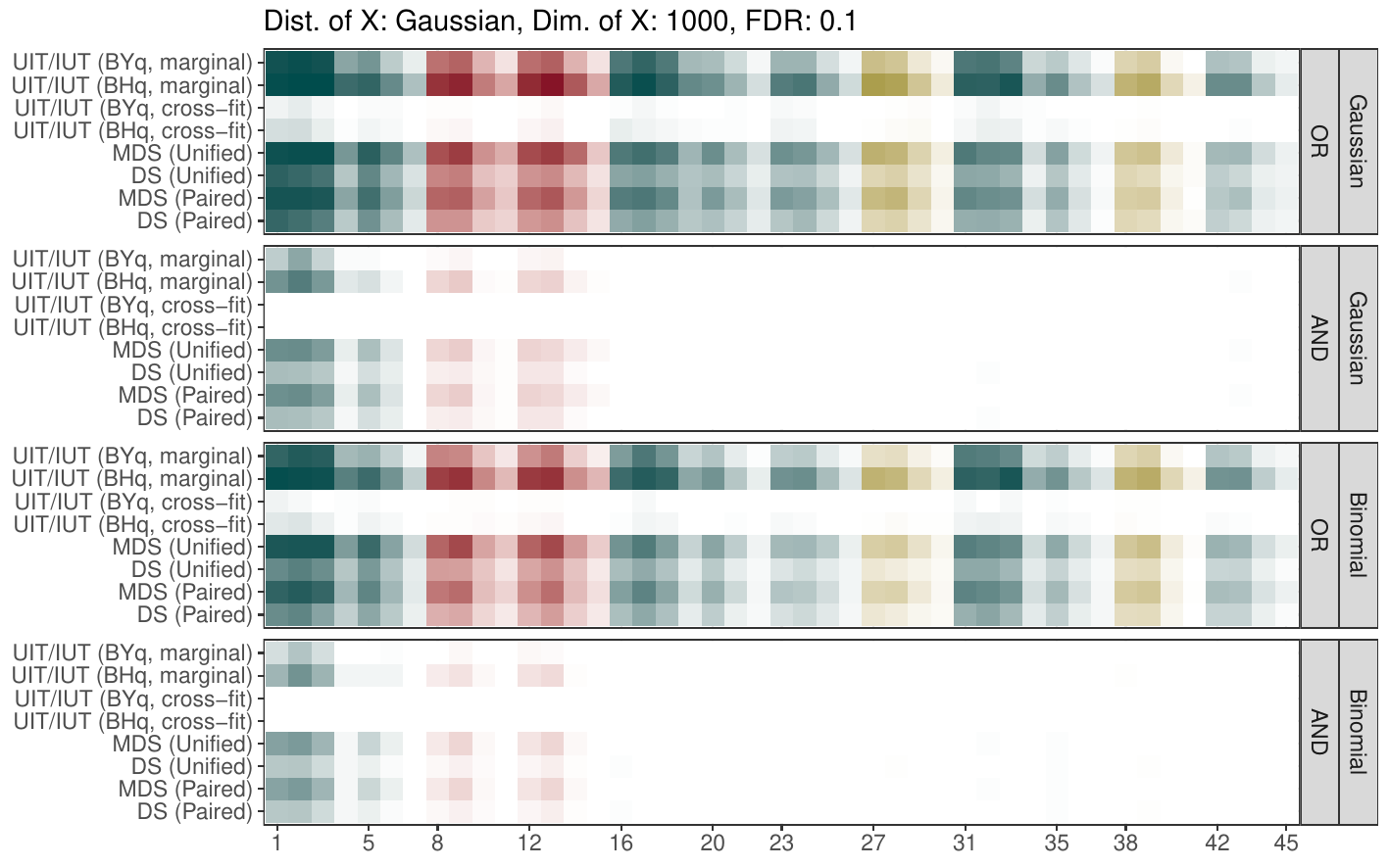}
    \caption{Selection rates over 100 simulations for each true predictor in the high-dimensional Gaussian setting ($p=1000$). The parameters were estimated using the lasso for the mirror statistics, and the designated FDR is 0.1.}
\end{figure}

\begin{figure}[ht]
    \centering
    \includegraphics[width=\textwidth]{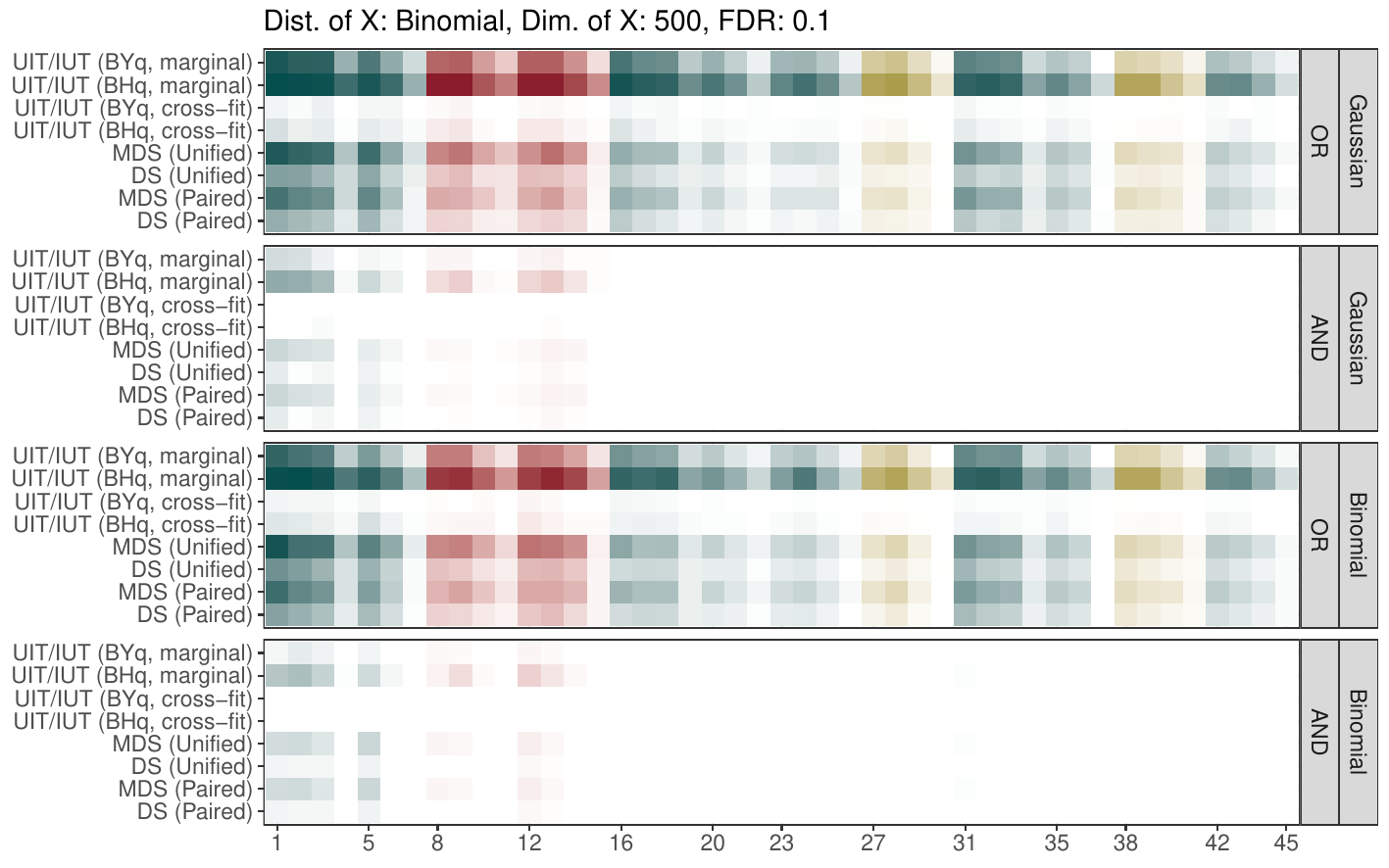}
    \caption{Selection rates over 100 simulations for each true predictor in the high-dimensional binomial setting ($p=500$). The parameters were estimated using cross-fitting for the mirror statistics, and the designated FDR is 0.1.}
\end{figure}

\begin{figure}[ht]
    \centering
    \includegraphics[width=\textwidth]{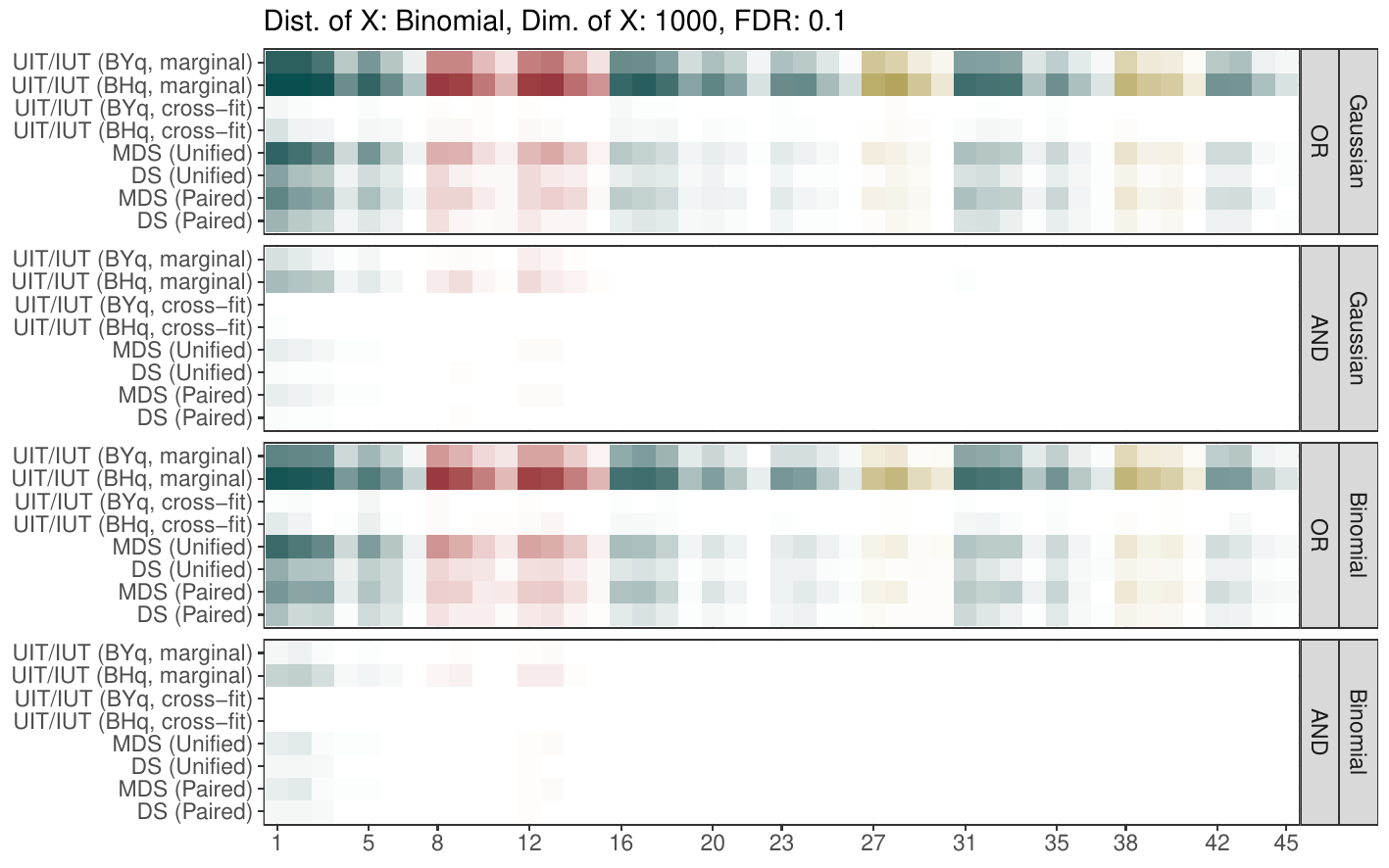}
    \caption{Selection rates over 100 simulations for each true predictor in the high-dimensional binomial setting ($p=1000$). The parameters were estimated using cross-fitting for the mirror statistics, and the designated FDR is 0.1.}
\end{figure}

\begin{figure}[ht]
    \centering
    \includegraphics[width=\textwidth]{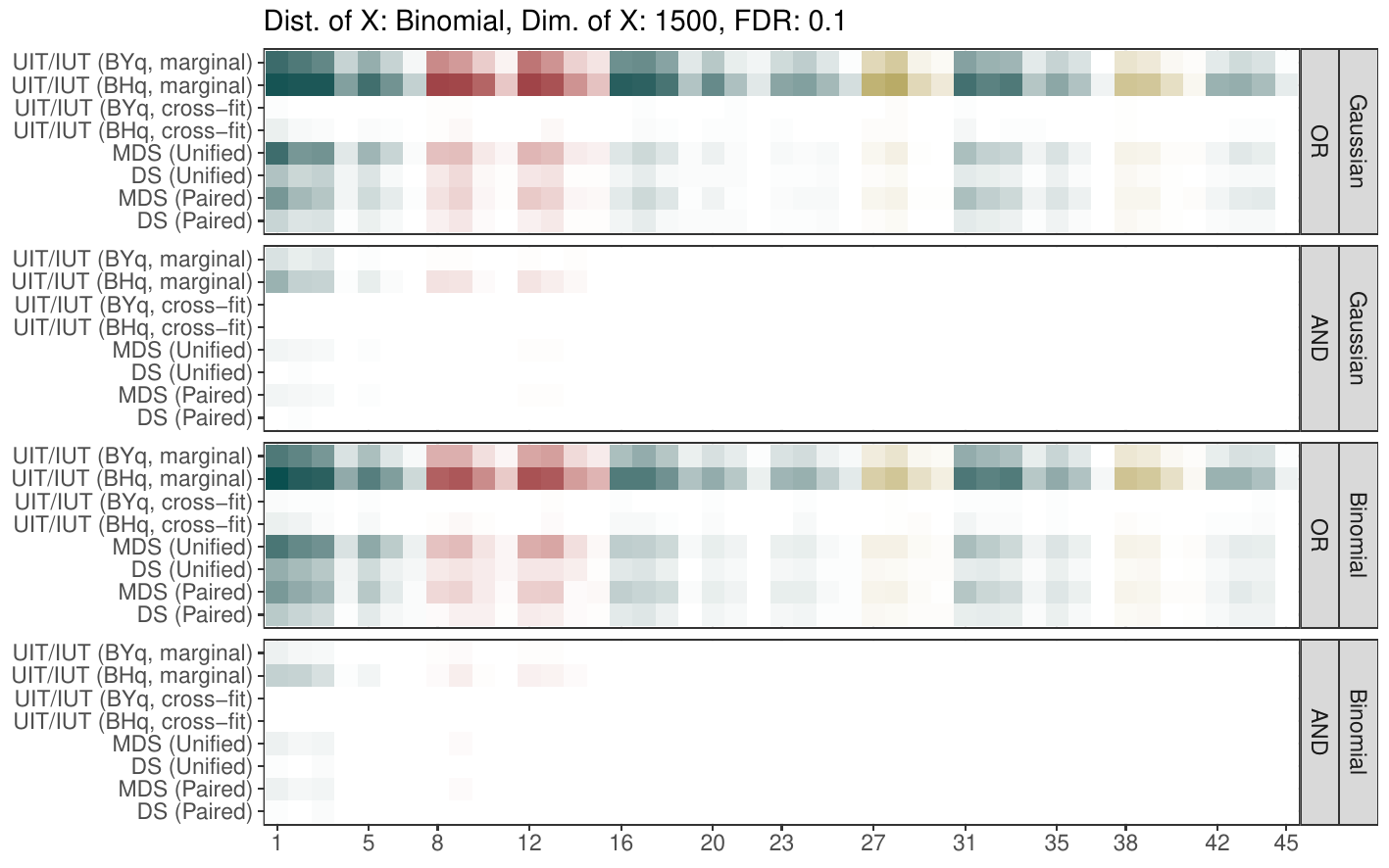}
    \caption{Selection rates over 100 simulations for each true predictor in the high-dimensional binomial setting ($p=1500$). The parameters were estimated using cross-fitting for the mirror statistics, and the designated FDR is 0.1.}
\end{figure}

\begin{figure}[ht]
    \centering
    \includegraphics[width=\textwidth]{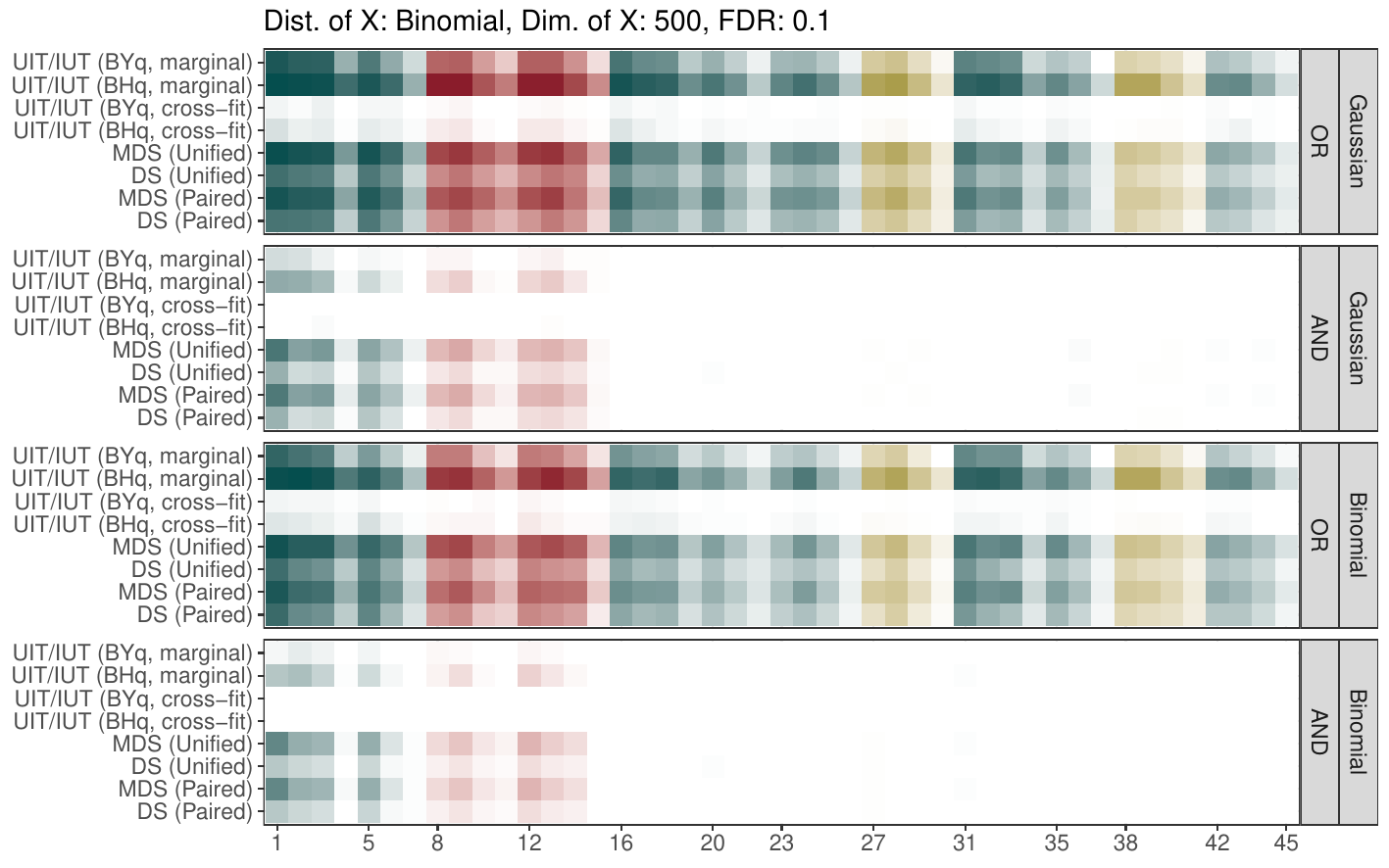}
    \caption{Selection rates over 100 simulations for each true predictor in the high-dimensional binomial setting ($p=500$). The parameters were estimated using the lasso for the mirror statistics, and the designated FDR is 0.1.}
\end{figure}

\begin{figure}[ht]
    \centering
    \includegraphics[width=\textwidth]{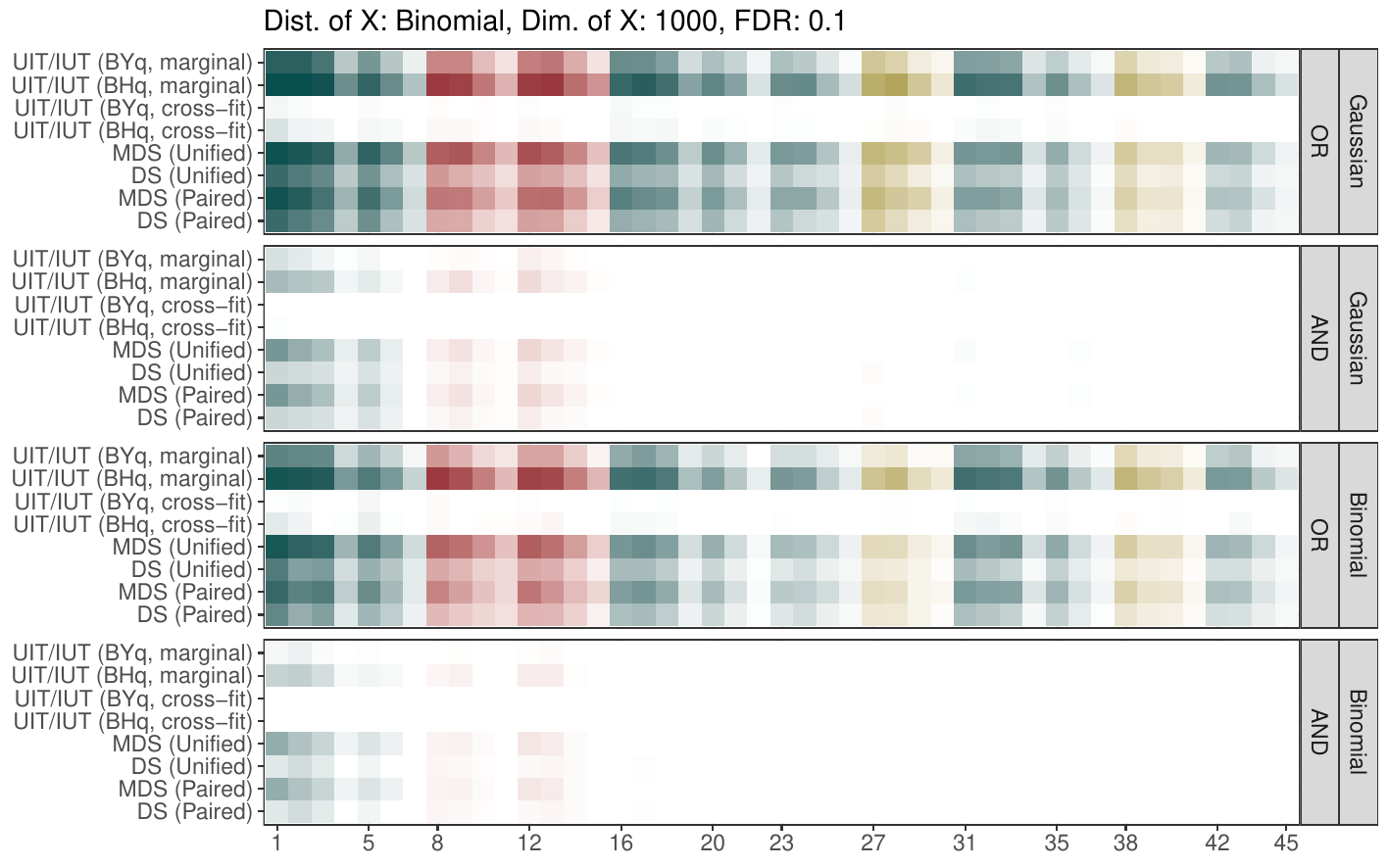}
    \caption{Selection rates over 100 simulations for each true predictor in the high-dimensional binomial setting ($p=1000$). The parameters were estimated using the lasso for the mirror statistics, and the designated FDR is 0.1.}
\end{figure}

\begin{figure}[ht]
    \centering
    \includegraphics[width=\textwidth]{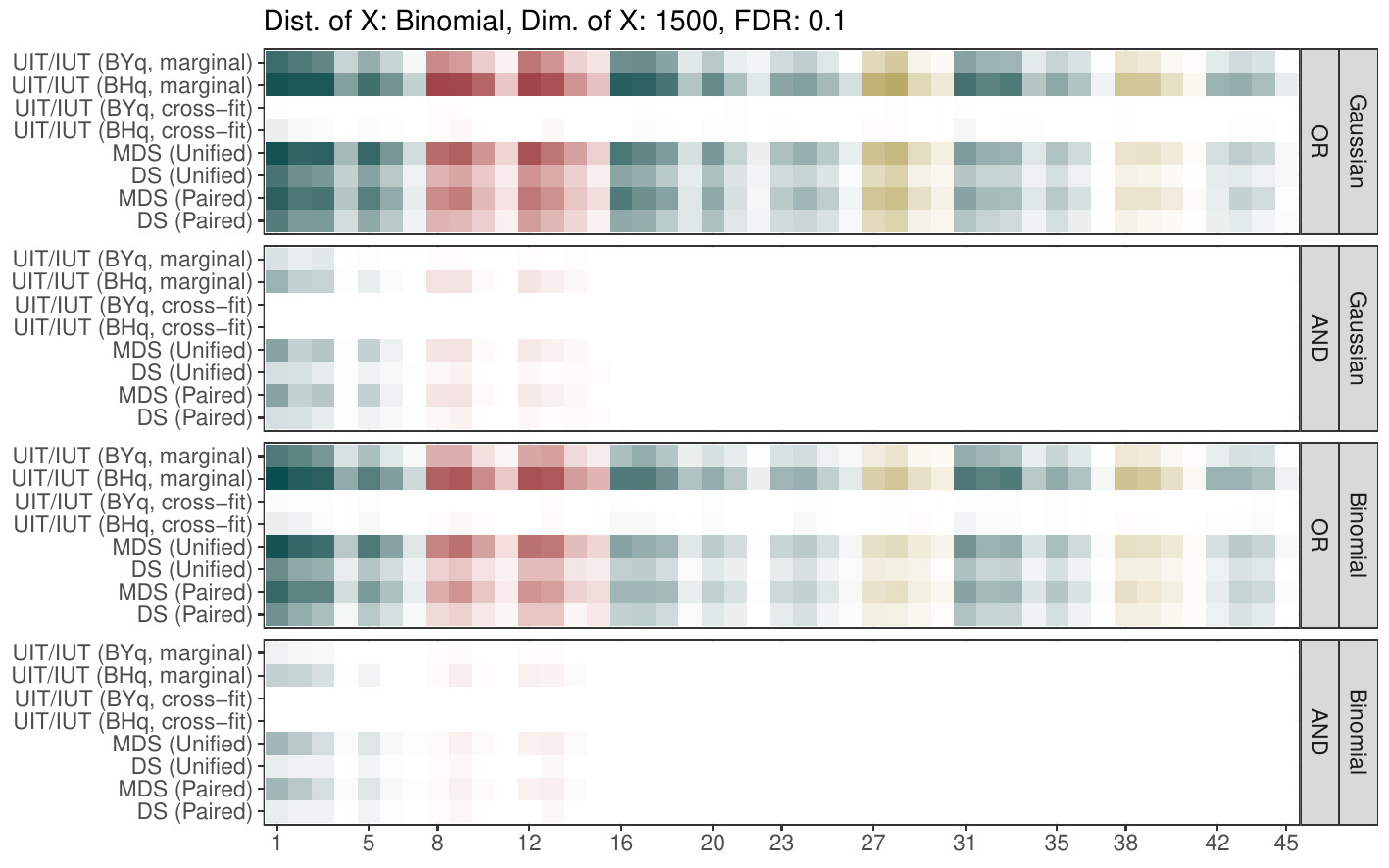}
    \caption{Selection rates over 100 simulations for each true predictor in the high-dimensional binomial setting ($p=1500$). The parameters were estimated using the lasso for the mirror statistics, and the designated FDR is 0.1.}
\end{figure}

\end{document}